\documentclass[reprint,amsmath,amssymb,aps,prb,superscriptaddress,floatfix,nofootinbib]{revtex4-2}

\pdfoutput=1
\usepackage{graphicx}
\usepackage{dcolumn}
\usepackage{bm}
\usepackage{braket}
\usepackage[mathlines]{lineno}
\usepackage{stackengine}
\usepackage{verbatim}
\usepackage{graphics}
\usepackage{hyperref}
\usepackage{xcolor}
\usepackage{appendix}

\newcommand{\rev}[1]{{\color{black}#1}}
\usepackage{soul}
\usepackage{ulem} 

\begin{document}
\preprint{APS/123-QED}
\title{Flat-band-based multifractality in the all-bands-flat diamond chain}

\author{Aamna Ahmed}
\affiliation{Department of Physics, Indian Institute of Science Education and Research, Bhopal, Madhya Pradesh 462066, India\\}
\author{Ajith Ramachandran}
\affiliation{Department of Physics, Christ College, Irinjalakuda, Kerala 680125, India\\}
\author{\rev{Ivan M. Khaymovich}}%
\affiliation{\rev{Nordita, Stockholm University and KTH Royal Institute of Technology Hannes Alfv{\'e}ns v{\"a}g 12, SE-106 91 Stockholm, Sweden}\\}
\author{Auditya Sharma}%
\affiliation{Department of Physics, Indian Institute of Science Education and Research, Bhopal, Madhya Pradesh 462066, India\\}

\date{\today}


\begin{abstract}
We study the effect of quasiperiodic Aubry-Andr\'e disorder on the
energy spectrum and eigenstates of a one-dimensional all-bands-flat
(ABF) diamond chain. The ABF diamond chain possesses three
dispersionless flat bands with all the eigenstates compactly localized
on two unit cells in the zero disorder limit. The fate of the compact
localized states in the presence of the disorder depends on the
symmetry of the applied potential. \rev{We consider two
  cases here: a symmetric one, where the same disorder is applied to
  the top and bottom sites of a unit cell and an antisymmetric one,
  where the disorder applied to the top and bottom sites are of equal
  magnitude but with opposite signs.} Remarkably, the symmetrically
perturbed lattice preserves compact localization, although the
degeneracy is lifted. When the lattice is perturbed antisymmetrically,
not only is the degeneracy is lifted but compact localization is also
destroyed. Fascinatingly, all eigenstates exhibit a multifractal
nature below a critical strength of the applied potential. A central
band of eigenstates continue to display an extended yet non-ergodic
behaviour for arbitrarily large strengths of the potential. All other
eigenstates exhibit the familiar Anderson localization above the
critical potential strength. \rev{We show how the
  antisymmetric disordered model can be mapped to a $\frac{\pi}{4}$
  rotated square lattice with nearest and selective next-nearest
  neighbour hopping and a staggered magnetic field - such models have
  been shown to exhibit multifractality.} Surprisingly, the
antisymmetric disorder (with an even number of unit cells) preserves
chiral symmetry - we show this by explicitly writing down the chiral
operator.
\end{abstract}

\maketitle


\section{INTRODUCTION}
Highly degenerate dispersionless or
\textit{flat band}
(FB)~\cite{PARAMESWARAN2013816,doi:10.1142/S021797921330017X,intro1,intro2,intro3}
systems, which support \textit{compact localized eigenstates}
(CLS)~\cite{intro4,Sathe_2021} have been of great interest over
the last decade. The localization properties
and associated repressed transport have been discussed in the context
of engineered lattices in one, two and three dimensions, such as diamond
~\cite{PhysRevLett.85.3906,PhysRevLett.88.227005,PhysRevLett.121.075502}, cross-stitch~\cite{cross1,cross2,cross3},
dice~\cite{dice1,dice2,dice3,dice4}, honeycomb~\cite{honey1,honey2},
kagome~\cite{kagome1,kagome2,kagome3} and pyrochlore
lattices~\cite{PhysRevB.86.241111,PhysRevB.99.235118}. The compact
localized states have been experimentally found to exist in a range of
setups such as Hubbard model systems~\cite{hubbard1,hubbard2},
photonic systems~\cite{Guzm_n_Silva_2014,photonic3}, exciton-polariton
condensates~\cite{kagome1} and Bose-Einstein condensates~\cite{BE1}.

In the most familiar type of localization, namely Anderson
localization~\cite{PhysRevLett.42.673, RevModPhys.80.1355}, which is
induced by on-site disorder, the `spread' of a state dies down
exponentially with a well-defined notion of a characteristic
localization length~\cite{PhysRev.109.1492,RevModPhys.57.287}. Compact
localization, in contrast, is much stronger with the span restricted
strictly to a few unit cells, with zero probability amplitude
elsewhere. Paradoxically the interplay of both these types of strong
localization results in a drop in localization. When a tiny amount of
uniform disorder is turned on in a compactly localized all-bands-flat
(ABF) diamond chain, the eigenstates exhibit an extremely weak
\textit{flat-band-based
localization}~\cite{PhysRevResearch.2.043395}. The disorder, in fact,
facilitates the hybridization of the large-scale degenerate network of
compact localized eigenstates, and we see weak localization that is on
the cusp of delocalization. In this work, we show that when the ABF
chain is subjected instead to a quasiperiodic Aubry-Andr\'e ($AA$)
disorder~\cite{AA,Modugno_2009}, the eigenstates are in fact extended
but non-ergodic thus exhibiting multifractality, a delicate phenomenon
that has attracted a wave of interest in recent
times~\cite{PhysRevB.82.104209,Kravtsov_2015,PhysRevE.98.032139,Facoetti_2016,Truong_2016,10.21468/SciPostPhys.6.1.014,Monthus_2017,Amini_2017,Monthus_2017,https://doi.org/10.48550/arxiv.2002.02979,PhysRevResearch.2.043346,10.21468/SciPostPhys.11.2.045,PhysRevB.100.195143,PhysRevLett.123.025301,Voliotis_2019,PhysRevB.96.104201,PhysRevB.99.155121,PhysRevLett.126.080602,PhysRevB.99.104203,PhysRevA.105.033716}. The
familiar $1$-D Aubry-Andr\'e Harper (AAH) tight-binding
model~\cite{PhysRevE.70.066203, PhysRevB.34.7367,
  PhysRevB.100.195143}, which is endowed with
self-duality~\cite{AA,PhysRevB.28.4272} exhibits
multifractality~\cite{RevModPhys.80.1355,Castellani_1986} only at the metal-insulator
transition that occurs at a critical potential strength. Here, we
report a robust \textit{flat-band-based multifractality}(FBM) that is
seen in an extensive region of the phase diagram.

Moreover, we find that the symmetry of the applied external potential
is
crucial~\cite{intro5,Leykam2017,Poli_2017,https://doi.org/10.48550/arxiv.2204.05198,PhysRevB.99.224208,https://doi.org/10.48550/arxiv.2112.09700,https://doi.org/10.48550/arxiv.2112.05066,10.21468/SciPostPhys.11.6.101,PhysRevB.99.224208,https://doi.org/10.48550/arxiv.2112.09856,PhysRevLett.110.176403,PhysRevLett.110.146404,PhysRevB.93.104504}. A
symmetric disorder\rev{, where the same disorder is applied to the top
  and bottom sites of a unit cell,} causes a complete lift of
degeneracy; however, remarkably we find that the CLS remain
robust~\cite{fbl2}. Although all the eigenstates are compactly
localized over two unit cells, these cannot be obtained by translating
the coefficients by an integer number of unit cells since the disorder
breaks the translation symmetry. In contrast, when we apply the $AA$
potential in an antisymmetric manner, \rev{where the disorder applied
  to the top and bottom sites of a unit cell are of equal magnitude
  but opposite sign,} we find that both the degeneracy and compact
localization are destroyed. A tiny disorder hybridizes the different
compact localized states, resulting in flat-band-based
multifractality. When the strength of the disorder is higher than a
critical value (where inter-band hybridization becomes possible), we
recover conventional Anderson localization for all the eigenstates
except those in a central band. Another of our striking findings is
that when the lattice is perturbed anti-symmetrically, the chiral
symmetry of the Hamiltonian is left intact despite the presence of
disorder. We show this with an explicit construction of the chiral
symmetry operator that anticommutes with the Hamiltonian independent
of the strength of the disorder.

The paper is organized as follows. In Section~\ref{sec:level2}, the
ABF diamond chain, as well as the Aubry-Andr\'e potential, are
described. \rev{In Section~\ref{sec:level3}, we discuss the
  effects of applying the $AA$ potential in the symmetric
  configuration that leaves the compact localization robust.  In
  Section~\ref{sec:level4}, we discuss the antisymmetric application
  of disorder, which supports multifractal states.  In
  Section~\ref{sec:level5}, we present an analytic treatment in
  support of the numerical results discussed in
  Section~\ref{sec:level4}.  We then summarize our results in
  Section~\ref{sec:level6}. The details of the lattice transformations
  used in Section~\ref{sec:level5} and numerical analysis of a variety
  of complementary quantities in support of the main findings are
  presented in the Appendix.}
\section{Model}\label{sec:level2}

\rev{We consider the ABF diamond
  lattice~\cite{PhysRevResearch.2.043395}, where each unit cell $n$
  consists of three sites labelled by $u_n$ (up), $c_n$ (center) and
  $d_n$ (down) respectively. The corresponding single particle states
  are conveniently represented in Dirac notation as $\ket{u_n}$,
  $\ket{c_n}$ and $\ket{d_n}$ respectively. The Hamiltonian is given
  by:
\begin{widetext}
\begin{align}
\hat{H}=-J \sum_{n=1}^{N / 3}\left(-\ket{{u}_{n}}\bra{{c}_{n}}+\ket{{d}_{n}} \bra{{c}_{n}}+\ket{{c}_{n}} \bra{{u}_{n+1}}+\ket{{c}_{n}}\bra{{d}_{n+1}}+\text { H.c. }\right)+\sum_{n=1}^{N / 3}\left(\zeta_{n}^{u} \ket{{u}_{n}} \bra{{u}_{n}}+\zeta_{n}^{c} \ket{{c}_{n}} \bra{{c}_{n}}+\zeta_{n}^{d} \ket{{d}_{n}}\bra{{d}_{n}}\right).
\label{eq1}
\end{align}
\end{widetext}}

Since each unit cell has $3$ sites, the total number of sites denoted
by $N$ should be a multiple of $3$. All the energy terms are
represented in hopping amplitude $J$ units, assuming $J=1$ for
simplicity. For each site of the $n^{\text{th}}$ unit cell, we include
on-site energy using independent Aubry-Andr\'e potentials
\begin{equation}
\zeta_n^\alpha=\lambda_{\alpha}\cos(2\pi nb+\theta_p),
\label{eq2}
\end{equation}
where the strength of the potential is $\lambda_{\alpha}$ and
\rev{$\alpha$ can take the values $u$, $c$ and $d$.  The
  quasi-periodicity parameter $b$ must be an irrational number which
  we set to be the golden mean $(\sqrt{5}-1)/2$.} $\theta_p$ is an
arbitrary global phase chosen randomly from a uniform distribution in
the range $[0,2\pi]$. Also, periodic boundary conditions have been
assumed~\rev{\cite{fn1}}.

The ABF diamond chain is a particular case of the Hamiltonian for
which the on-site energies $\zeta_{n}$ are zero. This system possesses
three flat bands at energies $\pm 2$ and $0$ and no dispersive
bands. Consequently, the system possesses only compact localized
eigenstates. Hence, the system is highly degenerate and also a good
insulator. The Hamiltonian in the zero-disorder
limit possesses chiral symmetry~\cite{Topology,Ryu_2010,Ludwig_2015},
which is represented by a $k$-independent operator $\Gamma_0=
\text{diag}(1,1,-1)$. This operator is unitary since
$\Gamma_0^2=\mathbb{I}$ and $\Gamma_0 \mathcal{H}^{\dagger}(k) \Gamma_0^{-1}= -\mathcal{H}(k)$.  Since the
chiral operator $\Gamma_0$ anti-commutes with the Hamiltonian, for each
eigenvalue $E$ with eigenvector $\ket{\phi_E}$ the negative $-E$ is
also an eigenvalue with eigenvector $\Gamma_0\ket{\phi_E}$.

In upcoming sections, we introduce the quasiperiodic Aubry-Andr\'e potential
on the ABF diamond chain, and investigate how it affects the spectrum
and the compact localized states. Broadly there are two natural ways
in which the on-site energies on the up and down sites may be correlated:
a symmetric configuration in which $\zeta_{n}^{u}=\zeta_{n}^{d}$ and an
antisymmetric configuration in which $\zeta_{n}^{u}=-\zeta_{n}^{d}$.
Furthermore, we discuss two sub-cases within the symmetric setup: one in which the Aubry-Andr\'e
potential is applied on the $u$ and $d$ sites, and the second one in which
the Aubry-Andr\'e potential is applied on the $c$ sites while keeping the
$u$ and $d$ sites at zero potential.
%
\section{SYMMETRIC APPLICATION OF AUBRY-ANDR\'E POTENTIAL}\label{sec:level3}

 %

\begin{figure}[b]
\centering
\hspace{-0.6cm}\stackunder{\hspace{-3.0cm}(a)}{\includegraphics[height=3.5cm, width=4.6cm]{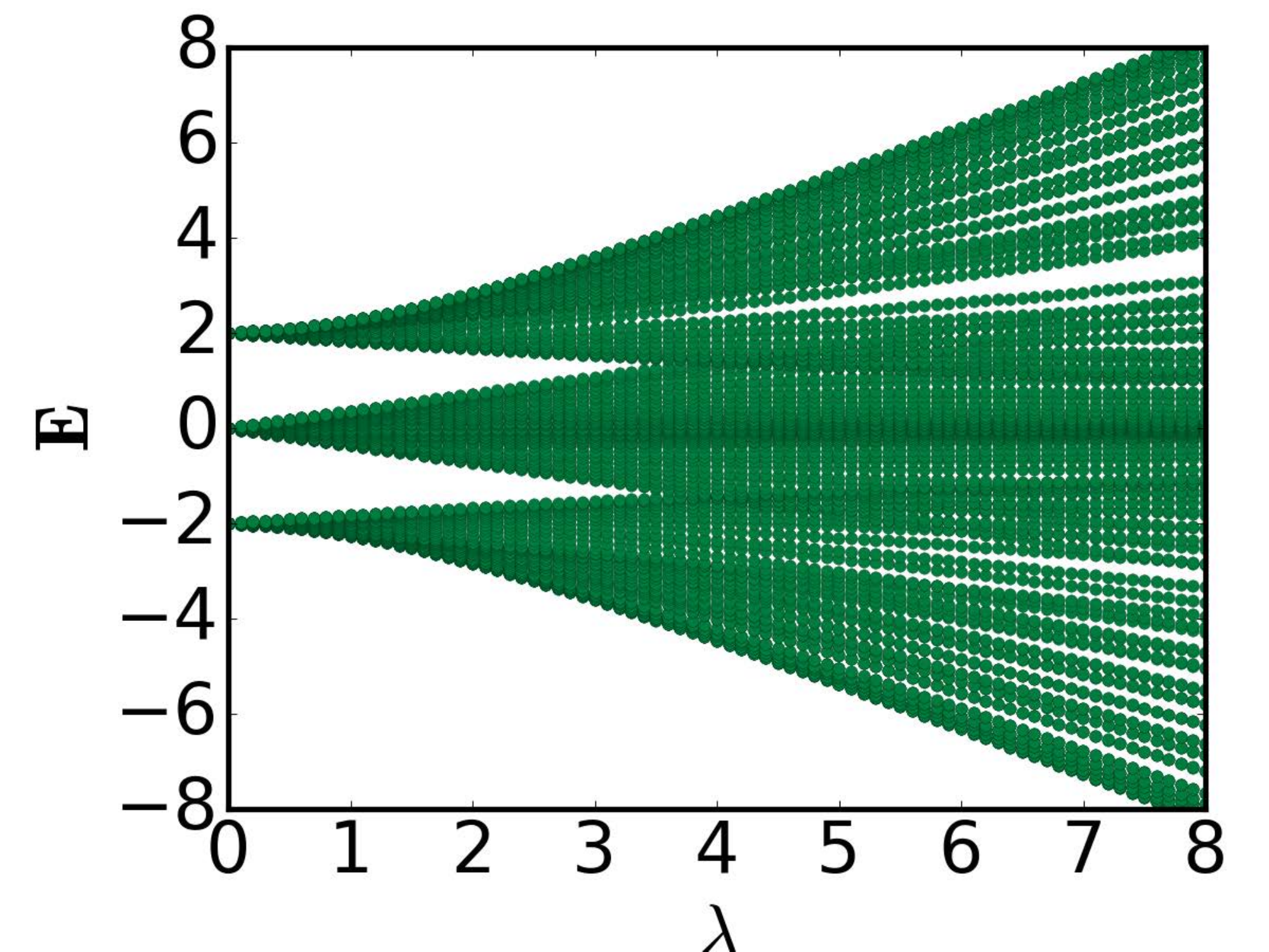}}\hspace{-0.2cm}
\stackunder{\hspace{-3.0cm}(b)}{\includegraphics[height=3.5cm, width=4.6cm]{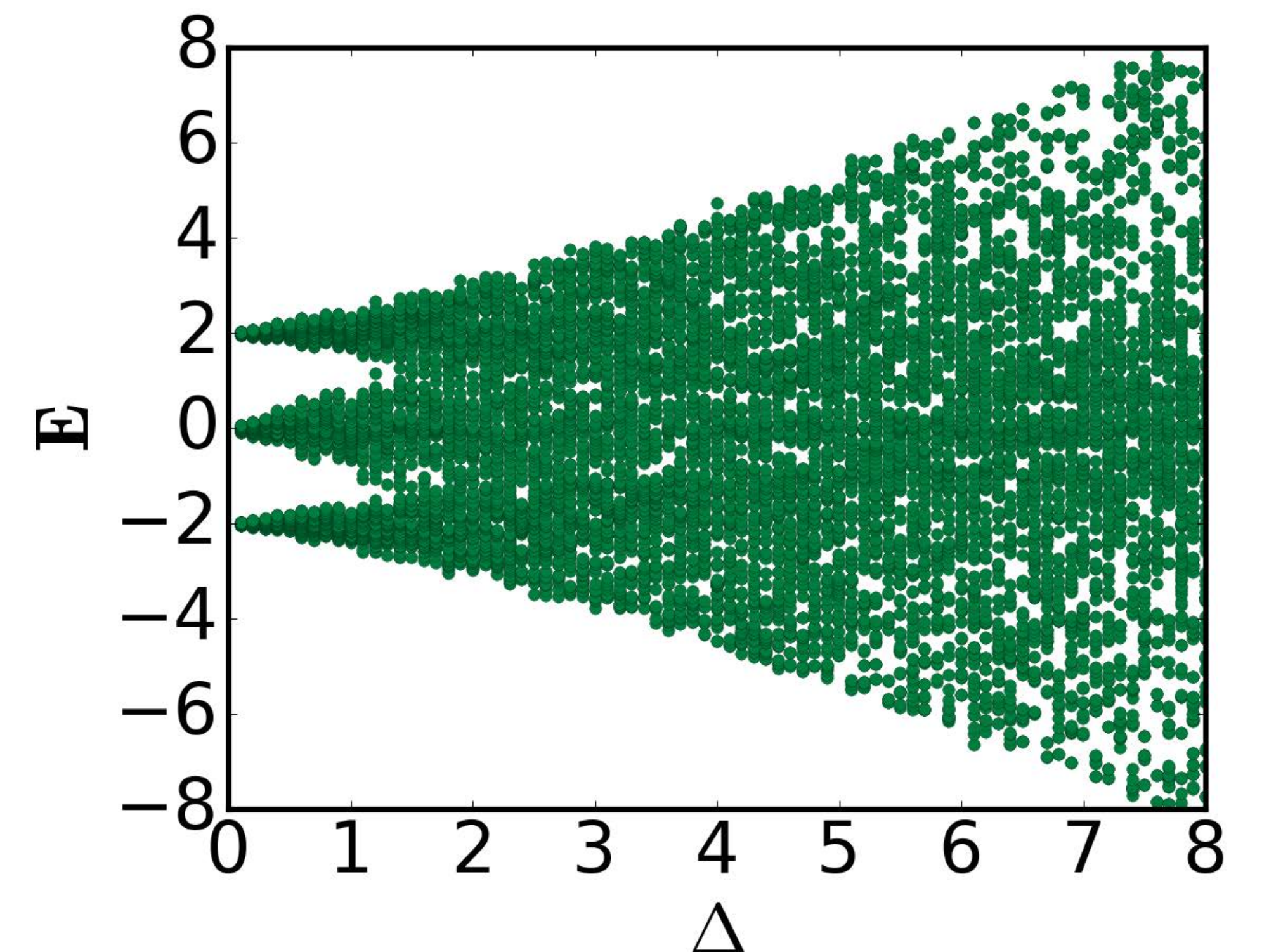}}

\stackunder{\hspace{-2.7cm}(c)}{\includegraphics[height=2.0cm, width=2.8cm]{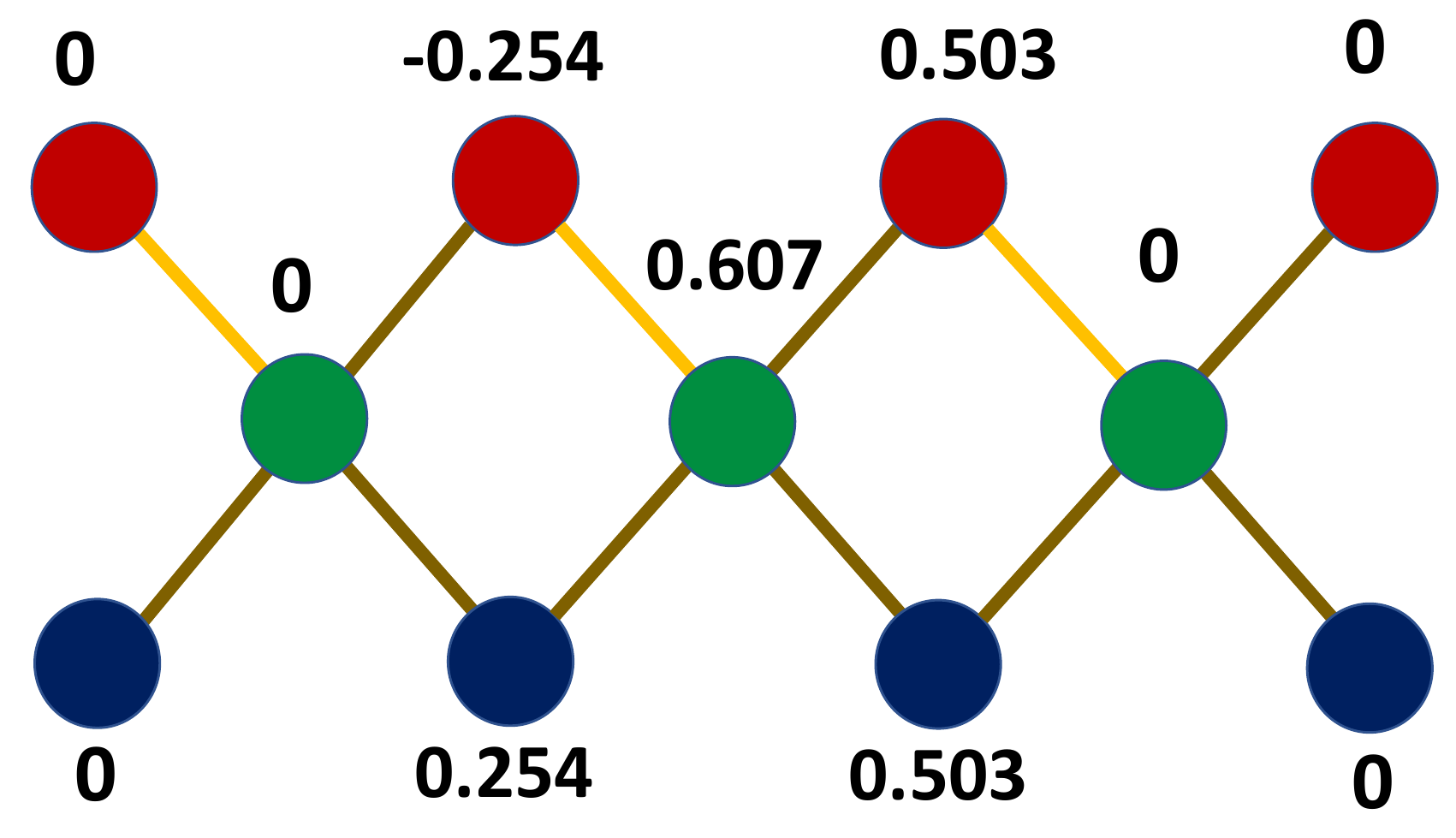}}\hspace{0.01cm}
\stackunder{\hspace{-2.7cm}(d)}{\includegraphics[height=2.0cm, width=2.8cm]{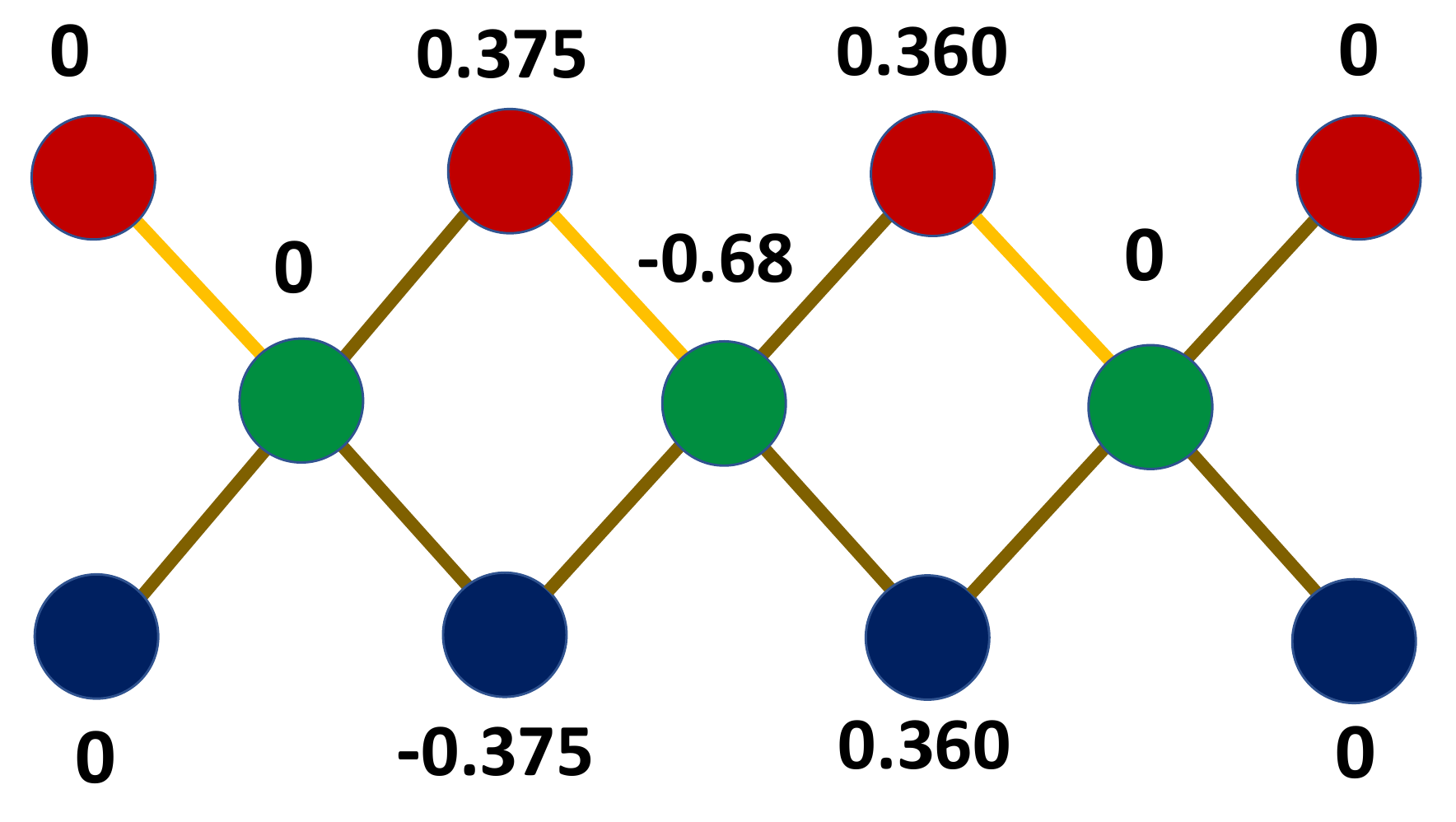}}\hspace{0.01cm}
\stackunder{\hspace{-2.7cm}(e)}{\includegraphics[height=2.0cm, width=2.8cm]{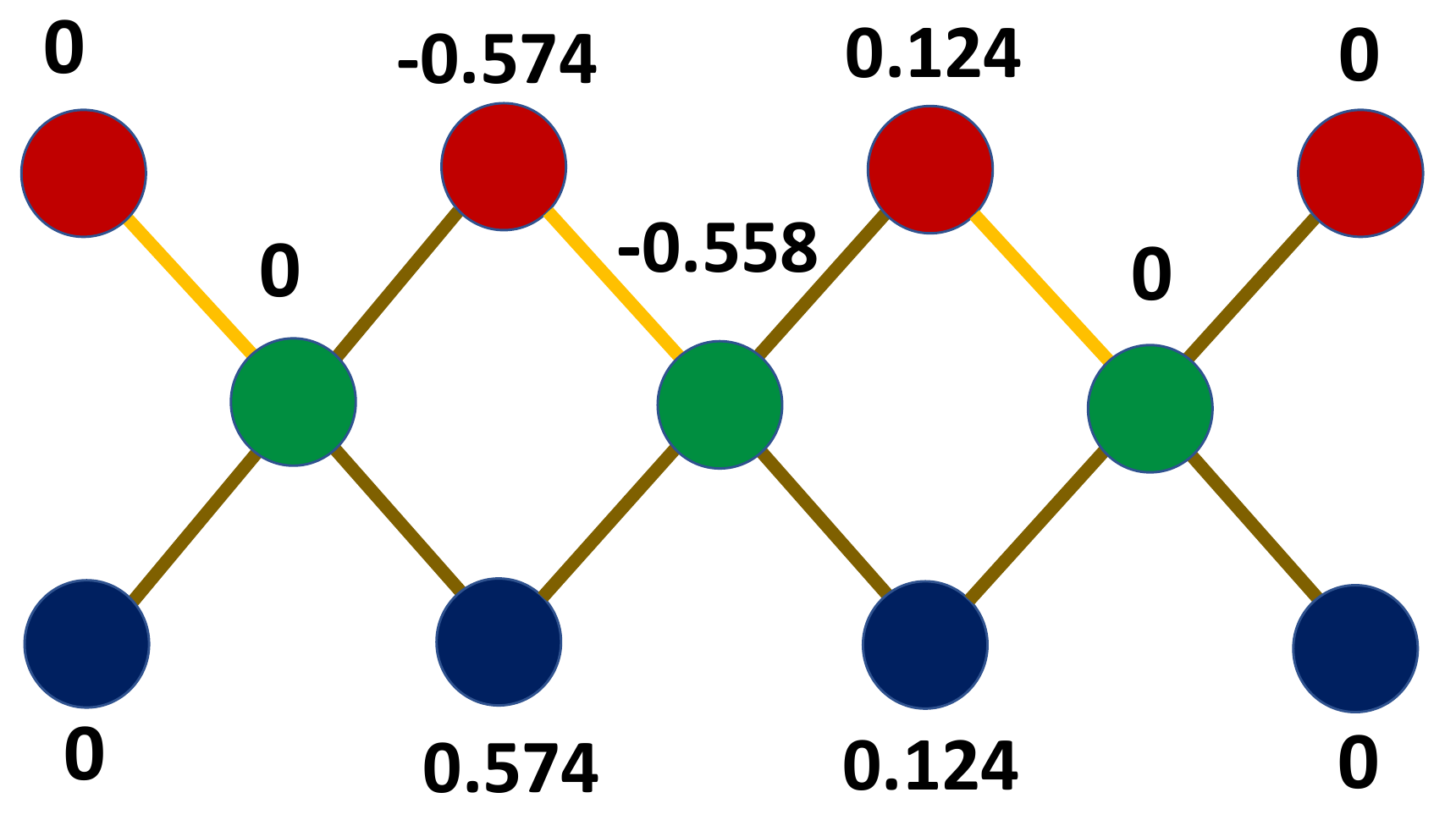}}
\caption{\label{fig1}The spectrum of the ABF diamond lattice in the
  symmetric case with (a)~increasing strength of the quasiperiodic
  potential $\lambda$ and \rev{(b)~uniform uncorrelated random
    disorder} with increasing strength $\Delta$ on the $u$ and $d$
  sites. Schematic representations of the diamond chain: \rev{compact
    localized states} from the (c) lower band \rev{($E=-2.49$)}, (d)
  middle band \rev{($E=-0.025$)} and (e) upper band \rev{($E=
    2.51$)} at $\lambda=2$ (the amplitudes are obtained
  numerically). The system size is $N=126$.}
\end{figure}

\subsection{AA potential on the $u$ and $d$ sites}
\rev{First, we} consider the symmetric configuration:
\begin{equation}
\zeta_{n}^{u}=\zeta_{n}^{d} \qquad \text{and} \qquad \zeta_{n}^{c}=0.
\label{eq3}
\end{equation}
The introduction of the $AA$ potential
i.e. $\zeta_{n}^{u}=\lambda\cos(2\pi n b+\theta_p)$ lifts the
degeneracy of all the flat bands, as can be seen from
Fig.~\ref{fig1}(a). The eigenstates of the Hamiltonian are found to
reside on two unit cells even in the presence of the $AA$ potential
(see Fig.~\ref{fig1}(c)--\ref{fig1}(e)). Remarkably, the compactness
of the eigenstates is preserved at higher strengths of the potential,
even after all the bands mix and the system exhibits a single-band
energy spectrum. Also, although the spectrum appears to be symmetric
\rev{with respect to} $E=0$, a closer look reveals that this is not quite
true. There is no requirement that every energy $E$ comes with its
negative counterpart $-E$ since the disorder breaks the chiral symmetry.

\rev{Interestingly, as shown in Fig.~\ref{fig1}(b), when
  the applied potential is drawn from a uniform uncorrelated random
  distribution $[-\Delta,\Delta]$, the results are similar (to those
  with quasiperiodic disorder of the same strength $\lambda$).}  We
conclude that the presence of the compact localized states is a
consequence of the symmetric application of the disorder and not the
details of the applied potential. There exists a useful transformation
to a new lattice (see Section~\ref{app:level2}), which in the
disorder-free limit, takes the system to a set of completely uncoupled
sites. From Fig.~\ref{fig8}(a), we observe that the transformation
results in a lattice made of three-site unit cells but with an absence
of inter-cell hopping, indicating the preservation of the CLSs (see
Section~\ref{app:subsec1}).

%
\begin{figure}
\centering
\hspace{-0.6cm}\stackunder{\hspace{-3.0cm}(a)}{\includegraphics[height=3.5cm, width=4.6cm]{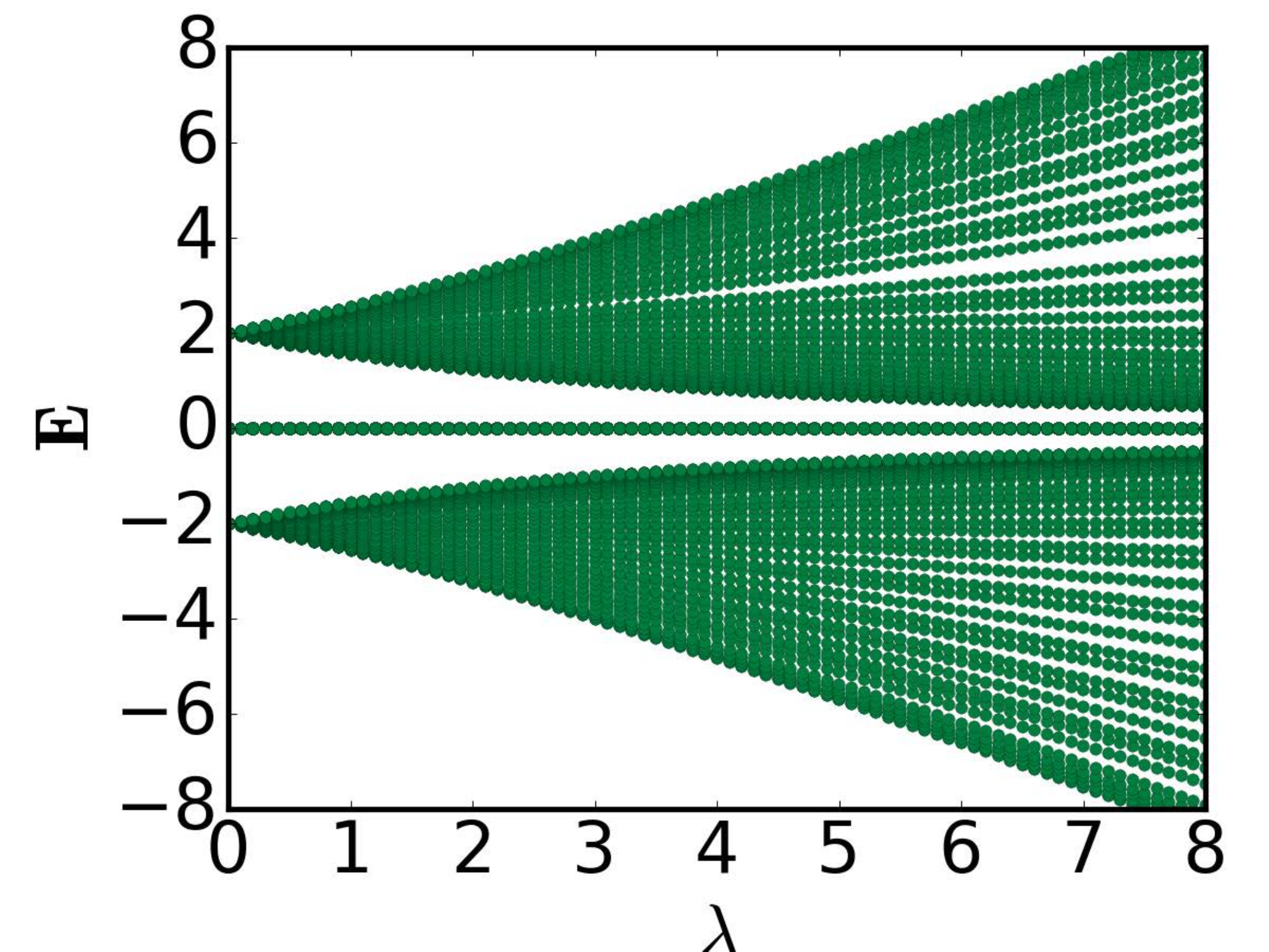}}\hspace{-0.2cm}
\stackunder{\hspace{-3.0cm}(b)}{\includegraphics[height=3.5cm, width=4.6cm]{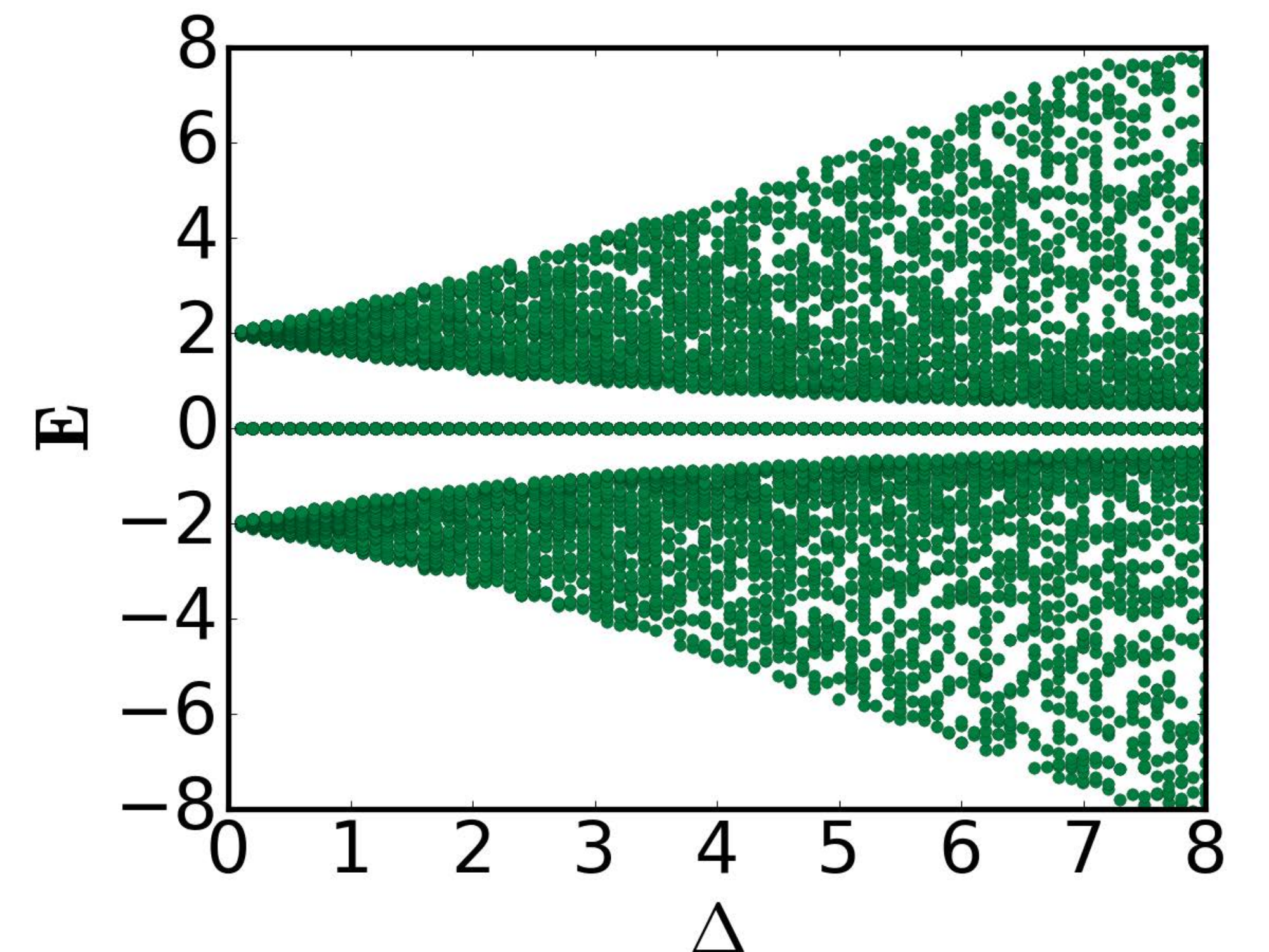}}

\stackunder{\hspace{-2.5cm}(c)}{\includegraphics[height=2.0cm, width=2.8cm]{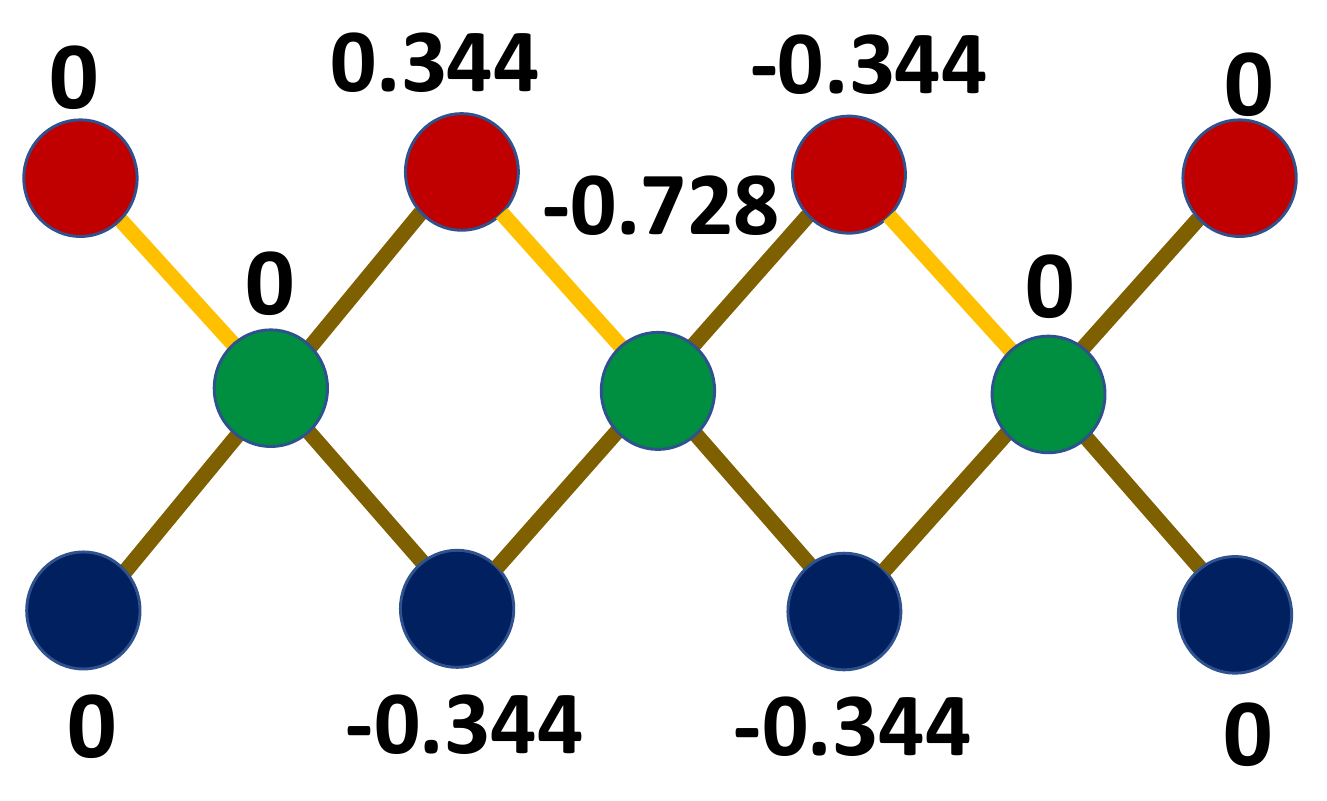}}\hspace{0.01cm}
\stackunder{\hspace{-2.5cm}(d)}{\includegraphics[height=2.0cm, width=2.8cm]{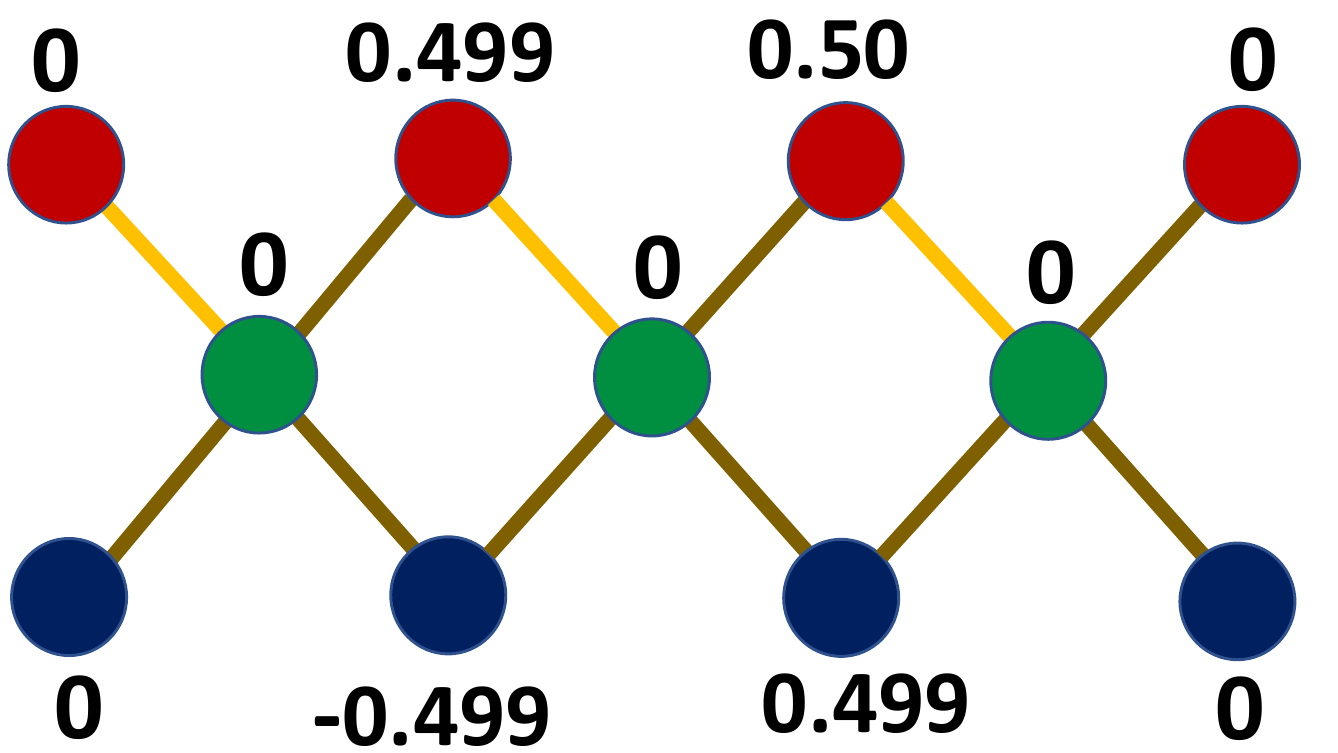}}\hspace{0.01cm}
\stackunder{\hspace{-2.5cm}(e)}{\includegraphics[height=2.0cm, width=2.8cm]{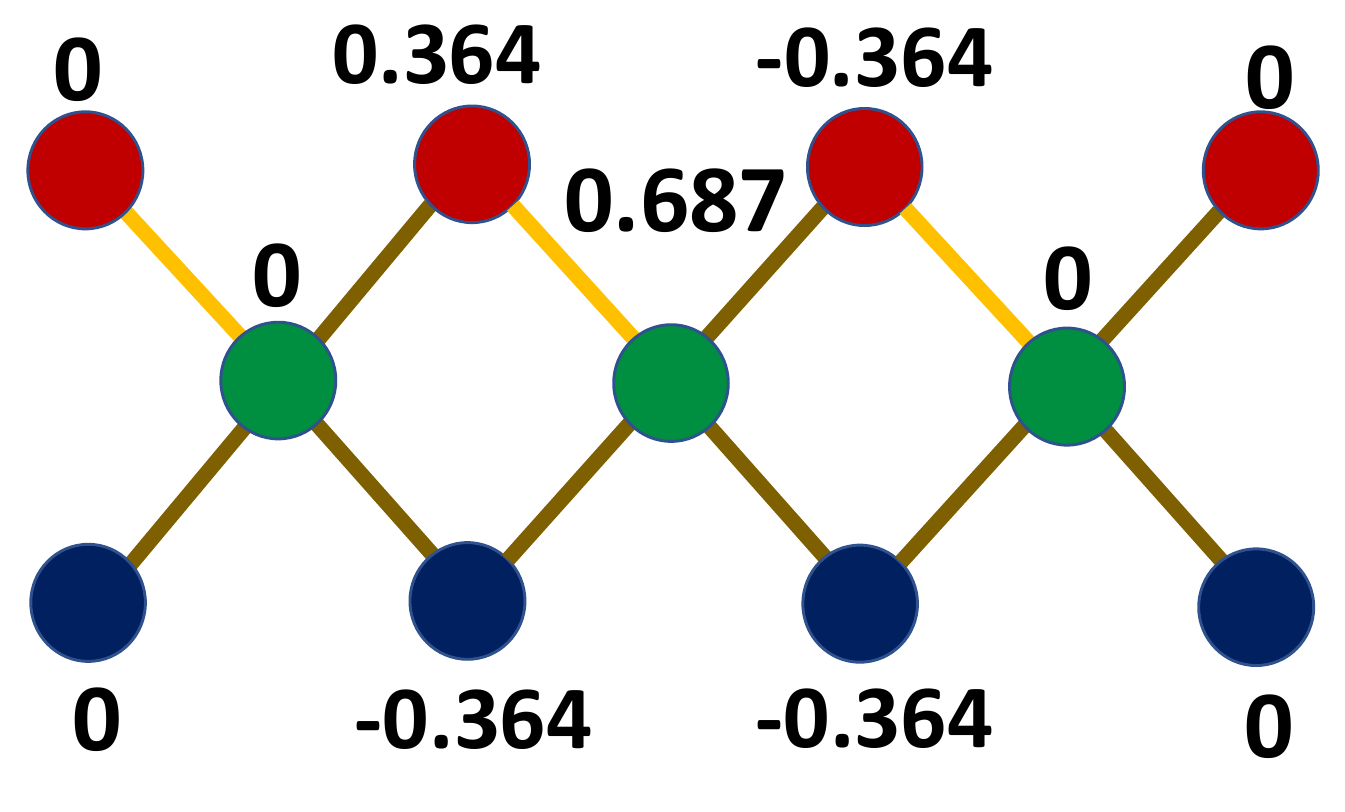}}
\caption{\label{fig2}The spectrum of the ABF diamond lattice with
  (a)~$AA$ potential with increasing quasiperiodic strength $\lambda$
  and \rev{(b)~uniform uncorrelated random disorder} with increasing
  strength $\Delta$ on the central $c$ sites. Schematic
  representations of the diamond chain with the $AA$ potential only on
  the $c$ sites. CLSs for the (c)~lower band \rev{($E=-2.12$)},
  (d)~middle band \rev{($E=0$)} (e)~upper band \rev{($E=1.89$)} at
  $\lambda=5$. The system size is $N=126$.}
\end{figure}
%
\subsection{AA potential on the $c$ sites}

 \rev{Next, we} consider the case when disorder is introduced only on the $c$ sites i.e.
\begin{equation}
\zeta_{n}^{u}=\zeta_{n}^{d}=0 \quad \text{and} \quad \zeta_{n}^{c}\neq 0.
\end{equation}
This is a special type of symmetric configuration. The energy spectrum
with increasing strength of potential has been plotted for both the
$AA$ potential i.e. $\zeta_{n}^{c}=\lambda\cos(2\pi nb+\theta_p)$ (see
Fig.~\ref{fig2}(a)) and for the uniform disorder case (see
Fig.~\ref{fig2}(b)) i.e. $\zeta_{n}^{c}$ drawn uniformly from
$[-\Delta,\Delta]$. We observe that in both the cases, the
degeneracies of the eigenstates are broken for the upper and lower
bands, while the flatband at $E=0$ remains robust even at higher
disorder strengths. Further, the associated eigenstates preserve
compact localization in both cases. We have numerically verified the
same, as shown in the case of $AA$ potential with $\lambda=5$ in
Figs.~\ref{fig2}(c)--\ref{fig2}(e).  With the help of the
transformation discussed in Section~\ref{app:level2}, we show that the
lattice is made of three-site unit cells but with an absence of
inter-cell hopping, indicating the preservation of the CLSs (see
Fig.~\ref{fig8}(b)). Additionally, each unit cell has a decoupled site
which strengthens the observation of a robust flatband even in the
presence of the disorder.

\section{Antisymmetric APPLICATION OF AUBRY-ANDR\'E POTENTIAL}\label{sec:level4}
\rev{As a further step, we} consider the impact of the application of
the $AA$ potential in an antisymmetric manner, defined by
\begin{equation}
\zeta_{n}^{u}=-\zeta_{n}^{d} = \lambda\cos(2\pi n b+\theta_p) \quad \text{and} \quad \zeta_{n}^{c}=0.
\end{equation}
 \rev{Here we observe that the introduction of the tinest of disorder results in the loss of compact localization of the energy eigenstates. Hence} we explore the localization
characteristics as a function of the disorder strength $\lambda$
with the aid of several measures: inverse
participation ratio (IPR), multifractal dimensions and level spacings.

\begin{figure*}
\centering
\stackunder{\hspace{-5.0cm}(a)}{\includegraphics[height=4.3cm, width=6.1cm]{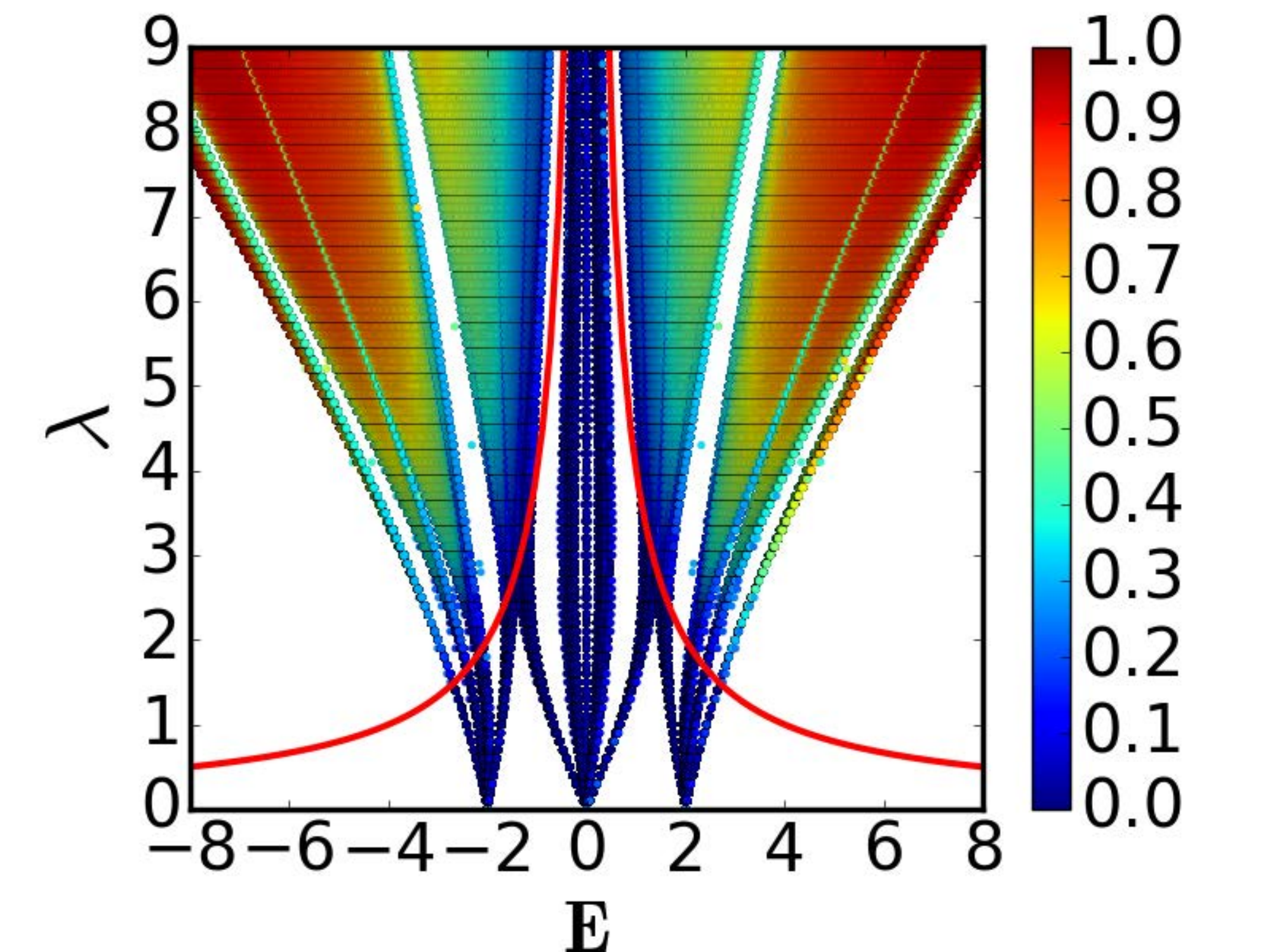}}\hspace{-4mm}
\stackunder{\hspace{-5.0cm}(b)}{\includegraphics[height=4.5cm, width=5.8cm]{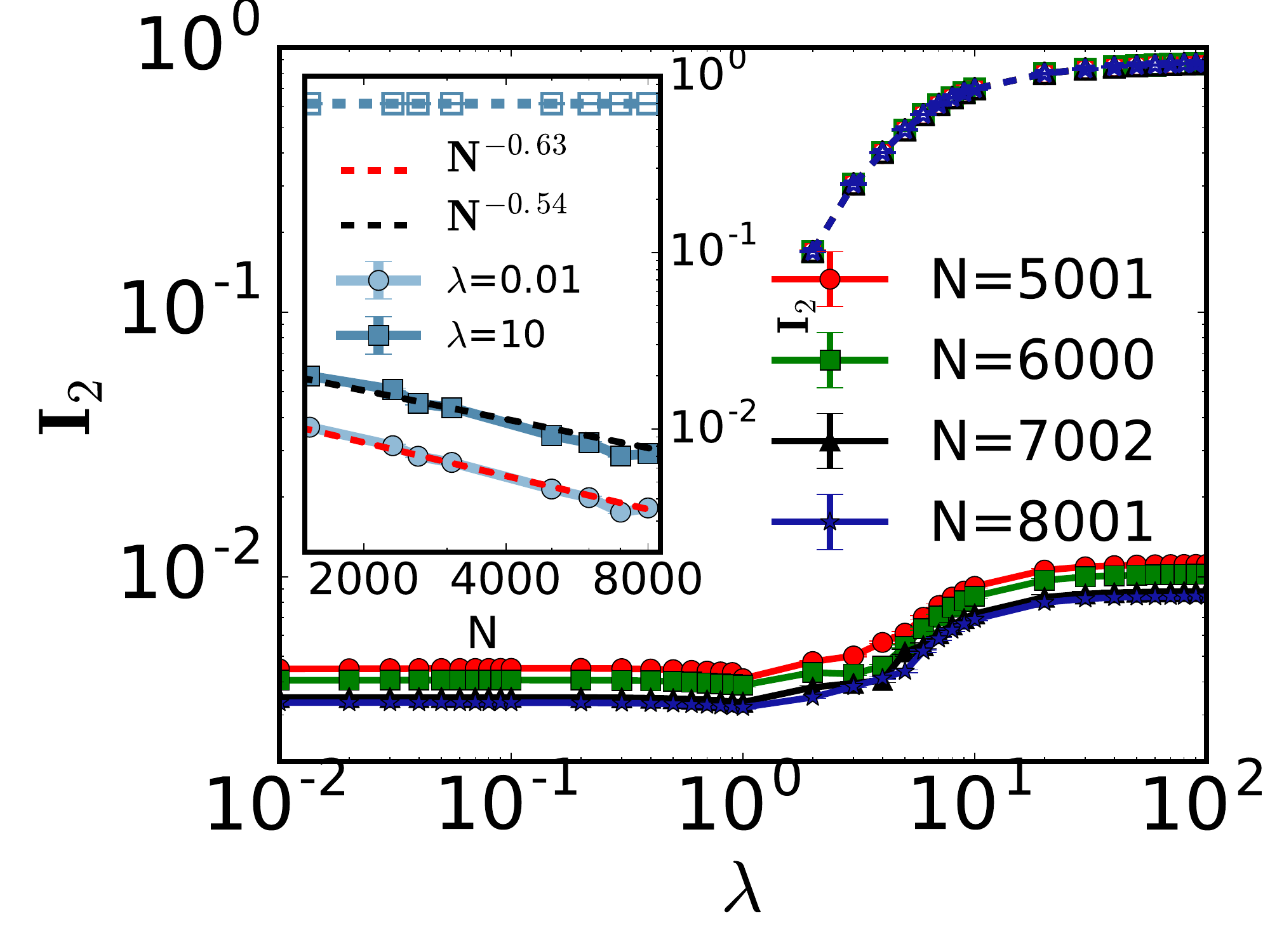}}\hspace{-1.0mm}
\stackunder{\hspace{-5.0cm}(c)}{\includegraphics[height=4.3cm, width=6.1cm]{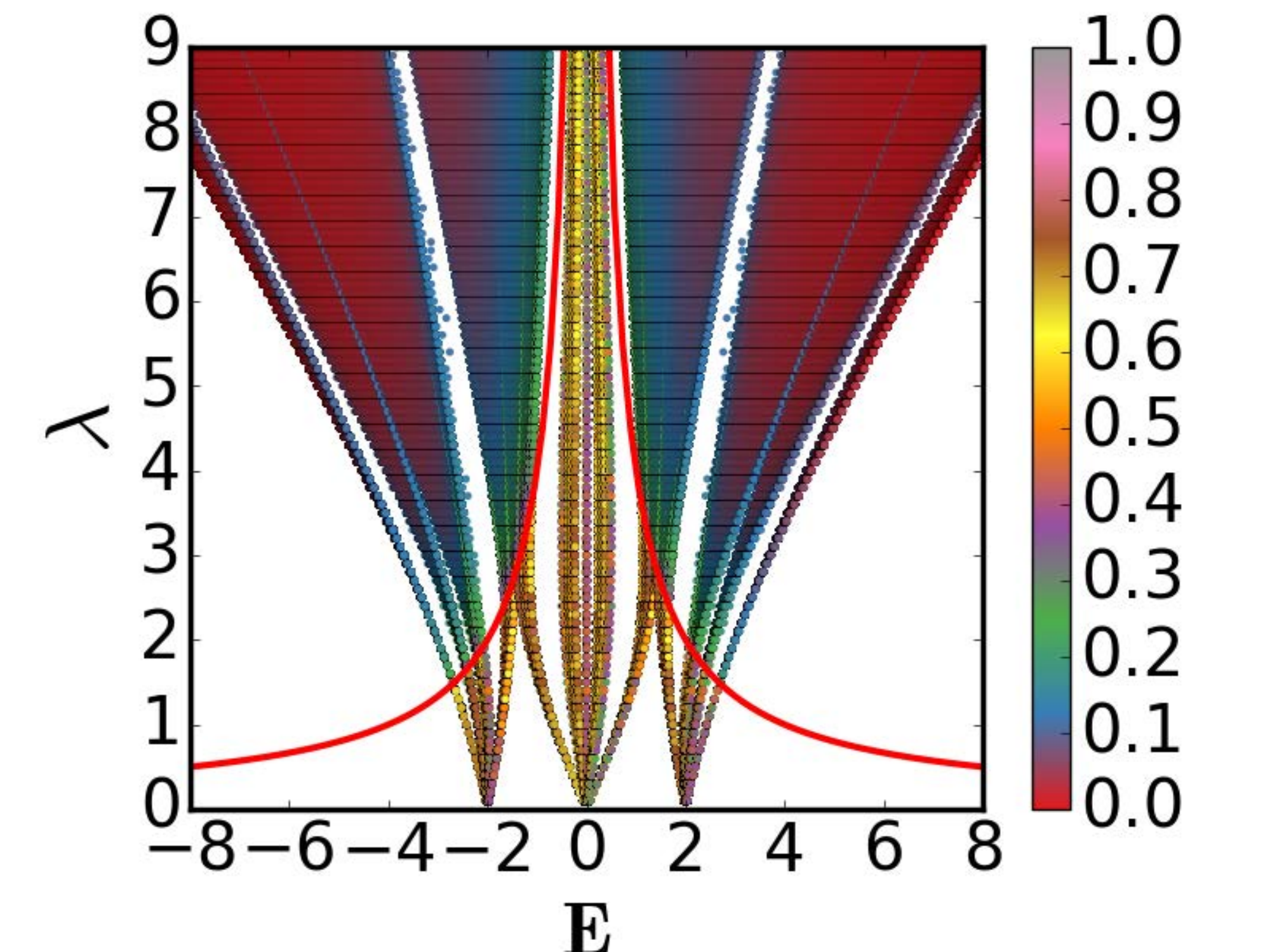}}
\caption{\label{fig3}\rev{$I_2$ and $D_2$ in the anti-symmetric
    case. (a) The spectrum as a function of increasing strength
    $\lambda$, where the colour denotes the value of $I_2$. The system
    size is $N = 6000$.  (b) $I_2$ averaged over the eigenstates with
    increasing strength of $AA$ potential $\lambda$ for various system
    sizes averaged over $50$ values of $\theta_p$. Inset of (b) shows
    the scaling of $\left\langle I_2 \right\rangle$ with system size
    for $\lambda=0.01$ and $10$. Here the fitting for the multifractal
    states at $\lambda=0.01$ ($\left\langle I_2
    \right\rangle_{0.01}\sim N^{-0.63}$) is shown with red color and
    the fitting for the states at $\lambda=10$ ($\left\langle I_2
    \right\rangle_{10}\sim N^{-0.54}$) is shown with black color. (c)
    The spectrum as a function of increasing strength $\lambda$, where
    the colour denotes the value of the fractal dimension $D_2$,
    defined in Eq.~\eqref{eq6}, for all the single particle
    eigenstates. The system size is $N = 6000$. The red solid line in
    panels (a) and (c) given by $\lambda = 4/|E|$ shows the transition
    between multifractal and localized states, conjectured
    in~\cite{https://doi.org/10.48550/arxiv.2208.11930} from the
    analogy to the extended Harper problem (see Sec~\ref{sec:level5}
    for more details). In (b) $I_2$ averaged over the eigenstates
    drawn from the inner and outer regions separated by the fractal
    mobility edge are shown separately. Solid lines with filled
    symbols correspond to states in the inner region and dashed lines
    with open symbols to states in the outer region. We observe that
    the open symbols corresponding to different system sizes overlap
    indicating that these states are localized. }}
\end{figure*}


The localization characteristics of the eigenstates can be understood
with the help of the inverse participation ratio (IPR), which is
defined as:
\rev{
\begin{equation}
\rev{I_{2}}=\sum_{n=1}^{\frac{N}{3}} \sum_{\alpha=u, c, d}\left|\psi_{k}(\alpha_n)\right|^{4}
\end{equation}}
where the $k^\text{th}$ normalized single-particle eigenstate
\rev{$\left|\psi_{k}\right\rangle=$ $\sum_{n, \alpha}
  \psi_{k}( \alpha_n)| \alpha_n\rangle$} is written in terms of the
Wannier basis\rev{ $| \alpha_n\rangle$}, representing the eigenstate of a
single particle localized at the site $\alpha$ $(\alpha=u, c, d)$ in
the $n^{th}$ unit cell of the lattice. For a completely localized
eigenstate $\rev{I_2} = O(1)$ (when the state is localized
on a few sites), while for a perfectly delocalized eigenstate
$\rev{I_2} =O(1) /N$. Figure~\ref{fig3}(a) shows the
disorder-averaged IPR as a function of the strength of the
antisymmetric $AA$ potential for the entire spectrum. The detuned
spectrum is confined within the limit $-2\lambda \leq E \leq2\lambda$,
with the bandwidth showing a roughly linear relation with
$\lambda$. The introduction of the potential lifts the degeneracy of
all the bands and also modifies the localization properties of the
eigenstates. We observe that all the eigenstates for
\rev{$\lambda \lesssim 1.5$} as well as those associated
with the central band for any strength of the potential are extended
while the remaining eigenstates show localization at higher strengths
of the potential.

\rev{It turns out that the central band of extended eigenstates are
  well-described by a fractal mobility edge $|E|<4/\lambda$
  conjectured in a recent
  preprint~\cite{https://doi.org/10.48550/arxiv.2208.11930}. Figure~\ref{fig3}(b)
  shows the IPR averaged over the eigenstates separated by the fractal
  mobility edge $\lambda =
  4/|E|$~\cite{https://doi.org/10.48550/arxiv.2208.11930}. The IPR of
  states constituting the inner section shows system-size dependence
  and is $\simeq 10^{-3}$, which is a signature of the extended nature
  of the eigenstates. However for the states comprising the outer
  section, IPR is independent of the system size and is close to
  unity, which is an indication of Anderson localization.}

\subsection{Multifractal analysis}
We also analyze a quantity called the fractal dimension
$D_q$~\cite{RevModPhys.80.1355,PhysRevLett.122.106603, PhysRevLett.123.180601} which is
defined as:
\begin{equation}
D_q=\frac{S_q}{\ln(N)},
\label{eq6}
\end{equation}
where $N$ is the dimension of the Hilbert space and $S_{q}$ are the
participation entropies obtained from the $k^\text{th}$ eigenstate\rev{
$\left|\psi_{k}\right\rangle=$ $\sum_{n, \alpha}
\psi_{k}(\alpha_n)|\alpha_n\rangle$} using the relation \rev{ $S_{q}=\frac{1}{1-q} \ln I_{q}$ where
\rev{
\begin{equation}
I_{q}=\sum_{n=1}^{\frac{N}{3}} \sum_{\alpha=u, c, d}\left|\psi_{k}( \alpha_n)\right|^{2 q}
\end{equation}}
are the $q^{\text{th}}$ order moments.}
\begin{figure}
\centering
\stackunder{\hspace{-4.5cm}(a)}{\includegraphics[height=4.2cm,width=6cm]{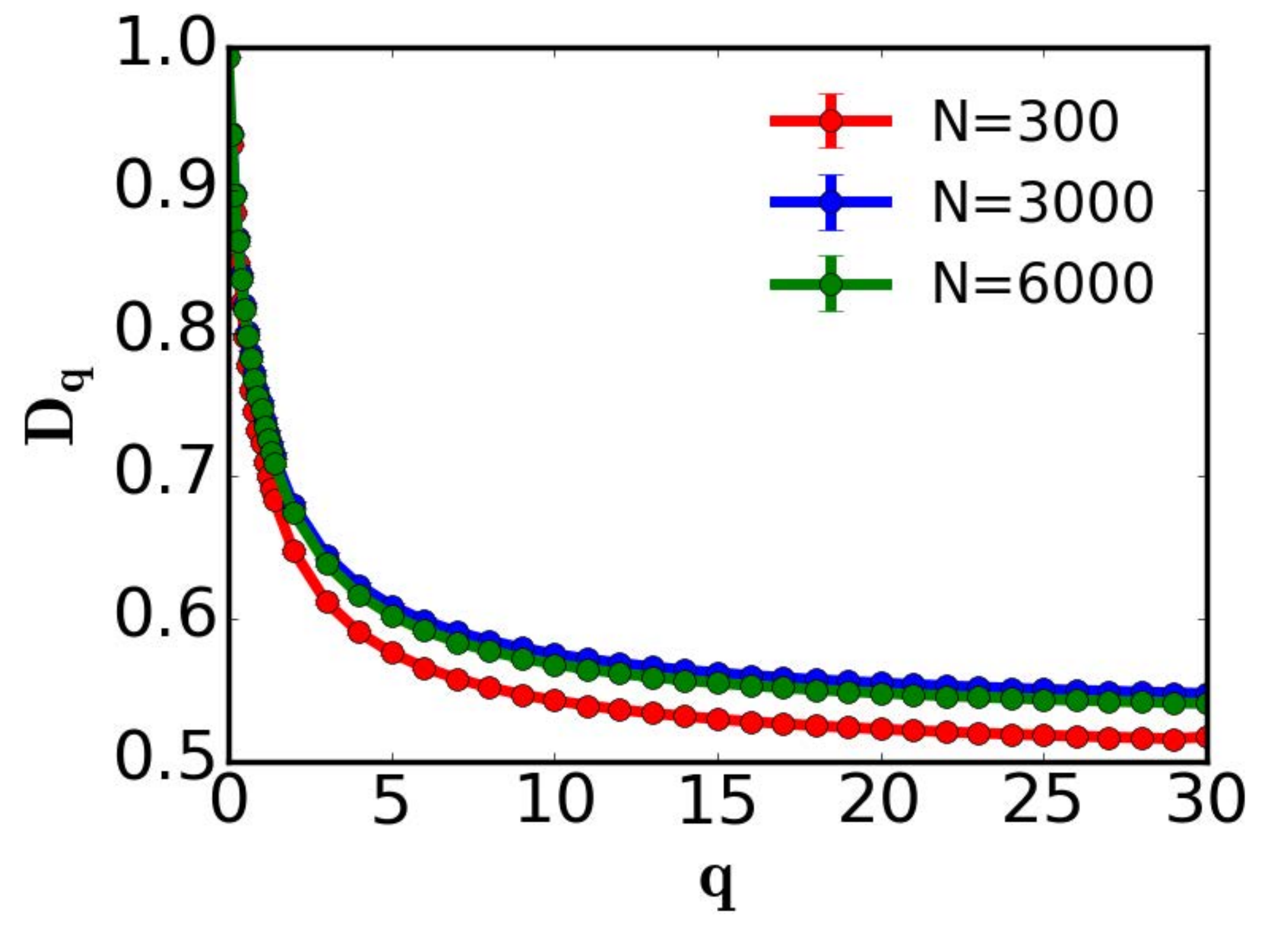}}
\vspace{-0.6cm}

\stackunder{\hspace{-4.5cm}(b)}{\includegraphics[height=4.2cm,width=6cm]{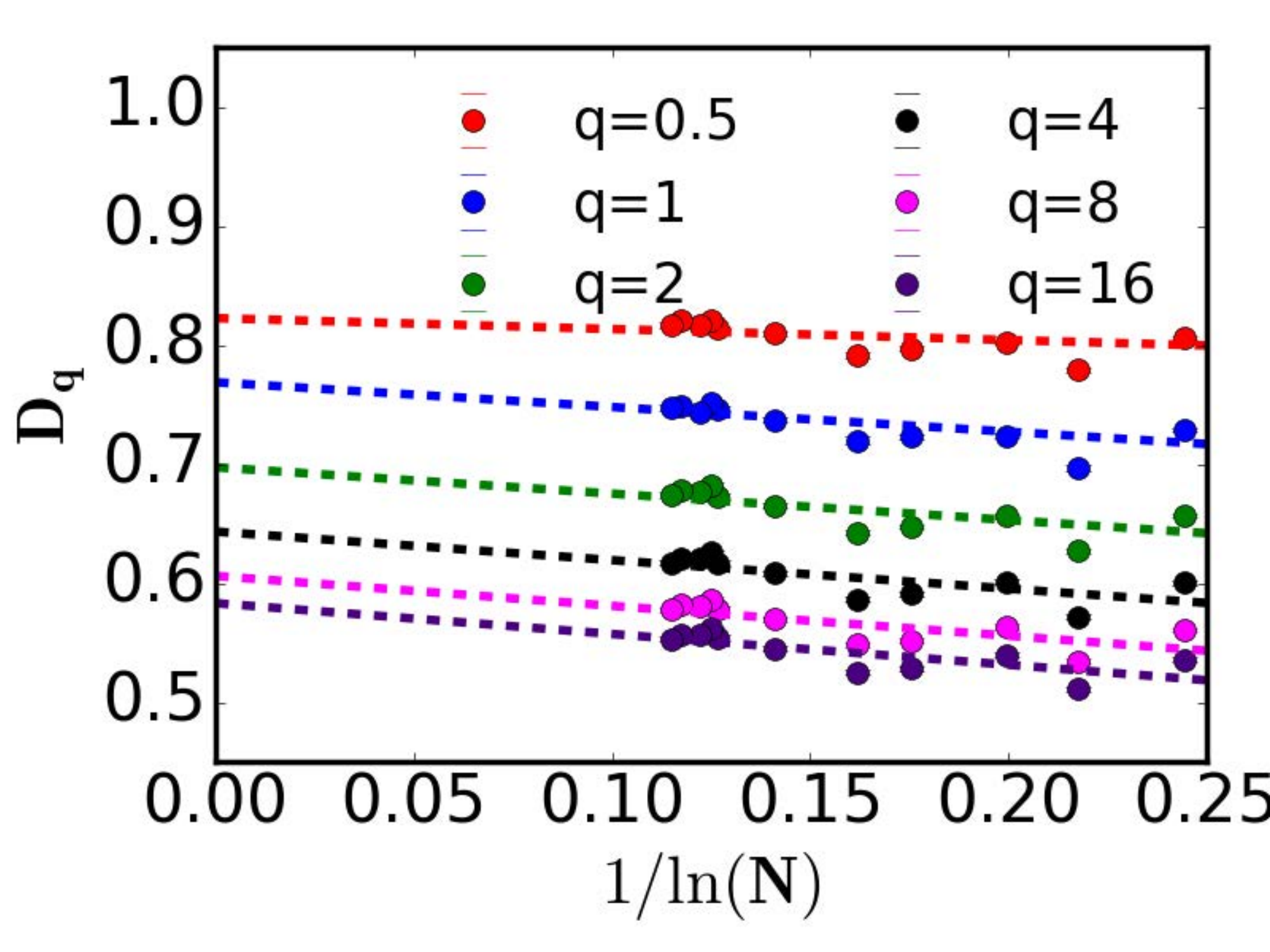}}
\caption{\label{fig4}In the antisymmetric case (a)~fractal dimension
  $D_q$ vs $q$ for various system sizes. (b)~Fractal dimension $D_q$
  vs $\frac{1}{\text{ln}(N)}$ for various $q$ values and system sizes
  ranging from $N=60$ to $N=6000$. Here quasiperiodic strength
  $\lambda=0.01$ and $D_q$ is averaged over all the eigenstates for at
  least $50$ values of $\theta_p$.}
\end{figure}
\begin{figure*}
\centering
\stackunder{\hspace{-4.5cm}(a)}{\includegraphics[height=4.0cm, width=5.7cm]{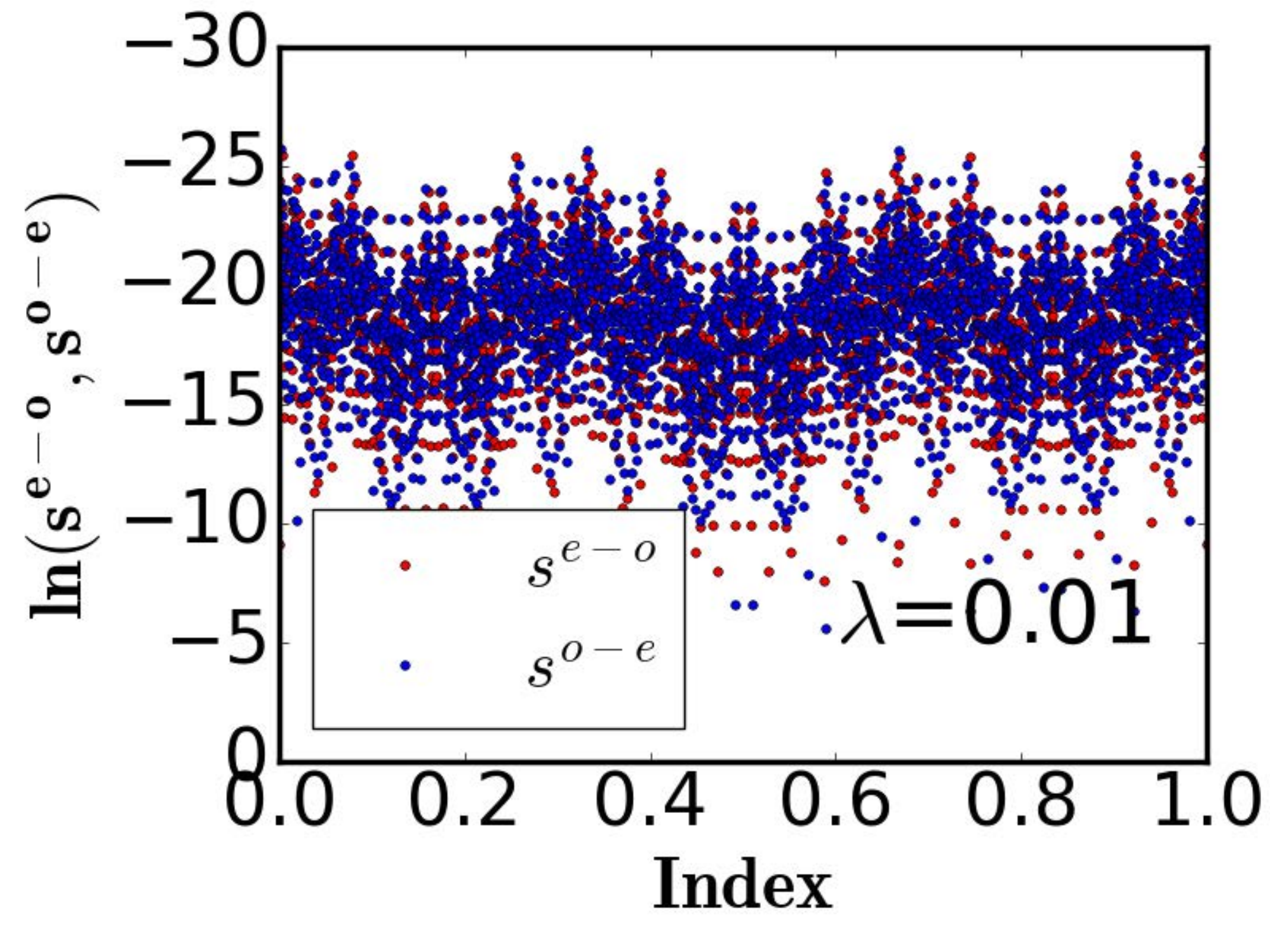}}\hspace{-0.1cm}
\stackunder{\hspace{-4.5cm}(b)}{\includegraphics[height=4.0cm, width=5.7cm]{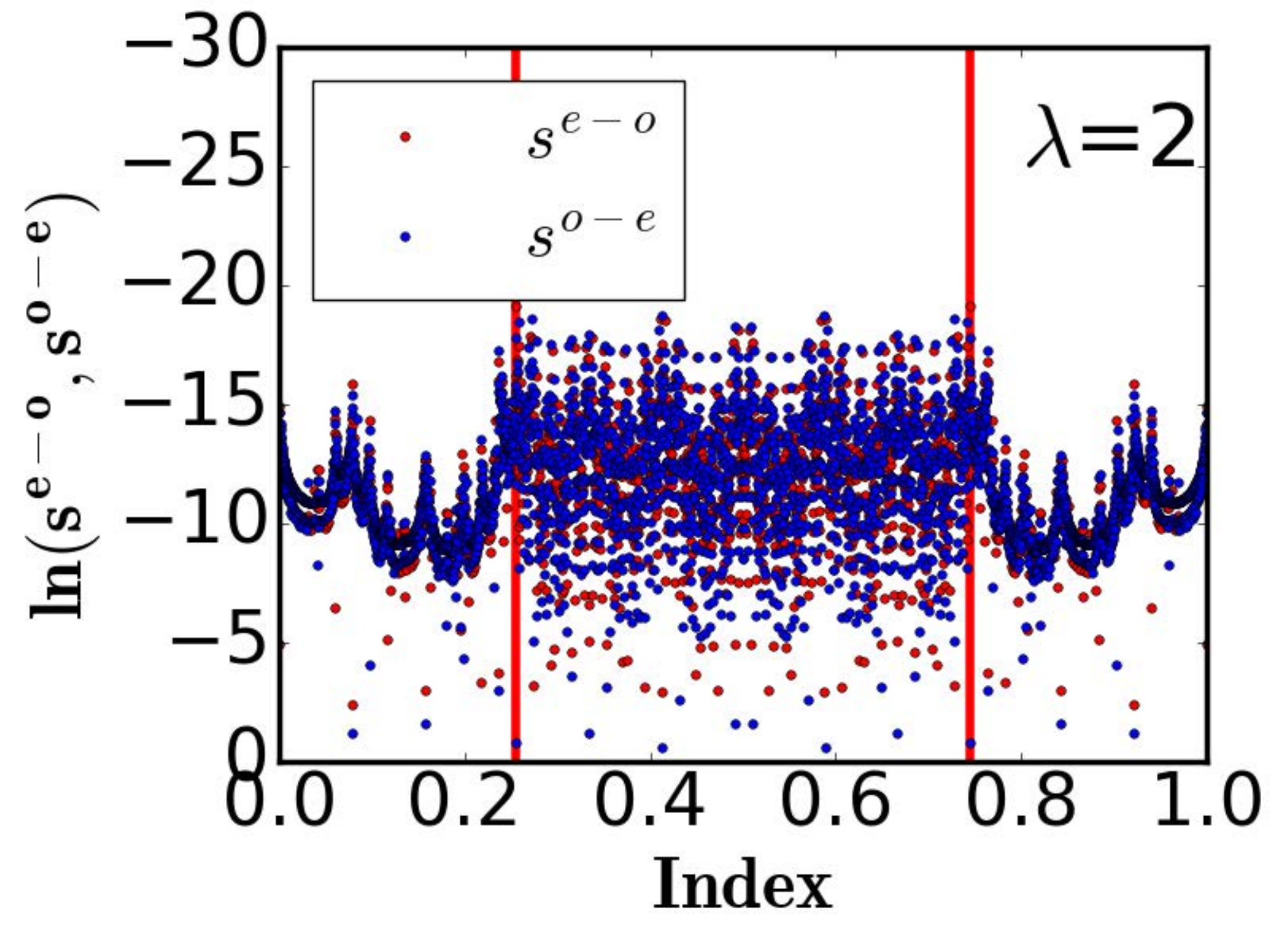}}\hspace{-0.1cm}
\stackunder{\hspace{-4.5cm}(c)}{\includegraphics[height=4.0cm, width=5.7cm]{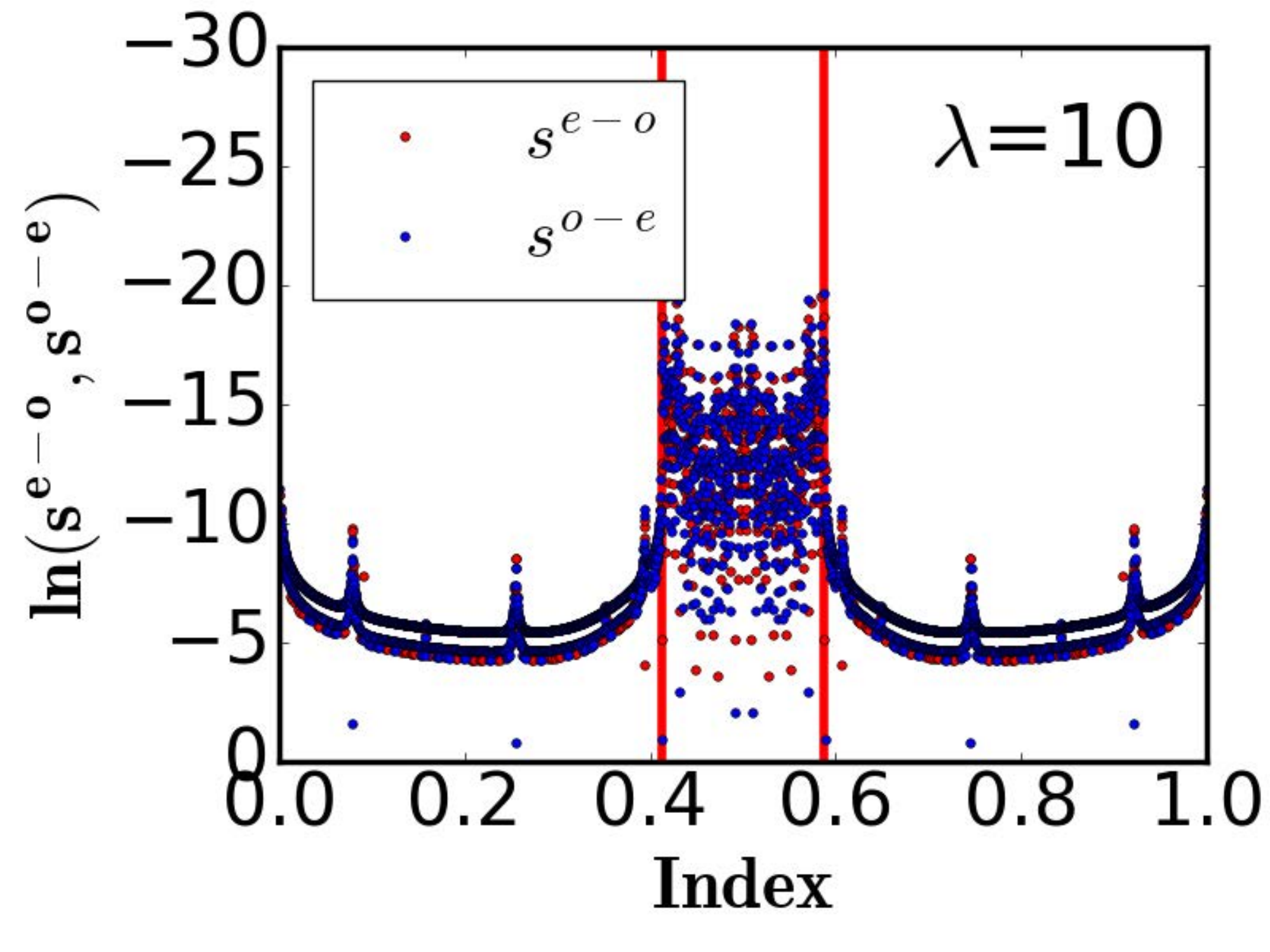}}\vspace{-0.5cm}

\stackunder{\hspace{-4.5cm}(d)}{\includegraphics[height=4.0cm, width=5.7cm]{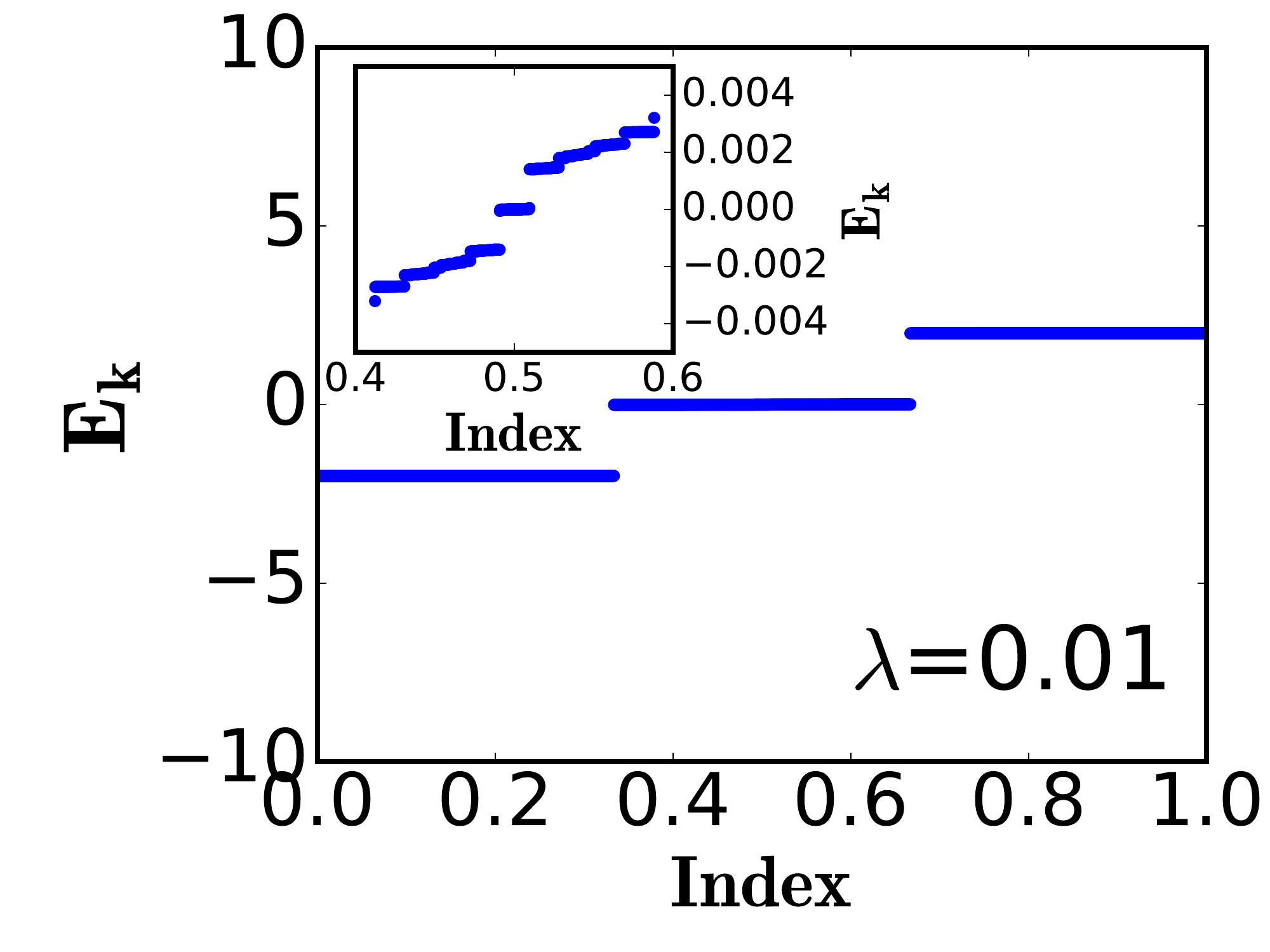}}\hspace{-0.1cm}
\stackunder{\hspace{-4.5cm}(e)}{\includegraphics[height=4.0cm, width=5.7cm]{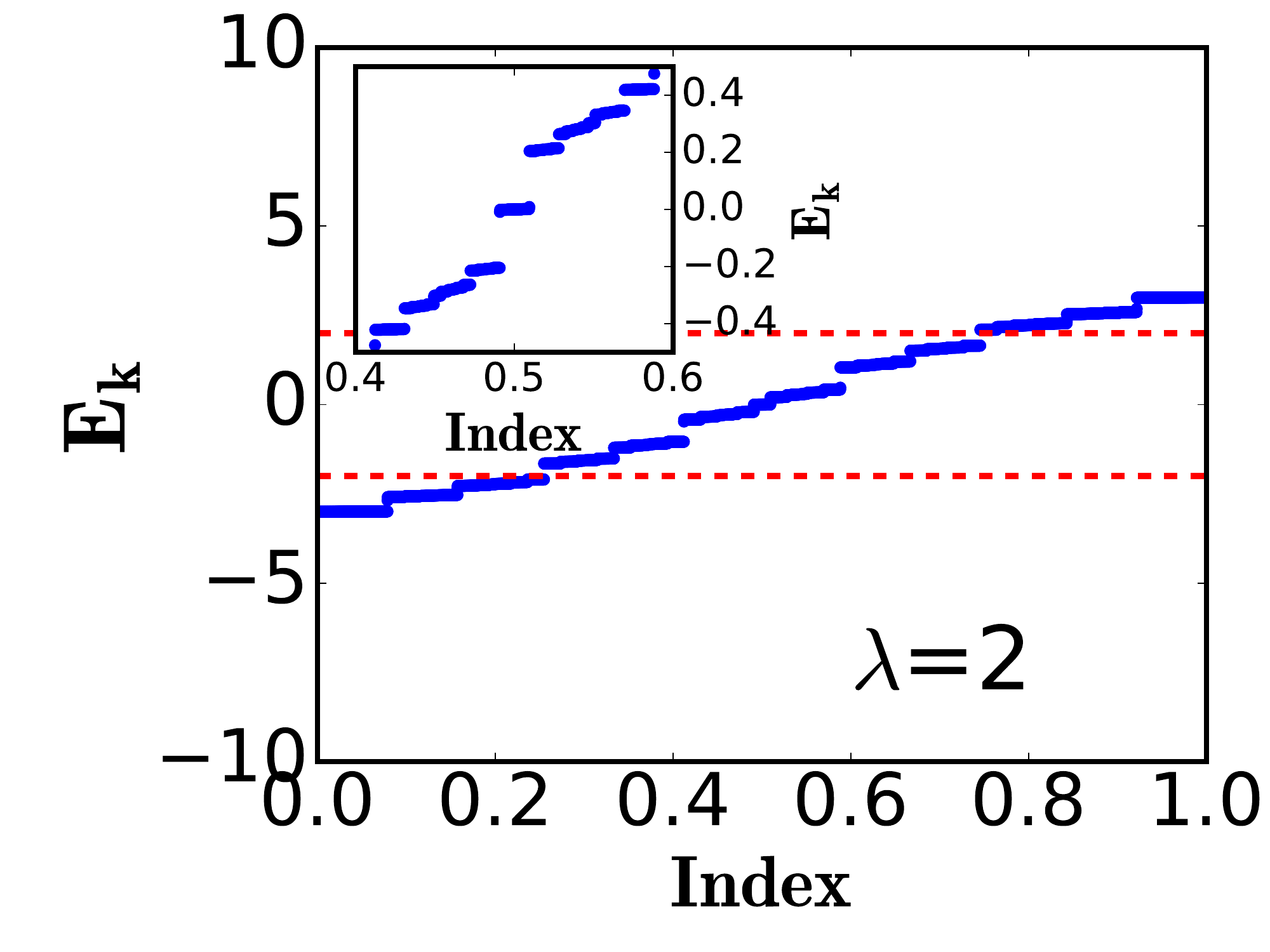}}\hspace{-0.1cm}
\stackunder{\hspace{-4.5cm}(f)}{\includegraphics[height=4.0cm, width=5.7cm]{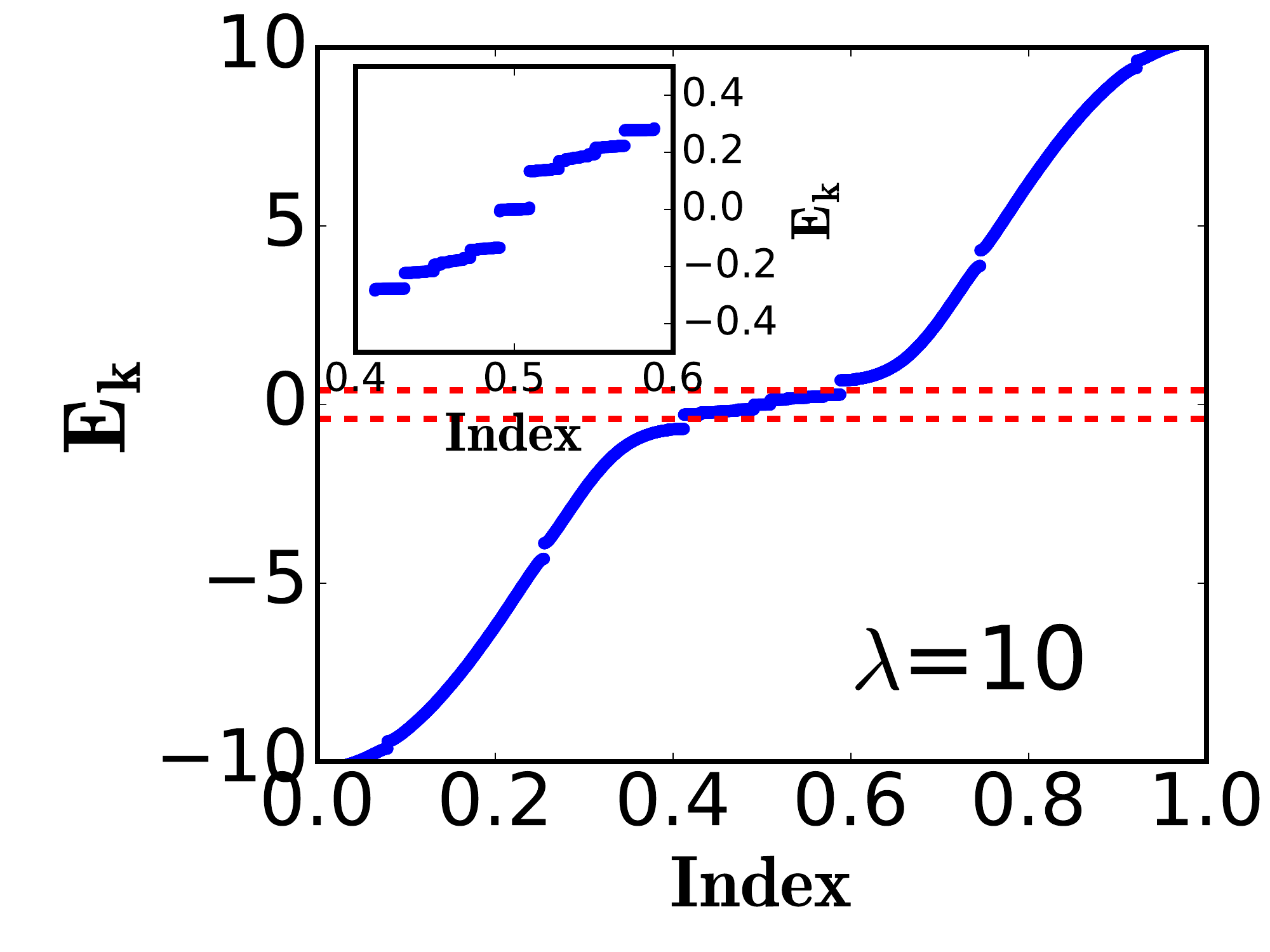}}\vspace{-0.5cm}

\stackunder{\hspace{-4.5cm}(g)}{\includegraphics[height=4.0cm, width=5.7cm]{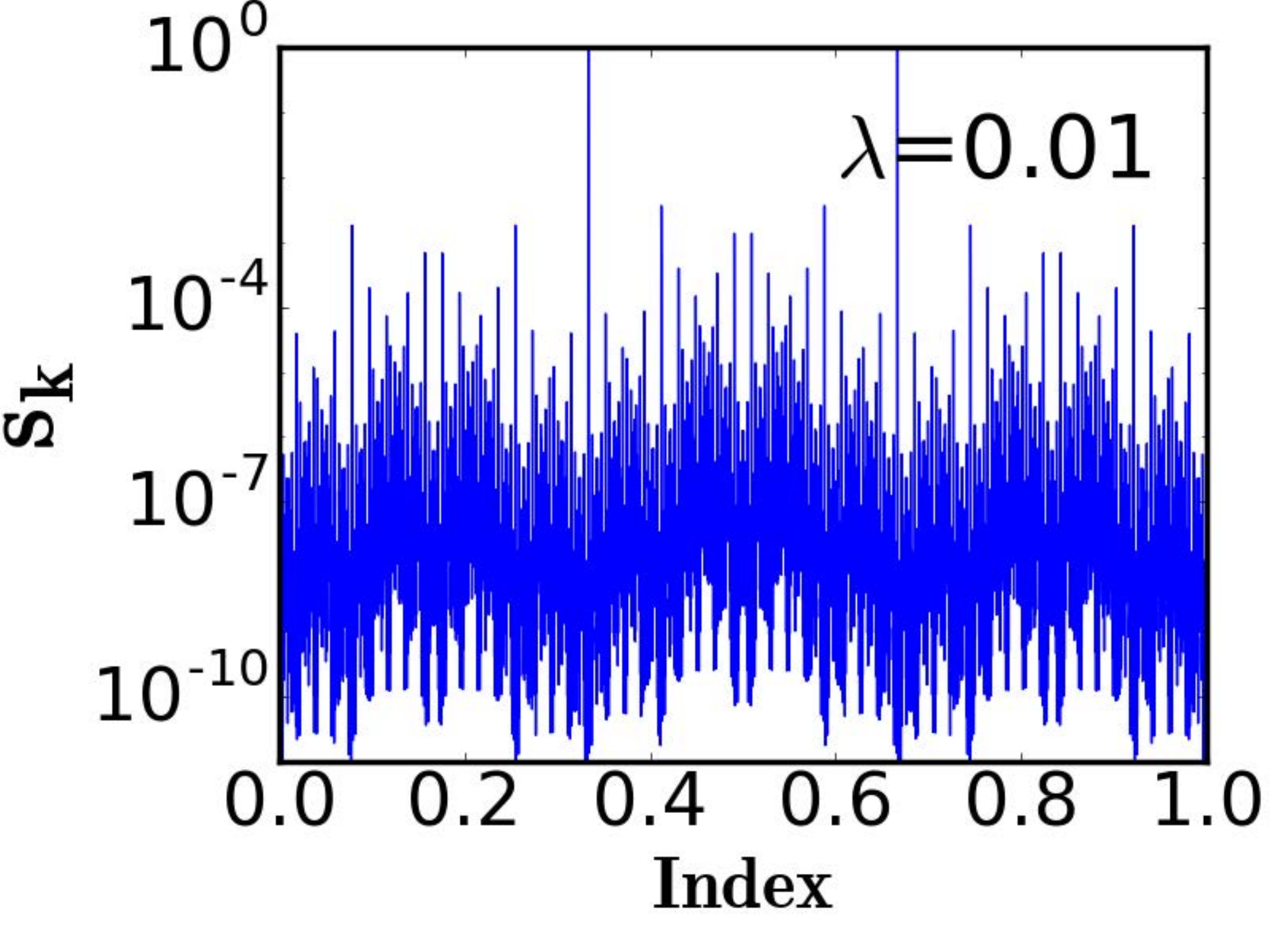}}\hspace{-0.1cm}
\stackunder{\hspace{-4.5cm}(h)}{\includegraphics[height=4.0cm, width=5.7cm]{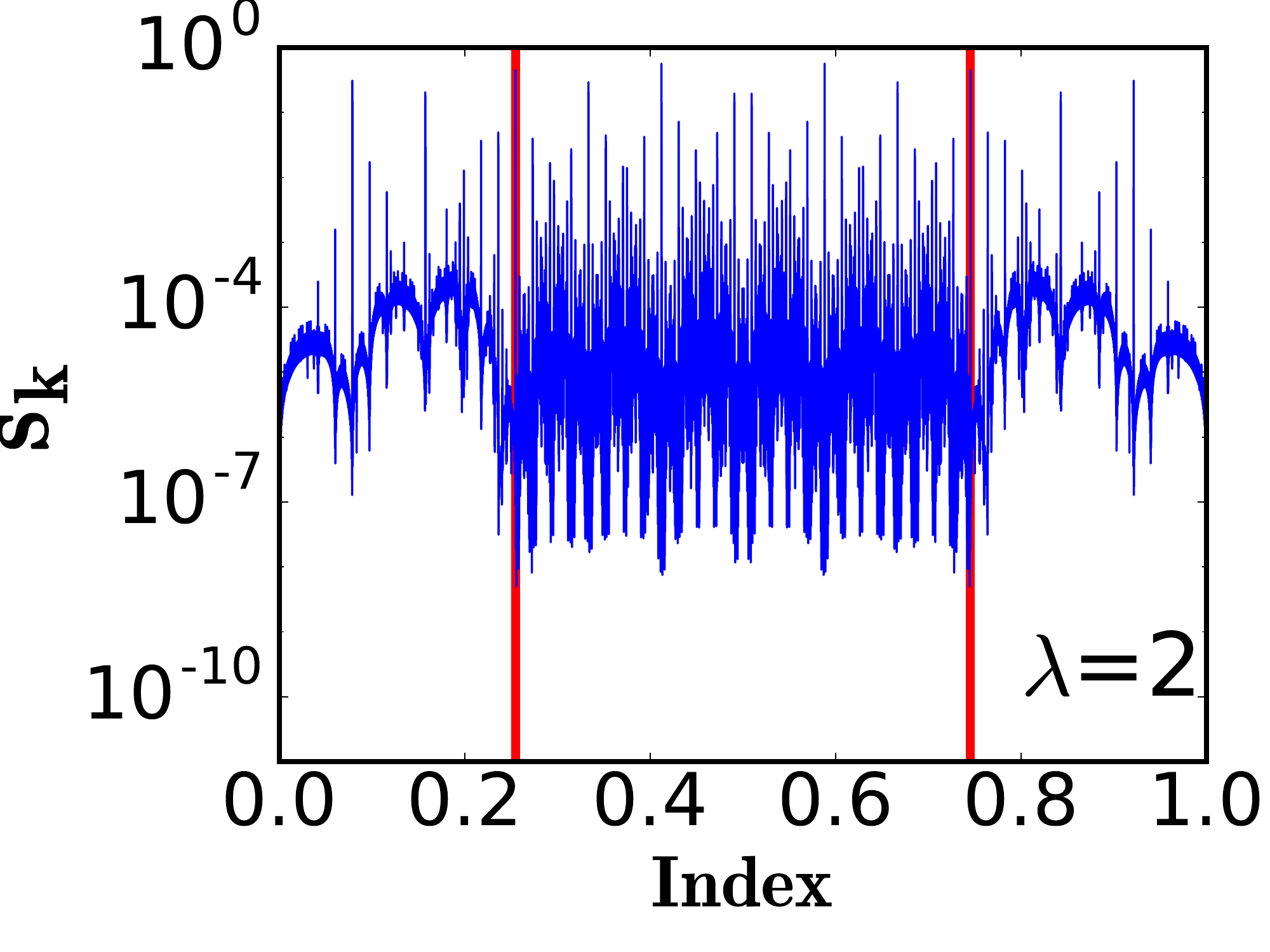}}\hspace{-0.1cm}
\stackunder{\hspace{-4.5cm}(i)}{\includegraphics[height=4.0cm, width=5.7cm]{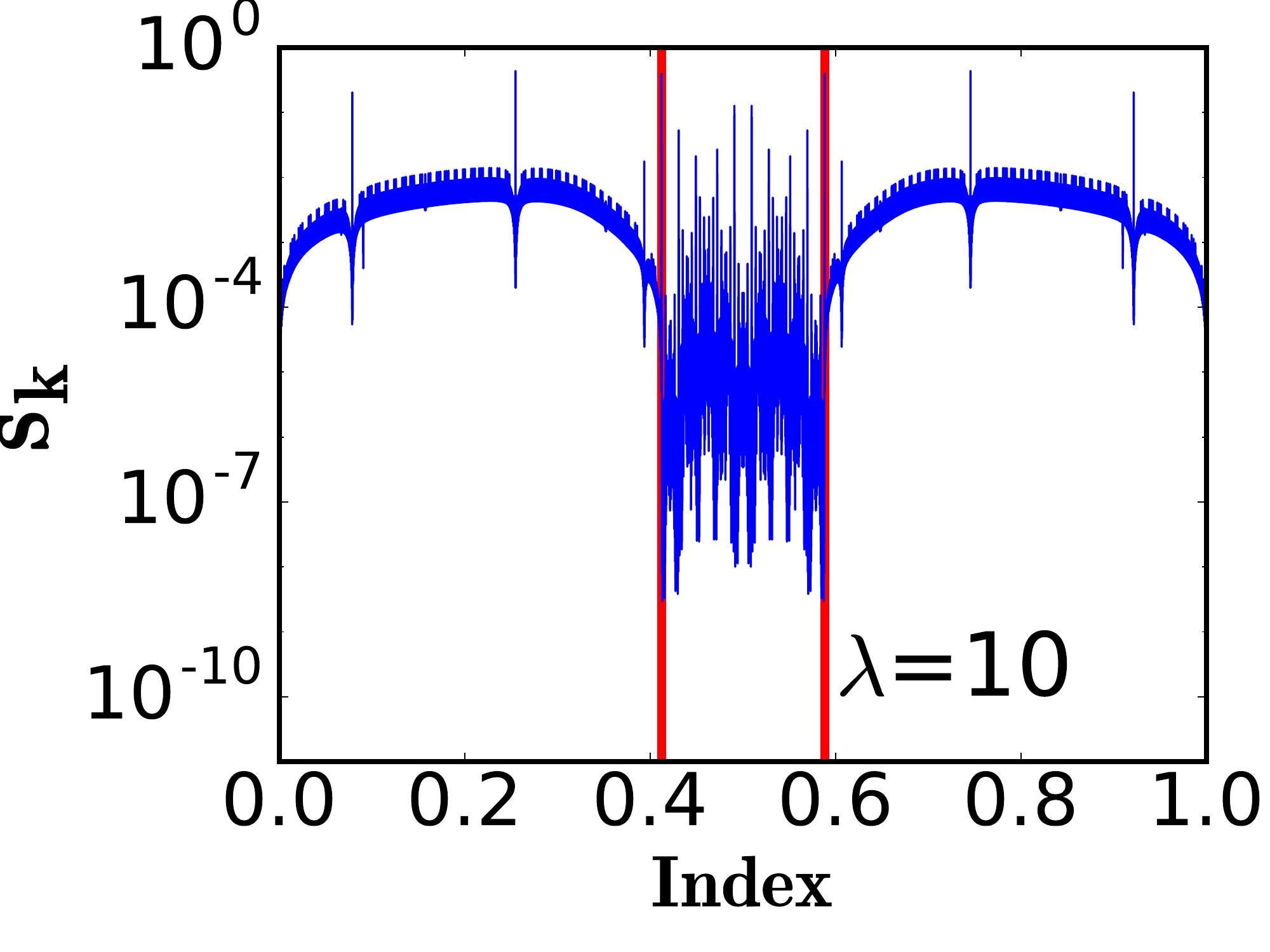}}
\caption{\label{fig5} In the antisymmetric case, (a)--(c) level
  spacing $s_k^{e-o}$ (red) and $s_k^{o-e}$ (blue) at quasiperiodic
  strength $\lambda=0.01, 2$ and $10$ respectively with averaging done
  over $50$ values of $\theta_p$. (d)--(f) Single-particle energy
  spectra $E_k$ for $\lambda = 0.01, 2$, and $10$ respectively. Inset
  shows the broken degeneracy for the states in the central band, and
  (g)--(i) are corresponding level-spacing $s_k$. Here index is the
  serial number of energy (gap) levels divided by the total number of
  gaps, and system size is $N=6000$. \rev{For $\lambda=2$ and $10$,
    the indices corresponding to the fractal mobility edges $|E| =
    4/\lambda$, are determined from (e)--(f). We have plotted them in
    the corresponding Figs.~(b)--(c) and Figs.~(h)--(i) using the
    \rev{vertical red lines}.}}
\end{figure*}

\rev{ While considering the ensemble average, the fractal
  dimension can be defined in two
  forms~\cite{PhysRevB.62.7920,PhysRevE.100.032117}; the first one
  uses \rev{arithmetically} ensemble-averaged moments
  $\left\langle I_q \right\rangle$ after which the logarithm is taken:
\begin{equation}
\tilde{D}_q = \frac{1}{1-q}\frac{1}{\ln N} \ln \left\langle I_q \right\rangle.
\label{d21}
\end{equation}
In the second approach, the averaging is done \rev{in a geometric fashion, i.e.,} after taking the logarithm:
\begin{equation}
D_q=\frac{1}{1-q}\frac{1}{\ln N}\left\langle  \ln I_q \right\rangle.
\label{d22}
\end{equation}
} \rev{ Also, $\tilde{D}_q$ is a lower bound to $D_q$ from
  Jensen's inequality i.e. $\tilde{D}_q \leq D_q$, since logarithm is
  a concave function.} For $q=2$, one obtains
$S_{2}=-\ln(I_2)$. Figure~\ref{fig3}\rev{(c)} shows $D_2$
as a function of the strength of the antisymmetric $AA$ potential for
the entire spectrum. \rev{We observe the presence of the
  fractal mobility edge conjectured
  recently~\cite{https://doi.org/10.48550/arxiv.2208.11930} using an
  analogy to the extended Harper problem (see
  Sec~\ref{sec:level5}). The fractal mobility edge ($\lambda\leq
  4/|E|$) is shown by the red curves in Fig~\ref{fig3}(c), separating
  the multifractal wave functions (extended but non-ergodic) with
  $0<D_{2}<1$, from the Anderson localized eigenstates (red colour
  signifies $D_2 \approx 0$).} Here we also study the scaling of IPR
\rev{($\left\langle I_2 \right\rangle$)} with system size
(see inset of Fig.~\ref{fig3}(b)). The IPR scales as \rev{
  $N^{-\tilde{D}_2}$}~\cite{PhysRevE.71.065303}. We observe from the
inset of Fig.~\ref{fig3}(b) that for $\lambda=0.01$, IPR scales as
$N^{-0.63}$, which reaffirms the multifractal nature of the
eigenstates at low potential strengths.

The fractal dimension in the limit $N\rightarrow \infty$ is given
by~\cite{RevModPhys.80.1355}:
\begin{equation}
D_q^{\infty}=\text{lim}_{N\rightarrow \infty}D_q.
\end{equation}
For a perfectly delocalized state $D_q^{\infty}=1$ while for a
localized state $S_{q}$ is a constant, as observed for Anderson
localization and results in vanishing $D_q^{\infty}$ for all
$q>0$. For intermediate cases, $0<D_{q}^{\infty}<1$ , which is a sign
that the state is extended but non-ergodic. Further, the eigenstates
are multifractal if $D_q^{\infty}$ depends non-trivially on $q > 0$
while for a constant $0<D_q^{\infty}<1$, the states are fractal. The
$q$-dependence of the fractal dimension $D_q$ is shown in
Fig.~\ref{fig4}(a). We observe that for all $q>0$, $0<D_q<1$ with a
non-trivial dependence on the moment $q$. This indicates that all the
eigenstates in the low-$\lambda$ region exhibit multifractal nature.

\rev{For large enough $N$ the IPR is given by the expression
\begin{gather}
I_q = c_q N^{-D_q^\infty} \ ,
\end{gather}
with a certain $c_q$, weakly-dependent on $N$. This leads to the
finite-size $D_q$ being linear in $1/\ln N$ (see Fig.~\ref{fig4}(b))
and allows one to extract $D_q^\infty$ via a linear extrapolation in
$1/\ln N$~\cite{PhysRevE.100.032117}.}  We observe that in the limit
of $N\rightarrow \infty$, the fractal dimension tends to a value
significantly lower than unity. It can be concluded that the
multifractality seen here is robust against increasing system sizes.
Moreover, we have also verified that the system size dependence of
$D_q$ seen here is very similar to what is displayed by the AAH model
at the critical point.

Another useful method to distinguish between localized, multifractal
and delocalized phases is to carry out an analysis of the even-odd
(odd-even) spacings of the energy eigenvalues $E_k$ (arranged in
ascending order)~\cite{PhysRevLett.123.025301}. They are defined as
$s_k^{e-o}=E_{2k}-E_{2k-1}$ and $s_k^{o-e}=E_{2k+1}-E_{2k}$ for
even-odd and odd-even cases respectively. For a localized state, the
gap vanishes as both the spacings exhibit the same form. In the
multifractal case, the distributions of both spacings are strongly
scattered. We observe from Fig.~\ref{fig5}(a) that at $\lambda=0.01$,
both the spacings are scattered for the entire energy spectrum,
indicating that all the eigenstates are multifractal in the low
potential regime. From Fig.~\ref{fig5}(b), we observe that while there
is strong scattering corresponding to the states at the centre
\rev{($|E|<4/\lambda$)}, the gap begins to vanish as one
move toward the edges. At higher potential strengths, for example
$\lambda=100$ (see Fig.~\ref{fig5}(c)), except for the level spacing
at the centre, the gap completely disappears, indicating localization.

It is well known that multifractal states are characterized by a broad
distribution in energy
gaps~\cite{PhysRevLett.84.1643,doi:10.1142/S021797929400049X}. To
study this aspect, we show plots of $E_k$ and $s_k=E_{k+1}-E_k$ for
various $\lambda$, in Fig.~\ref{fig5}(d)--\ref{fig5}(i). We observe
that at low potential strength, i.e. $\lambda=0.01$, the energy
spectrum has a large number of subbands (see inset of
Fig.~\ref{fig5}(d)), and fluctuations are observed in the spacing of
these gaps (see Fig.~\ref{fig5}(g)). We also observe that the level
spacing distribution of all the eigenstates at $\lambda=0.01$ follows
an inverse power law~\cite{PhysRevE.70.066203}, indicating that all
the eigenstates are multifractal in the low potential regime. From
Fig.~\ref{fig3}(c), the transition point from the
\rev{fully extended regime to the mixed one with the
  mobility edge, $|E|=4/\lambda$, is is observed to be around
  $\lambda\simeq 1.5$}. The magnitude of the gap between the energy
levels becomes larger as one moves toward the edges
(Fig.~\ref{fig5}(e)), which is accompanied by a decrease in the
magnitude of fluctuations in those gaps (see Fig.~\ref{fig5}(h)). At
higher potential strengths, for example $\lambda=10$ (see
Fig.~\ref{fig5}(f)), the spectrum is completely pure point-like, and
all the eigenstates except those in the central band show reduced
fluctuations, indicating localization (see Fig.~\ref{fig5}(i)). We
conclude that the presence of the $AA$ potential in the antisymmetric
case transforms the CLSs into multifractal states at low potential
strengths below a critical value (i.e. below
\rev{$\lambda\simeq 1.5$}, which is about the gap between
the flat bands in the zero disorder limit). At higher potential
strengths where the bands hybridize, we observe that all the
eigenstates localize except those in the central \rev{part
  of the spectrum}, which display multifractal nature.

\subsection{Chiral Symmetry}
The addition of diagonal disorder in the ABF diamond lattice breaks
translational invariance. One would also expect diagonal disorder to
break the chiral
symmetry~\cite{PhysRevResearch.2.023118,PhysRevA.99.062107,Leumer_2020}
as observed in the symmetric case. However, remarkably in the
antisymmetric case, when $N$ is even, we observe pairs of eigenvalues
${\pm E}$, despite the on-site disorder. We infer that the chiral
symmetry of the lattice is not broken. Indeed this is confirmed
explicitly by the identification of the chiral operator $\Gamma$ \rev{which is required
to be a local operator~\cite{Topology}:}
\rev{
\begin{equation}
\Gamma=\gamma_1\oplus-\gamma_2\oplus\gamma_3\cdots= \bigoplus_{i=1}^n (-1)^{i-1}\gamma_i,
\end{equation}
where each of the matrices $\gamma_i=\begin{pmatrix}
0 & 1 & 0 \\
1 & 0 & 0 \\
0 & 0 & 1
\end{pmatrix}_i$ acts on the $i^\text{th}$ unit cell}
and $n$ is the total number of unit cells. We can verify that
$\Gamma^{-1} H^{\dagger} \Gamma=-H$, and that $\Gamma$ is involutory since
$\Gamma \Gamma^{\dagger}=\mathbb{I}$. When periodic boundary
conditions are imposed, the chiral symmetry is valid only for an even
number of unit cells.

\rev{Before we conclude, we remark that multifractal states
  have been reported to exist in systems with chiral symmetry. For the
  Anderson model in $2-$D~\cite{PhysRevB.62.12775} and
  $3-$D~\cite{PhysRevB.74.113101}, chiral symmetry is known to induce
  multifractal states near the band centre, the origin of which can be
  traced back to the power-law decay of the eigenstates. It has been
  observed that chiral symmetry tends to delocalize eigenstates close
  to the origin~\cite{PhysRevB.62.8249}. In $1-$D
  ~\cite{PhysRev.92.1331}, it has been proved that eigenstates at the
  band centre are not localized exponentially. Chiral symmetry
  combined with strong disorder induces power-law localization and
  multifractal states near the origin~\cite{PhysRevB.74.113101}.  }

\rev{\section{Multifractality in antisymmetric case: Analytical Treatment}\label{sec:level5}}

\rev{In this section, we analytically uncover the origin of the
  multifractality in the antisymmetric case. The diamond lattice in
  the disorder-free limit can be described through matrices $V$ and
  $T$ which capture the intra-cell and inter-cell information
  respectively:
\begin{equation}
 V=
  \begin{pmatrix}
    0 & 0 & -1 \\
    0& 0 &1\\
    -1&1&0
  \end{pmatrix},\hskip1cm
  T= \begin{pmatrix}
     0 & 0 & 0 \\
    0& 0 &0\\
    1&1&0
  \end{pmatrix}.
\end{equation}
For the on-site disorder, we introduce another matrix $W_n$ given by:
\begin{equation}
W_n=
  \begin{pmatrix}
    \zeta_n^u & 0& 0 \\
    0 &\zeta_n^d& 0\\
       0&  0 &\zeta_n^c
  \end{pmatrix}.
\end{equation}
The unitary matrix
\begin{equation}
\rev{U_1}=\frac{1}{\sqrt{2}}\begin{pmatrix}
 \frac{1}{\sqrt{2}}&-\frac{1}{\sqrt{2}}&1 \\
  -\frac{1}{\sqrt{2}}&\frac{1}{\sqrt{2}}&1 \\
  1&1&0
 \end{pmatrix},
\end{equation}
diagonalizes the $V$ matrix when we carry out the transformation
$V_1=\rev{U_1} V\rev{U_1}^\dagger$. The same transformation can be
applied to the $T$ matrix to obtain
$T_1=\rev{U_1}T\rev{U_1}^\dagger$. The transformed matrices are:
\begin{equation}
V_1=\left(
\begin{array}{ccc}
 -\sqrt{2} & 0 & 0 \\
 0 & \sqrt{2} & 0 \\
 0 & 0 & 0 \\
\end{array}
\right),\hskip0.25cm
T_1=\left(
\begin{array}{ccc}
 0 & 0 & 1 \\
 0 & 0 & 1 \\
 0 & 0 & 0 \\
\end{array}
\right).
\end{equation}
Applying the same transformation on the matrix $W_n$ we obtain
$(W_1)_n$ given by $\rev{U_1}W_n\rev{U_1}^\dagger$:
\begin{equation}
(W_{1})_n=\left(
\begin{array}{ccc}
 \frac{\zeta_n^u}{4}+\frac{\zeta_n^d}{4}+\frac{\zeta_n^c}{2} & \frac{-\zeta_n^u}{4}+\frac{-\zeta_n^d}{4}+\frac{\zeta_n^c}{2} & \frac{\zeta_n^u}{2\sqrt{2}}-\frac{\zeta_n^d}{2\sqrt{2}} \\
  \frac{-\zeta_n^u}{4}+\frac{-\zeta_n^d}{4}+\frac{\zeta_n^c}{2} & \frac{\zeta_n^u}{4}+\frac{\zeta_n^d}{4}+\frac{\zeta_n^c}{2} & \frac{-\zeta_n^u}{2\sqrt{2}}+\frac{\zeta_n^d}{2\sqrt{2}}  \\
 \frac{\zeta_n^u}{2\sqrt{2}}-\frac{\zeta_n^d}{2\sqrt{2}} & \frac{-\zeta_n^u}{2\sqrt{2}}+\frac{\zeta_n^d}{2\sqrt{2}} & \frac{\zeta_n^u}{2}+\frac{\zeta_n^d}{2} \\
\end{array}
\right).
\end{equation}
In the antisymmetric case, $ \zeta_n^c=0,$
$\zeta_n^u=-\zeta_n^d=\zeta_{n}$, yielding a simplification:
\begin{equation}
(W_{1})_n=\frac{\zeta_n}{\sqrt{2}}\left(
\begin{array}{ccc}
0 & 0 & 1 \\
0 & 0 & -1 \\
1 & -1 & 0
\end{array}
\right).
\end{equation}
The above transformation is a part of a series of lattice
transformations shown in Appendix~\ref{app:level2}.  Using the lattice
equation:
\begin{equation}
(W_1)_n\psi_n-V_1\psi_n - T_1 \psi_{n+1} - T_1^\dagger \psi_{n-1} = E \psi_n,
\end{equation}
}
\rev{the corresponding equation for the components $\psi_n=(u_n, d_n, c_n)^T$ can be written as:}
\rev{
\begin{align}
E u_n=\frac{\zeta_n}{\sqrt{2}} c_n-\left(-\sqrt{2} u_n+c_{n+1}\right),\\
E d_n=-\frac{\zeta_n}{\sqrt{2}} c_n-\left(\sqrt{2} d_n+c_{n+1}\right),\\
E c_n=\frac{\zeta_n}{\sqrt{2}}\left(u_n-d_n\right)-\left(u_{n-1}+d_{n-1}\right).
\label{eqn:g3}
\end{align}
Substituting $u_n$ and $d_n$ into Eq.~\ref{eqn:g3} we get,
\begin{equation}\label{eqn:1d_zeta^2}
E\left[E^2-4\right] c_n=\zeta_n^2 E c_n-2 \zeta_n c_{n+1}-2 \zeta_{n-1} c_{n-1}.
\end{equation}
Substituting $\zeta_n=\lambda \cos (2 \pi n b+\theta)$, we have}
\rev{
\begin{align}
\nonumber \frac{2}{\lambda^2}\left[E^2-4-\frac{\lambda^2}{2}\right] c_n=& \cos (4 \pi n b+2 \theta) c_n-\\
\nonumber -&\frac{4}{\lambda E} \Bigl(\lambda \cos (2 \pi n b+\theta) c_{n+1}+\\
+& \lambda \cos (2 \pi(n-1) b+\theta) c_{n-1}\Bigr).
\label{eqn:g4}
\end{align}
The above equation resembles the extended Harper model~\cite{JOUR},
but with a doubled frequency of the on-site potential.  Following the
recent preprint~\cite{https://doi.org/10.48550/arxiv.2208.11930}, we
compare the hopping amplitude $4/(\lambda|E|)$ with the level spacing
amplitude $\max_n\left[\cos (4 \pi n b+2 \theta) - \cos (4 \pi (n+1)
  b+2 \theta)\right] \simeq 2\sin (2 \pi b)$ and conjecture that the
transition between non-ergodic extended and localized states should be
at~\rev{\cite{fn2}}
\begin{gather}
\lambda_c = \frac{4}{|E|} \ .
\end{gather}
}

\begin{figure}
\centering
\stackunder{\hspace{-4.5cm}}{\includegraphics[height=4.2cm,width=6cm]{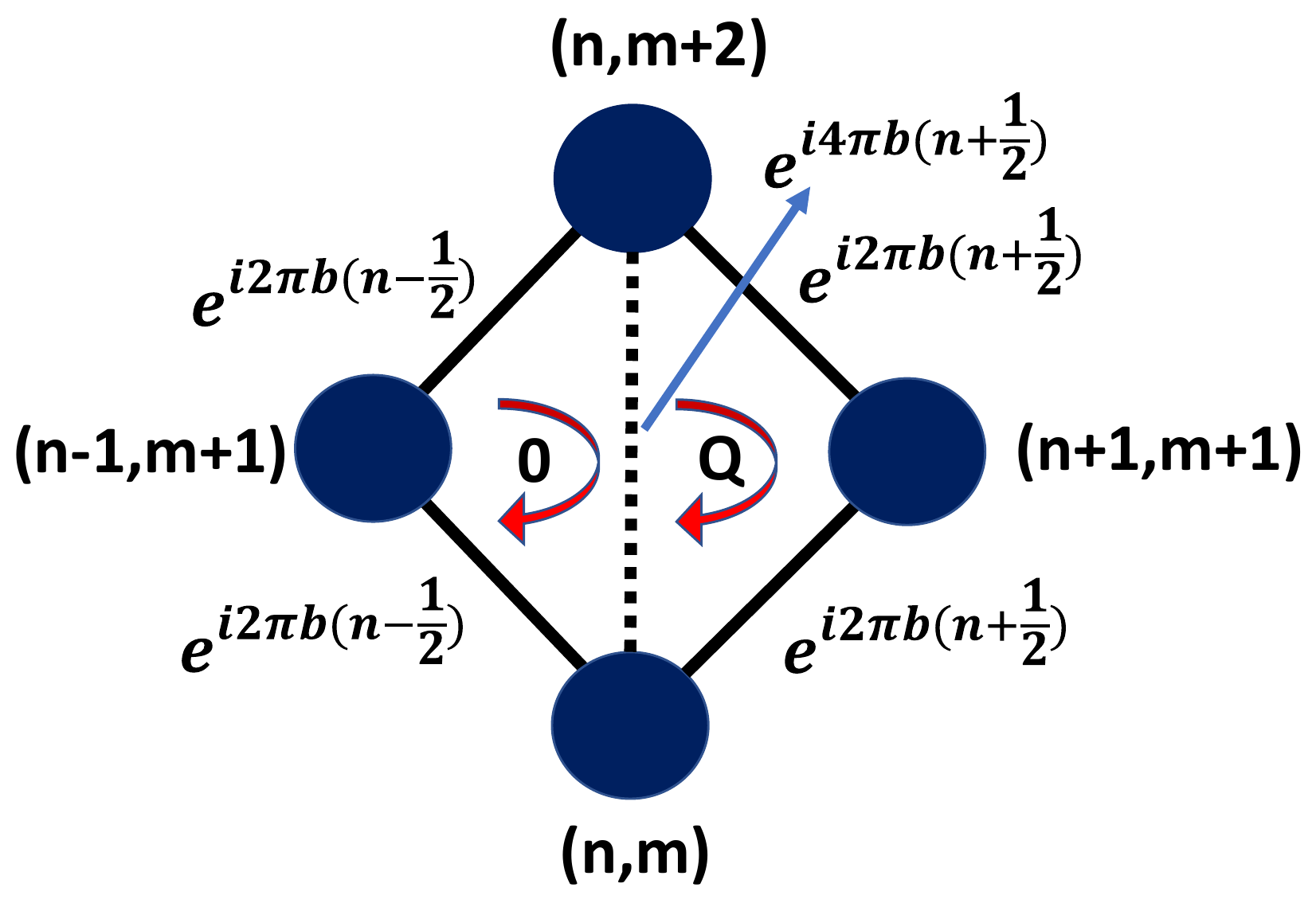}}
\caption{\label{fig6}\rev{A unit cell of the $\frac{\pi}{4}$ rotated
    square lattice with nearest-neighbour and selected
    next-nearest-neighbour hopping and staggered magnetic field showing
    the phases of the hopping on each link. The alternating fluxes $0$
    and $Q =4\pi b$ are derived by summing the phases in the clockwise
    direction over each separate loop.}}
\end{figure}

\rev{\rev{An alternative way to see the emergence of multifractality
    in this model is provided by Thouless in his consideration of the
    Harper model~\cite{PhysRevB.28.4272,PhysRevB.50.11365}.  Indeed,
    by taking the discrete Fourier transform:}
  \begin{equation}
    c_{m, n}=\frac{1}{\sqrt{N}} \sum_k {c_n}_{\left(\theta=\frac{2 \pi k}{N}+\pi b\right)} e^{\frac{2 \pi i}{N} k m},
  \end{equation}
  we get a $2-$D model \rev{without on-site potential, but with an
  effective magnetic flux, penetrating some of the plaquettes}:
  \begin{widetext}
    \begin{align}
      \nonumber E\left[E^2-4-\frac{\lambda^2}{2}\right] c_{n,m}=&\frac{E \lambda^2}{4}\left(e^{4 i\pi b(n+\frac{1}{2})} c_{n,m+2}+e^{-4 i\pi b(n+\frac{1}{2})} c_{n,m-2} \right)-\lambda \left(e^{2 i\pi b(n+\frac{1}{2})} c_{n+1,m+1} +e^{-2 i\pi b(n+\frac{1}{2})} c_{n+1,m-1} \right)\\
      -& \lambda \left(e^{2 i\pi b(n-\frac{1}{2})}c_{n-1,m+1}+e^{-2 i\pi b(n-\frac{1}{2})}c_{n-1,m-1}\right),
    \end{align}
  \end{widetext}
  which is the $\frac{\pi}{4}$ rotated square lattice with
  nearest-neighbour and selected next-nearest-neighbour hoppings and a
  staggered magnetic field (see Fig.~\ref{fig6}). \rev{Such models are
    also known to host multifractal states in an extended region of
    parameter
    space~\cite{PhysRevLett.105.046403,PhysRevB.55.12971,PhysRevB.91.014108}.}}

\rev{While our manuscript was under review, we came across
  a preprint~\cite{https://doi.org/10.48550/arxiv.2208.11930} where
  the effect of quasiperiodic perturbations on one-dimensional
  all-bands-flat lattice models has been investigated. \rev{There the
    authors discuss} the presence of an energy-dependent
  critical-to-insulating transition where ``fractality edges" separate
  localized states from critical states. In the case of the ABF
  diamond lattice with an antisymmetric application of disorder, they
  observed a critical-to-insulating transition at \rev{$|E| =
    4/\lambda$. In the current version, we have confirmed the above
    conjecture with the aid of our numerical data as well as by a
    mapping of the Hamiltonian to other models known to host
    multifractal states from the literature before.}
Further, the substitution of $E=0$ in Eq.~\ref{eqn:g4} gives:
\begin{equation}
0=-2 \lambda \cos (2 \pi n b+\theta) c_{n+1}-2 \lambda \cos (2 \pi(n-1) b+\theta) c_{n-1}.
\end{equation}
This model is equivalent to the off-diagonal Harper
model~\cite{PhysRevB.50.11365, PhysRevLett.109.116404} where the zero
energy modes remain critical at all disorder strengths.}

\section{Conclusion}\label{sec:level6}

This paper explores the effects of a quasiperiodic $AA$ potential on a
one-dimensional ABF diamond lattice. We find that the fate of the
compact localized states is strongly dependent on the manner in which
the potential is applied. We discuss the consequences when the
potential is applied in two specific ways: symmetric and
antisymmetric. When the diagonal disorder is applied symmetrically,
the chiral symmetry of the lattice is broken. Although the large-scale
degeneracy is destroyed, all the eigenstates remain compactly
localized. It is also observed that the compact localization is
independent of the precise nature of the applied perturbation, as long
as it is applied in a symmetric manner. The inclusion of the potential
only on the $c$ sites is a particular case where CLSs are observed and
the degeneracy of the central flatband is preserved.

In the antisymmetric case, the tiniest perturbation lifts the
degeneracy, and the eigenstates no longer remain compactly
localized. An exploration of the nature of the eigenvalues and
eigenfunctions through various observables shows the appearance of
flatband-based multifractal states. Here all the wave functions are
extended but non-ergodic in the low disorder regime. All the bands
start to hybridize at a critical potential strength, leading to
conventional Anderson localization at higher magnitudes of
$\lambda$. However, a central band is observed whose states continue
to display extended behaviour at all strengths of $\lambda$. A
systematic study of the spectrum and the spacing between consecutive
energy values is performed and compared with the AAH model. The robust
existence of multifractal states with increasing disorder strength is
a remarkable finding. Another interesting finding is that the chiral
symmetry of the Hamiltonian is preserved in the presence of
antisymmetric disorder when the total number of unit cells is even. In
support of this finding, we are able to write down explicitly the
chiral operator $\Gamma$.

\rev{The diamond lattice model in the zero-disorder limit can be
  converted through a series of transformations into a new lattice
  with decoupled sites. We study the effect of the same set of
  transformations for the diamond lattice model in the presence of
  disorder. We find that in the symmetric case, these transformations
  yield a lattice which displays an absence of inter-cell hopping,
  indicating the preservation of compact localization. On the other
  hand, the antisymmetric configuration of disorder supports the loss
  of CLSs owing to inter-cell hopping in the transformed lattice.
  Further, we demonstrate that our lattice transformations convert our
  Hamiltonian (with antisymmetric disorder) into a close relative of
  the extended Harper model and into a $\frac{\pi}{4}$ rotated square
  lattice with nearest-neighbour and selective next-nearest-neighbour
  hopping and staggered magnetic field, both of which support the
  multifractality observed in the antisymmetric case.}

We have seen that the introduction of the $AA$ potential in a
flat-band diamond chain yields interesting results within a
single-particle setup. An exciting direction for research would be to
explore the physics of such systems in the presence of interactions,
and in particular, to look for flat-band-based \textit{many-body}
localization phenomena. We also look forward to future studies that
can extend these ideas to two and three dimensional systems.


\begin{acknowledgments}
  We are grateful to Carlo Danieli and Nilanjan Roy for
  discussions. A.A. is grateful to the Council of Scientific and
  Industrial Research (CSIR), India, for her PhD
  fellowship. A.R. acknowledges the financial support from Christ
  College Irinjalakuda via CCRSF-2021. A.S. acknowledges financial
  support from SERB via the grant (File Number: CRG/2019/003447) and
  from DST via the DST-INSPIRE Faculty Award
  [DST/INSPIRE/04/2014/002461].
  I.~M.~K.~acknowledges the support by the European Research Council under the European Union's Seventh Framework Program Synergy No. ERC-2018-SyG HERO-810451.
\end{acknowledgments}

\appendix

\section{Lattice transformation}\label{sec:level7}
In this section, a series of transformations are presented, whose
application to the Hamiltonian of the diamond chain gives a new
lattice which has decoupled sites in the zero-disorder limit. We
utilize a similarity transformation to obtain a new Hamiltonian whose
spacial arrangement of the sites significantly differs from the
original lattice following the work of Danieli et
al.~\cite{PhysRevB.104.085131}. On the one hand, this process
identifies new flat band lattices and, on the other hand, helps in
understanding the influence of the symmetry of the applied
potential. We first present the transformation of the unperturbed ABF
diamond chain. We then discuss the effect of the symmetric and
antisymmetric configuration of disorder on the ABF diamond chain with
the help of these transformations.

\begin{figure*}
\centering
\stackunder{\hspace{-4.0cm}}{\includegraphics[height=2.5cm, width=8cm]{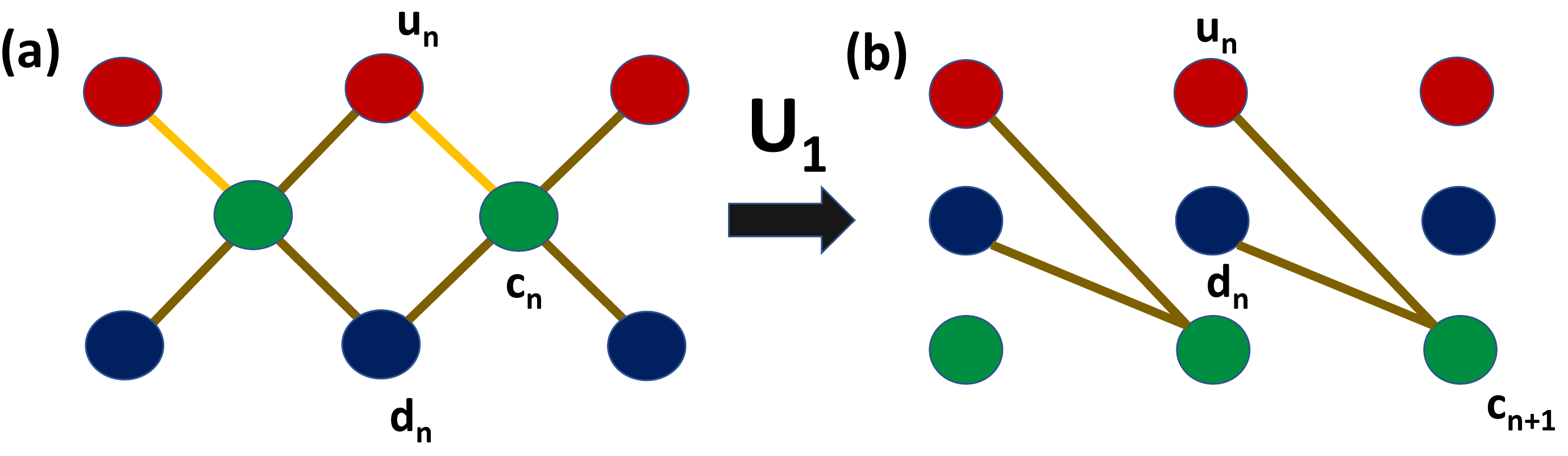}}\hspace{-0.1cm}
\stackunder{\hspace{-4.0cm}}{\includegraphics[height=2.5cm, width=8cm]{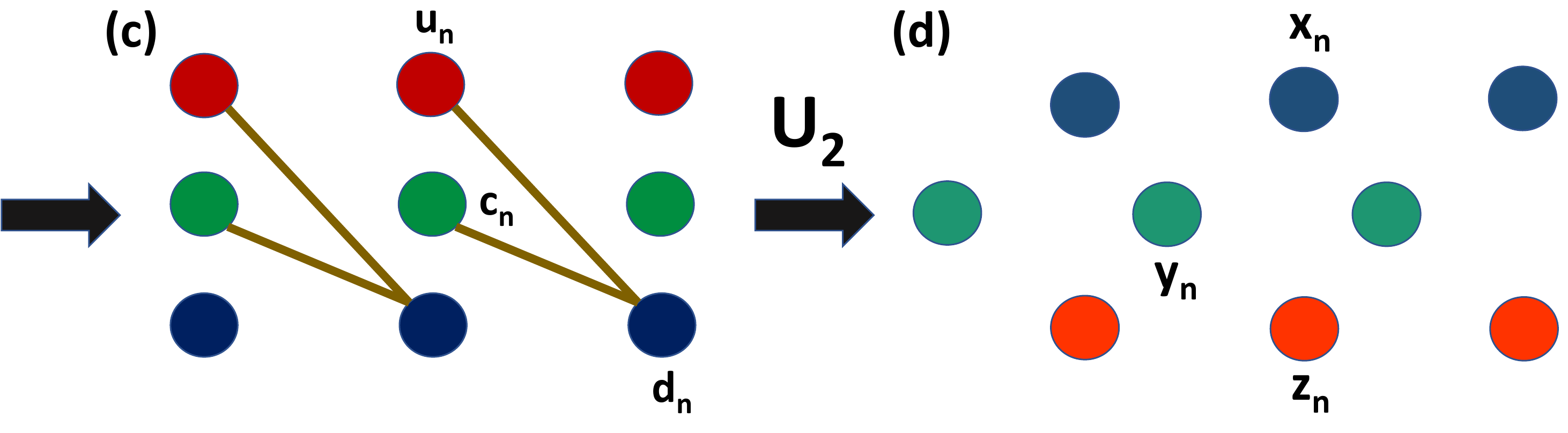}}
\caption{\label{fig7} Schematic representations of the transformation of the diamond chain into the new lattice with decoupled sites.}
\end{figure*}

\subsection{Transformation of unperturbed ABF diamond chain}\label{app:level2}
The unperturbed diamond chain shown in Fig.~\ref{fig7}(a) consists of
three sites per unit cell. The intra-cell and inter-cell information
of the Hamiltonian can be represented by $V$ and $T$ respectively and
is given by:

\begin{equation}
 V=
  \begin{pmatrix}
    0 & 0 & -1 \\
    0& 0 &1\\
    -1&1&0
  \end{pmatrix},\hskip1cm
  T= \begin{pmatrix}
     0 & 0 & 0 \\
    0& 0 &0\\
    1&1&0
  \end{pmatrix}.
\end{equation}
Thus, the lattice equation is
\begin{equation}
-V\psi_n - T \psi_{n+1} - T^\dagger \psi_{n-1} = E \psi_n,
\end{equation}
with \rev{$\psi_n=(u_n, d_n, c_n)^T$} being the tight binding
representation of the wavefunction for the $n^{th}$ unit cell.
Considering the unitary matrix
\begin{equation}
\rev{U_1}=\frac{1}{\sqrt{2}}\begin{pmatrix}
 \frac{1}{\sqrt{2}}&-\frac{1}{\sqrt{2}}&1 \\
  -\frac{1}{\sqrt{2}}&\frac{1}{\sqrt{2}}&1 \\
  1&1&0
 \end{pmatrix},
\end{equation}
we perform the transformations $V_1=\rev{U_1} V\rev{U_1}^\dagger$ and
$T_1=\rev{U_1} T\rev{U_1}^\dagger$ yielding new matrices
\begin{equation}
V_1=\left(
\begin{array}{ccc}
 -\sqrt{2} & 0 & 0 \\
 0 & \sqrt{2} & 0 \\
 0 & 0 & 0 \\
\end{array}
\right),\hskip0.25cm
T_1=\left(
\begin{array}{ccc}
 0 & 0 & 1 \\
 0 & 0 & 1 \\
 0 & 0 & 0 \\
\end{array}
\right).
\label{eqn:g1}
\end{equation}
The resulting lattice is shown in Fig.~\ref{fig7}(b).
A new unit cell can be identified considering the connected lattice sites \rev{$(u_n, c_{n+1}, d_n)$}, which affirms that the CLS stays in one unit cell and the class of the CLS is $U=1$ in this representation. The corresponding lattice is shown in Fig.~\ref{fig7}(c), and the matrices $V_2$ and $T_2$ are given by
\begin{equation}
V_2=\left(
\begin{array}{ccc}
 -\sqrt{2} & 1 & 0 \\
 1 & 0 & 1 \\
 0 & 1 & \sqrt{2} \\
\end{array}
\right),\hskip0.25cm
T_2=\left(
\begin{array}{ccc}
 0 & 0 & 0 \\
 0 & 0 & 0 \\
 0 & 0 & 0 \\
\end{array}
\right).
\end{equation}
Identifying the CLS in the new lattice, an inverse transformation can
be performed to obtain the CLS in the diamond chain, which matches
with Danieli et al.~\cite{PhysRevB.104.085131}. Hence the class of CLS
is not unique under a transformation.

A further transformation can be performed to represent the lattice
into a Fano defect form with decoupled sites. For this, the matrix
$H_2$ will be defined as $H_2=-V_2-T_2e^{ik}-T_2^{\dagger}e^{-ik}$ for
the lattice in Fig.~\ref{fig7}(c). We obtain the transformation matrix
$\rev{U_2}$ here from the eigenvectors of the matrix $H_2$, which is
given by
\begin{equation}
\rev{U_2}=\left(
\begin{array}{ccc}
 \frac{3-2\sqrt{2}}{2\sqrt{6-4\sqrt{2}}} & \frac{2-\sqrt{2}}{2\sqrt{6-4\sqrt{2}}}   & \frac{1}{2\sqrt{6-4\sqrt{2}}}\\
 \frac{3+2\sqrt{2}}{2\sqrt{6+4\sqrt{2}}} & \frac{-2-\sqrt{2}}{2\sqrt{6+4\sqrt{2}}}  & \frac{1}{2\sqrt{6+4\sqrt{2}}}  \\
 -\frac{1}{2} & -\frac{1}{\sqrt{2}} & \frac{1}{2}  \\
\end{array}
\right).
\end{equation}
The transformed matrices $V_3=\rev{U_2} V_2\rev{U_2}^\dagger$ and $T_3=\rev{U_2} T_2\rev{U_2}^\dagger$ are:
\begin{equation}
\label{eqn:v}
V_3=\left(
\begin{array}{ccc}
 2 & 0 & 0 \\
 0 & -2 & 0 \\
 0 & 0 & 0 \\
\end{array}
\right)
,\hskip0.25cm
T_3=\left(
\begin{array}{ccc}
 0 & 0 & 0 \\
 0 & 0 & 0 \\
 0 & 0 & 0 \\
\end{array}
\right).
\end{equation}
The resulting lattice (see Fig~\ref{fig7}(d)) consists of three linear
chains \rev{$(x_n, y_n, z_n)$} without any hopping between sites. The
isolated sites represent the flat band lattice in this representation,
with the eigenstates strictly localized on only one site.

\subsection{Transformation of ABF diamond chain with on-site potential}\label{app:level3}
In the presence of on-site disorder, in addition to the intra-cell and intercell matrices, we introduce another matrix $W_n$ given by
\begin{equation}
\rev{W_n}=
  \begin{pmatrix}
    \zeta_n^u & 0& 0 \\
    0 &\zeta_n^d& 0\\
       0&  0 &\zeta_n^c
  \end{pmatrix},
\end{equation}
which is added in the general Hamiltonian: $H = \rev{W} - V - T e^{ik} -
T^{\dagger} e^{-ik}$. The transformation of the matrix \rev{$W_n$} given
by $\rev{U_1}\rev{W_n}\rev{U_1}^\dagger$ results in:
\begin{equation}
(\rev{W_{1}})_n=\left(
\begin{array}{ccc}
 \frac{\zeta_n^u}{4}+\frac{\zeta_n^d}{4}+\frac{\zeta_n^c}{2} & \frac{-\zeta_n^u}{4}+\frac{-\zeta_n^d}{4}+\frac{\zeta_n^c}{2} & \frac{\zeta_n^u}{2\sqrt{2}}-\frac{\zeta_n^d}{2\sqrt{2}} \\
  \frac{-\zeta_n^u}{4}+\frac{-\zeta_n^d}{4}+\frac{\zeta_n^c}{2} & \frac{\zeta_n^u}{4}+\frac{\zeta_n^d}{4}+\frac{\zeta_n^c}{2} & \frac{-\zeta_n^u}{2\sqrt{2}}+\frac{\zeta_n^d}{2\sqrt{2}}  \\
 \frac{\zeta_n^u}{2\sqrt{2}}-\frac{\zeta_n^d}{2\sqrt{2}} & \frac{-\zeta_n^u}{2\sqrt{2}}+\frac{\zeta_n^d}{2\sqrt{2}} & \frac{\zeta_n^u}{2}+\frac{\zeta_n^d}{2} \\
\end{array}
\right),
\label{eqn:g2}
\end{equation}
in addition to $V_1$ and $T_1$ for the transformed lattice. As before, re-arranging the unit cell, new matrices incorporating the on-site contributions can be obtained as:
\begin{subequations}
\begin{align}
V_{(\rev{W_{2}})_n}=&  \left(
\begin{array}{ccc}
 \frac{\zeta_n^u}{4}+\frac{\zeta_n^d}{4}+\frac{\zeta_n^c}{2}  & 0 & \frac{\zeta_n^u}{2\sqrt{2}}-\frac{\zeta_n^d}{2\sqrt{2}}  \\
 0 & \frac{\zeta_{n+1}^u}{2}+\frac{\zeta_{n+1}^d}{2} & 0 \\
 \frac{\zeta_n^u}{2\sqrt{2}}-\frac{\zeta_n^d}{2\sqrt{2}} & 0 & \frac{\zeta_n^u}{4}+\frac{\zeta_n^d}{4}+\frac{\zeta_n^c}{2} \\
\end{array}
\right)\\
T_{(\rev{W_{2}})_n}=&  \left(
\begin{array}{ccc}
0 & 0 & 0 \\
\frac{\zeta_n^u}{2\sqrt{2}}-\frac{\zeta_n^d}{2\sqrt{2}} & 0  & -\frac{\zeta_n^u}{2\sqrt{2}}+\frac{\zeta_n^d}{2\sqrt{2}} \\
0 & 0 & 0 \\
\end{array}
\right) ,
\end{align}
\end{subequations}
in addition to $V_2$ and $T_2$.

Similarly,  we can obtain $V_{(\rev{W_{3}})_n}=\rev{U_2} V_{(\rev{W_{2}})_n}\rev{U_2}^\dagger$ and $T_{(\rev{W_{3}})_n}=\rev{U_2} T_{(\rev{W_{2}})_n}\rev{U_2}^\dagger$ in addition to $V_3$ and $T_3$ as:
\begin{align}
\label{eqn:vd3}
V_{(\rev{W_{3}})_n}=& \left(
\begin{array}{ccc}
\zeta_{1}   &  \zeta_{2} & \zeta_{3} \\
 \zeta_{2} & \zeta_{1}  & -\zeta_{3}\\
\zeta_{3} & -\zeta_{3}  &\zeta_{4}  \\
\end{array}
\right),\\
T_{\rev{(W_{3})_n}}=&  \left(
\begin{array}{ccc}
\zeta_{5} & -\zeta_{5}  &  \zeta_{6}\\
 -\zeta_{5}  & \zeta_{5} &  - \zeta_{6}\\
 \zeta_{6} &- \zeta_{6} & -\frac{1}{2}\zeta_{5} \\
\end{array}
\right),
\label{eqn:td3}
\end{align}
where
 $\zeta_{1} =  \frac{1}{8}(\zeta_n^u+\zeta_n^d+4\zeta_n^c+\zeta_{n+1}^u+\zeta_{n+1}^d)$,
 $\zeta_{2} =  \frac{1}{8}(-\zeta_n^u-\zeta_n^d+4\zeta_n^c-\zeta_{n+1}^u-\zeta_{n+1}^d)$,
 $\zeta_{3} =   \frac{1}{4\sqrt{2}}(\zeta_n^u+\zeta_n^d-\zeta_{n+1}^u-\zeta_{n+1}^d)$,
 $\zeta_{4} = \frac{1}{4}(\zeta_n^u+\zeta_n^d+\zeta_{n+1}^u+\zeta_{n+1}^d)$,
 $\zeta_{5} =  \frac{1}{8}(-\zeta_{n+1}^u+\zeta_{n+1}^d)$,
 $\zeta_{6} = \frac{1}{4\sqrt{2}}(-\zeta_{n+1}^u+\zeta_{n+1}^d)$.
\begin{figure*}
\centering
\stackunder{\hspace{-5.0cm}(a)}{\includegraphics[height=2.9cm, width=5cm]{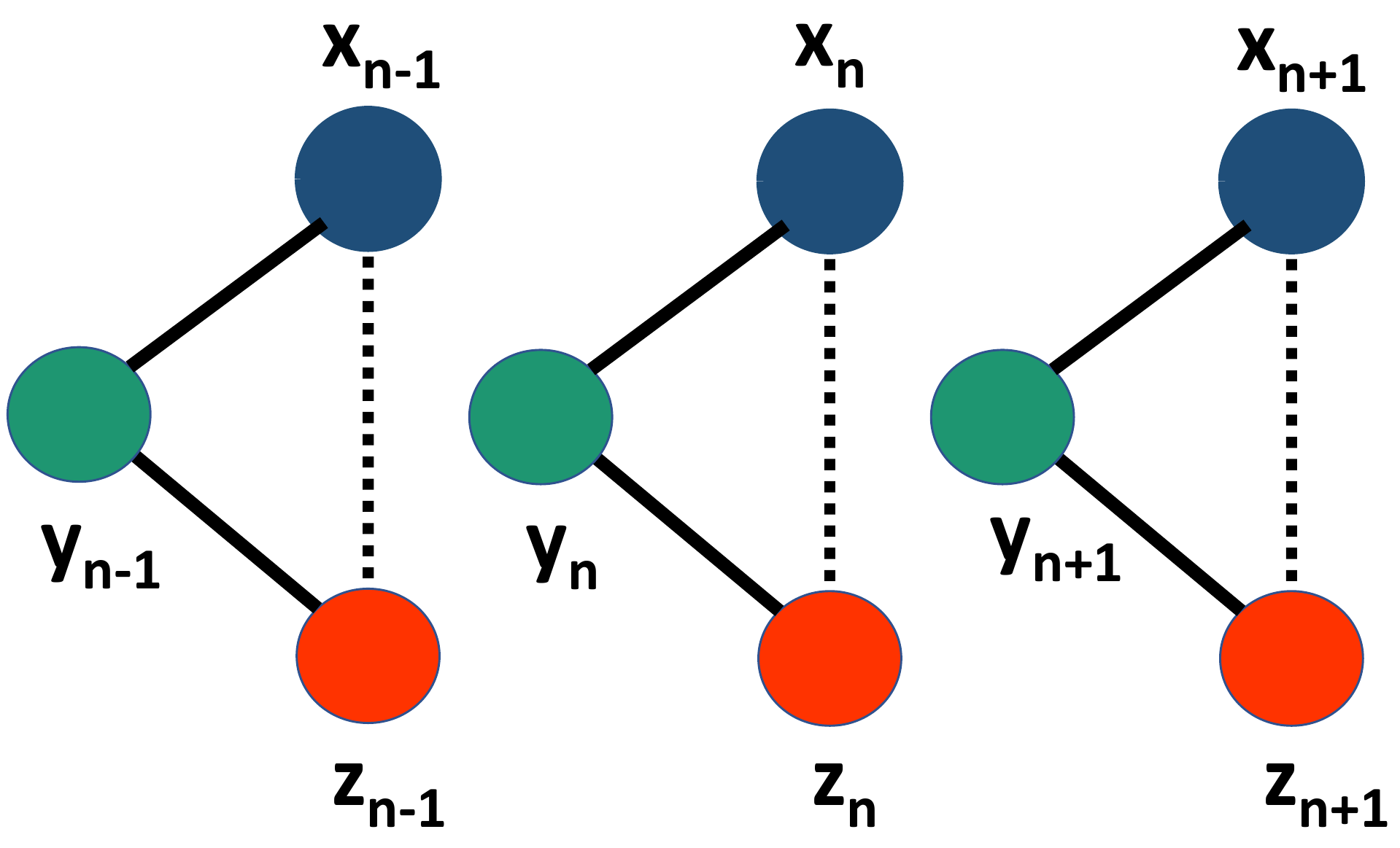}}
\hspace{0.5cm}\stackunder{\hspace{-5.0cm}(b)}{\includegraphics[height=2.9cm, width=5cm]{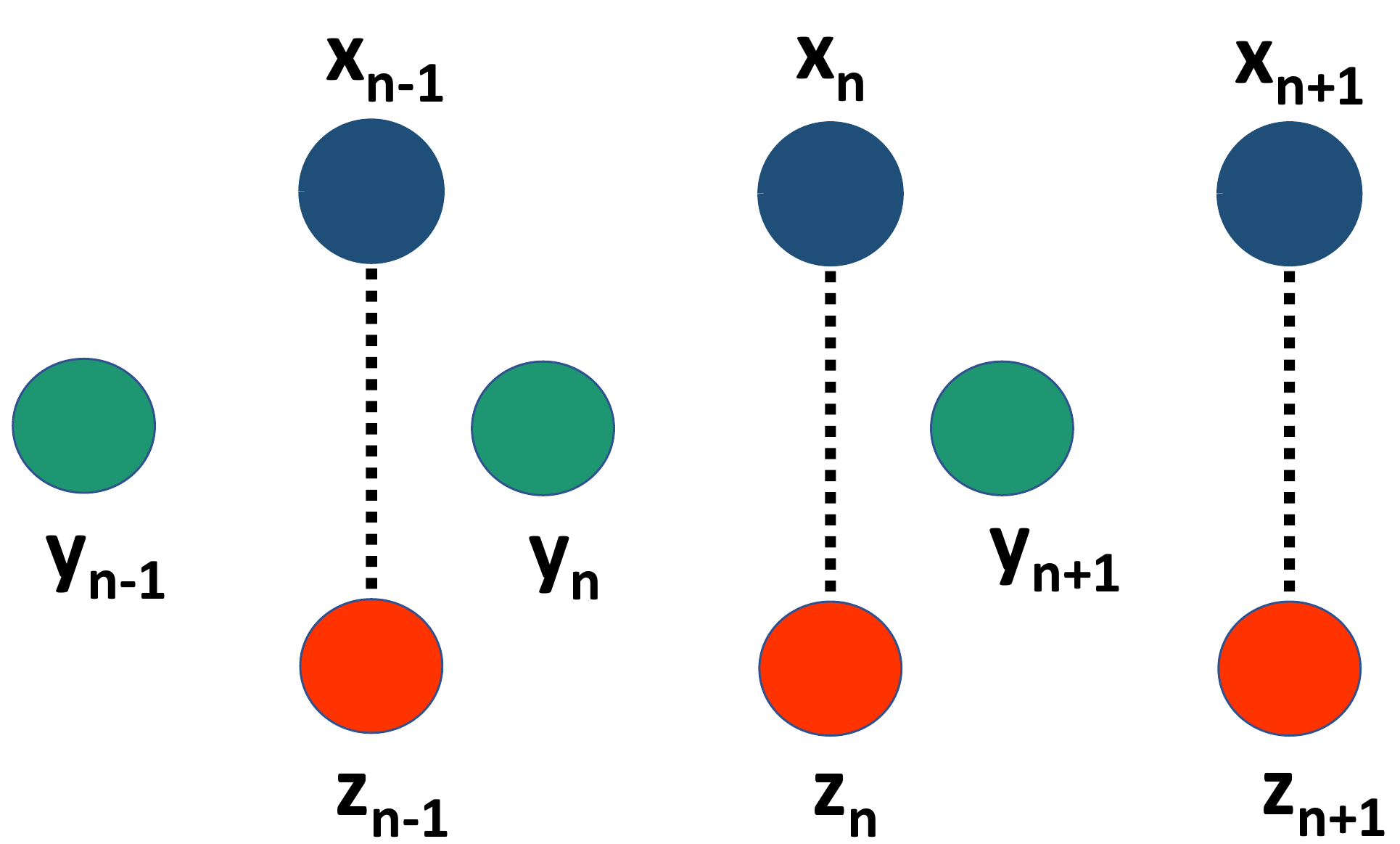}}
\hspace{0.5cm}\stackunder{\hspace{-5.0cm}(c)}{\includegraphics[height=2.9cm, width=5cm]{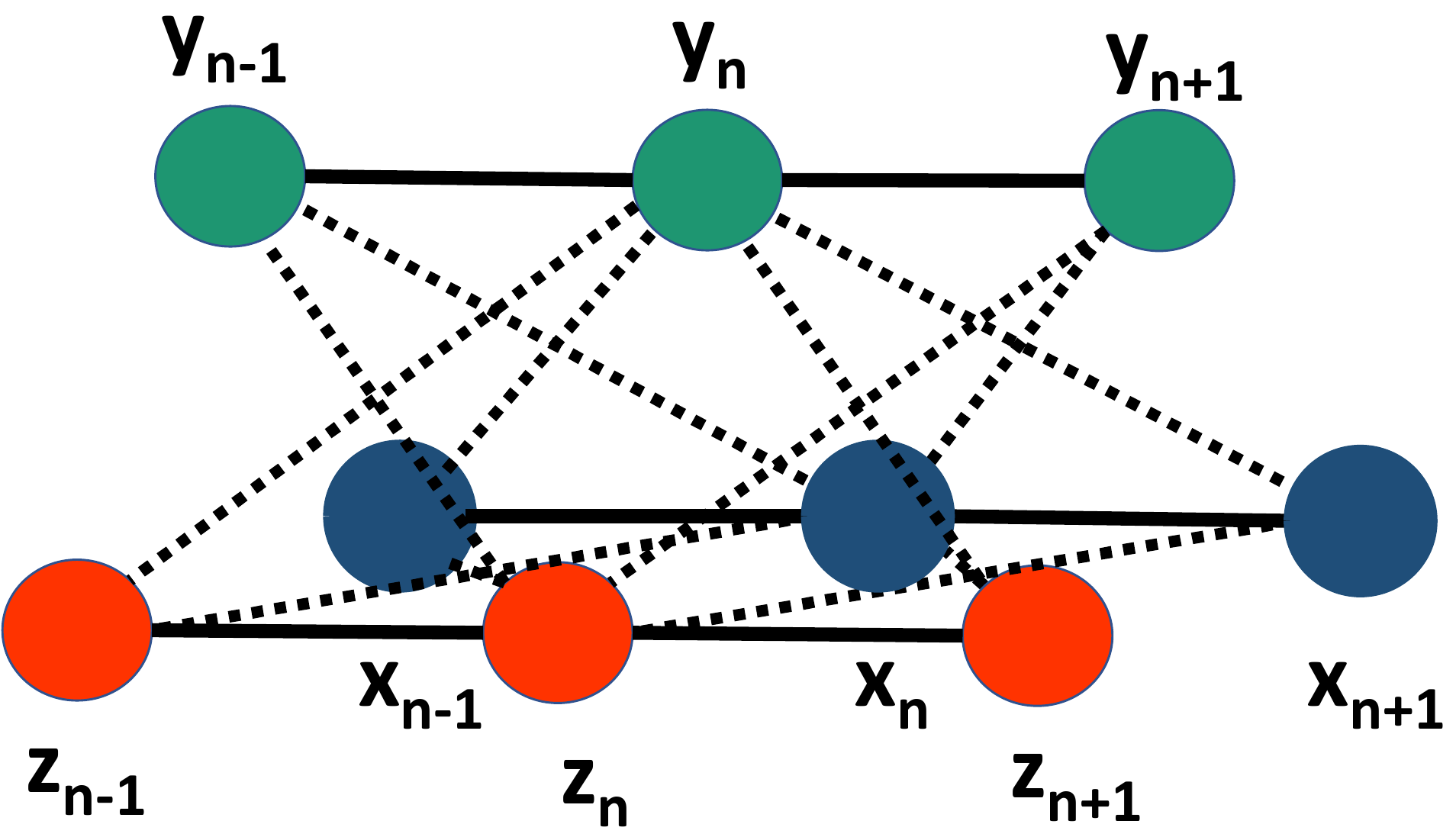}}
\caption{\label{fig8} Schematic representations of the new lattice with disorder in the symmetric case: (a) $\zeta_{n}^{u}=\zeta_{n}^{d} \neq 0, \zeta_{n}^{c}=0$, (b) $\zeta_{n}^{u}=\zeta_{n}^{d} = 0, \zeta_{n}^{c}\neq 0$ and (c) the antisymmetric case: $\zeta_{n}^{u}=-\zeta_{n}^{d} \neq 0 , \zeta_{n}^{c}=0$}
\end{figure*}

\subsubsection*{Symmetric case}\label{app:subsec1}

\paragraph{Disorder on u and d sites:}\label{para1}

\rev{First, we} consider the case with
$\zeta_{n}^{u}=\zeta_{n}^{d}\neq 0$ and $\zeta_{n}^{c}=0$.\\ \\ Here
the effects of on-site disorder incorporated through Eq.~\ref{eqn:vd3}
and Eq.~\ref{eqn:td3} together with Eq.~\ref{eqn:v} of the unperturbed
lattice, result in matrices $V_{S_1}$ and $T_{S_1}$~\rev{\cite{fn3}}
which represent the intracell and intercell hopping of the lattice:

\begin{widetext}
\begin{equation}
V_{S_1}= \left(
\begin{array}{ccc}
\frac{1}{8}(2\zeta_n^u+2\zeta_{n+1}^u)+2 &\frac{1}{8}(-2\zeta_n^u-2\zeta_{n+1}^u) &\frac{1}{4\sqrt{2}}(2\zeta_n^u-2\zeta_{n+1}^u)\\
\frac{1}{8}(-2\zeta_n^u-2\zeta_{n+1}^u) &\frac{1}{8}(2\zeta_n^u+2\zeta_{n+1}^u)-2& \frac{1}{4\sqrt{2}}(-2\zeta_n^u+2\zeta_{n+1}^u)\\
\frac{1}{4\sqrt{2}}(2\zeta_n^u-2\zeta_{n+1}^u) & \frac{1}{4\sqrt{2}}(-2\zeta_n^u+2\zeta_{n+1}^u) & \frac{1}{4}(2\zeta_n^u+2\zeta_{n+1}^u)\\
 \end{array}
\right),\hskip0.25cm
T_{S_1}=  \left(
\begin{array}{ccc}
0 & 0 & 0 \\
0 & 0 & 0 \\
0 & 0 & 0 \\
\end{array}
\right).
\end{equation}
\end{widetext}

The corresponding lattice (see Fig.~\ref{fig8}(a)) shows that the
symmetric nature of the disorder decouples the unit cells of the
system. The transformed lattice is a linear chain of uncoupled unit
cells, implying that the states do not hybridize. At the same time,
their probability amplitudes may be re-arranged along the sites to
satisfy the lattice equation. \\

\paragraph{Disorder on c sites:}\label{app:subsec3}

We next consider another type of symmetric configration where disorder
is introduced only on the $c$ sites
i.e. $\zeta_{n}^{u}=\zeta_{n}^{d}=0$ and $\zeta_{n}^{c}\neq 0$. Here
Eq.~\ref{eqn:vd3} and Eq.~\ref{eqn:td3} together with Eq.~\ref{eqn:v}
for the unperturbed lattice, give matrices $V_{S_2}$ and $T_{S_2}$:

\begin{equation}
V_{S_2}= \left(
\begin{array}{ccc}
 \frac{\zeta_n^c}{2}+2 &\frac{\zeta_n^c}{2} &0\\
\frac{\zeta_n^c}{2}&\frac{\zeta_n^c}{2}-2 & 0\\
 0&0& 0\\
 \end{array}
\right),\hskip0.25cm
T_{S_2}=  \left(
\begin{array}{ccc}
0 & 0 & 0 \\
0 & 0 & 0 \\
0 & 0 & 0 \\
\end{array}
\right).
\label{B14}
\end{equation}

As observed from Eq.~\ref{B14}, there is no intercell hopping
here. However, we find intracell hopping between two sites and an
uncoupled site in a unit cell (see Fig.~\ref{fig8}(b)).

\subsubsection*{Antisymmetric case}\label{app:subsec2}

In the antisymmetric case, $\zeta_{n}^{u}=-\zeta_{n}^{d}$ while
$\zeta_{n}^{c}=0$.  Then Eq.~\ref{eqn:vd3} and Eq.~\ref{eqn:td3} in
addition to $V_3$ and $T_3$ (see Eq.~\ref{eqn:v}) for the unperturbed
lattice, give matrices $V_{A}$ and $T_{A}$ which represent the
intracell and intercell hopping in the lattice:

\begin{equation}
V_{A}= \left(
\begin{array}{ccc}
2 & 0 & 0 \\
0 & -2 & 0 \\
0 & 0 & 0 \\
 \end{array}
\right),\hskip0.25cm
T_{A}=  \left(
\begin{array}{ccc}
 \frac{-\zeta_{n+1}^u}{4} & \frac{\zeta_{n+1}^u}{4} & \frac{-\zeta_{n+1}^u}{2\sqrt{2}}\\
 \frac{\zeta_{n+1}^u}{4} & \frac{-\zeta_{n+1}^u}{4} & \frac{\zeta_{n+1}^u}{2\sqrt{2}}\\
\frac{\zeta_{n+1}^u}{2\sqrt{2}} & \frac{-\zeta_{n+1}^u}{2\sqrt{2}} & \frac{\zeta_{n+1}^u}{2}\\
\end{array}
\right).
\label{B15}
\end{equation}

The lattice corresponding to it is shown in Fig.~\ref{fig8}(c). We observe that the resulting lattice is a $3-$leg
cross-stitch chain. The inter-cell hopping in the lattice results in the non-existence of the CLSs, unlike the symmetric case. We observe from the transformed lattice that the potential in the antisymmetric case leads to coupling between the adjacent unit cells. Still, the sites in a single unit cell remain decoupled. At low potential strengths, in the absence of intra-cell coupling, the nearest neighbour (NN) inter-cell coupling leads to the multifractal nature of the eigenstates. As the strength of $\lambda$ increases, the stronger coupling leads to Anderson localization.

\section{Complementary quantities}
In this section, we point out the usefulness of discussing several
complementary quantities when the Aubry-Andr\`e potential is applied
in an antisymmetric manner. We also analyze the effect of the
antisymmetric application of the uniform disorder through various
measures.
\begin{figure*}
\centering
\stackunder{\hspace{-4.2cm}(a)}{\includegraphics[height=4.0cm, width=5.8cm]{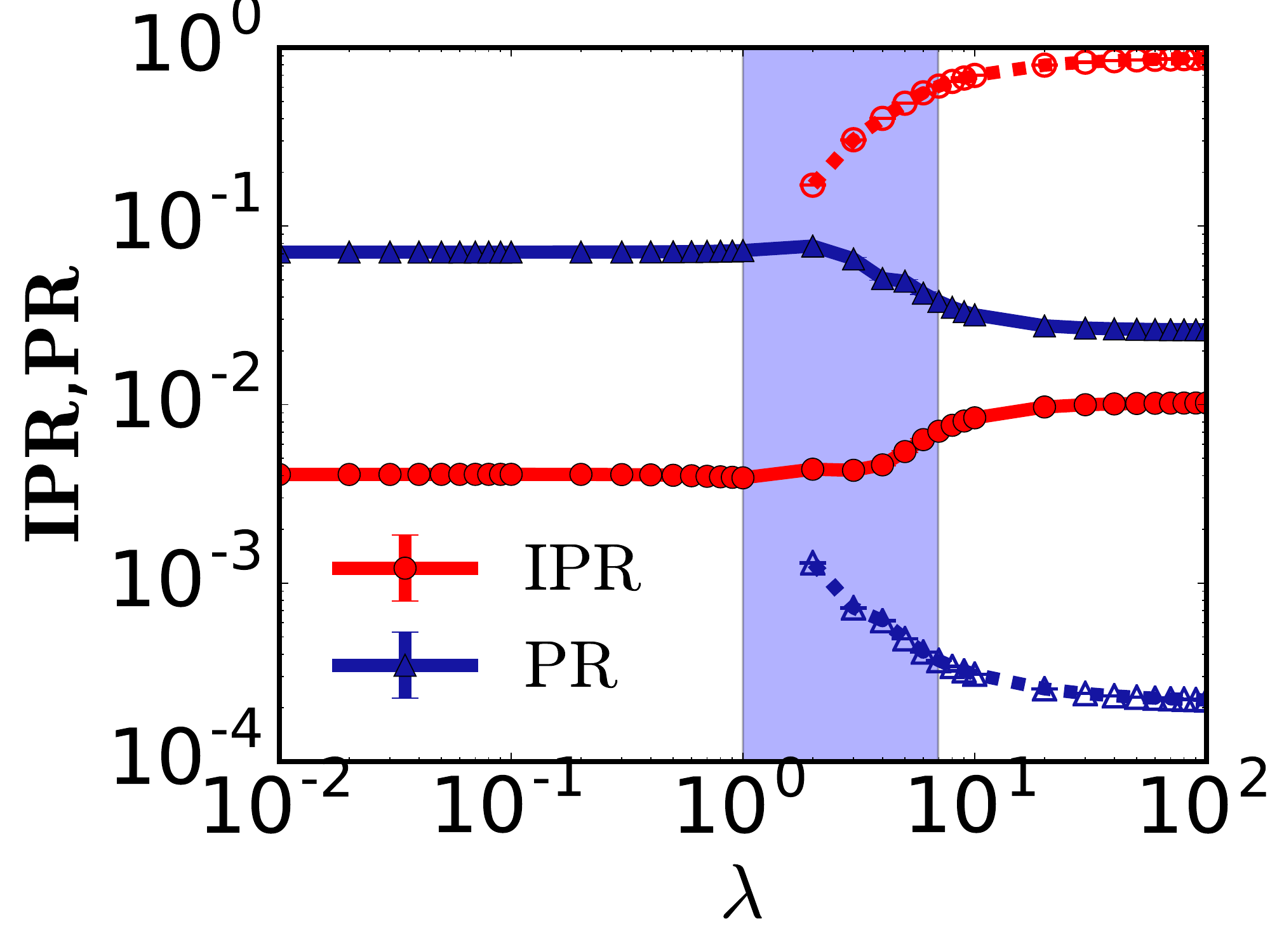}}\hspace{-0.1cm}
\stackunder{\hspace{-4.2cm}(b)}{\includegraphics[height=4.0cm, width=5.8cm]{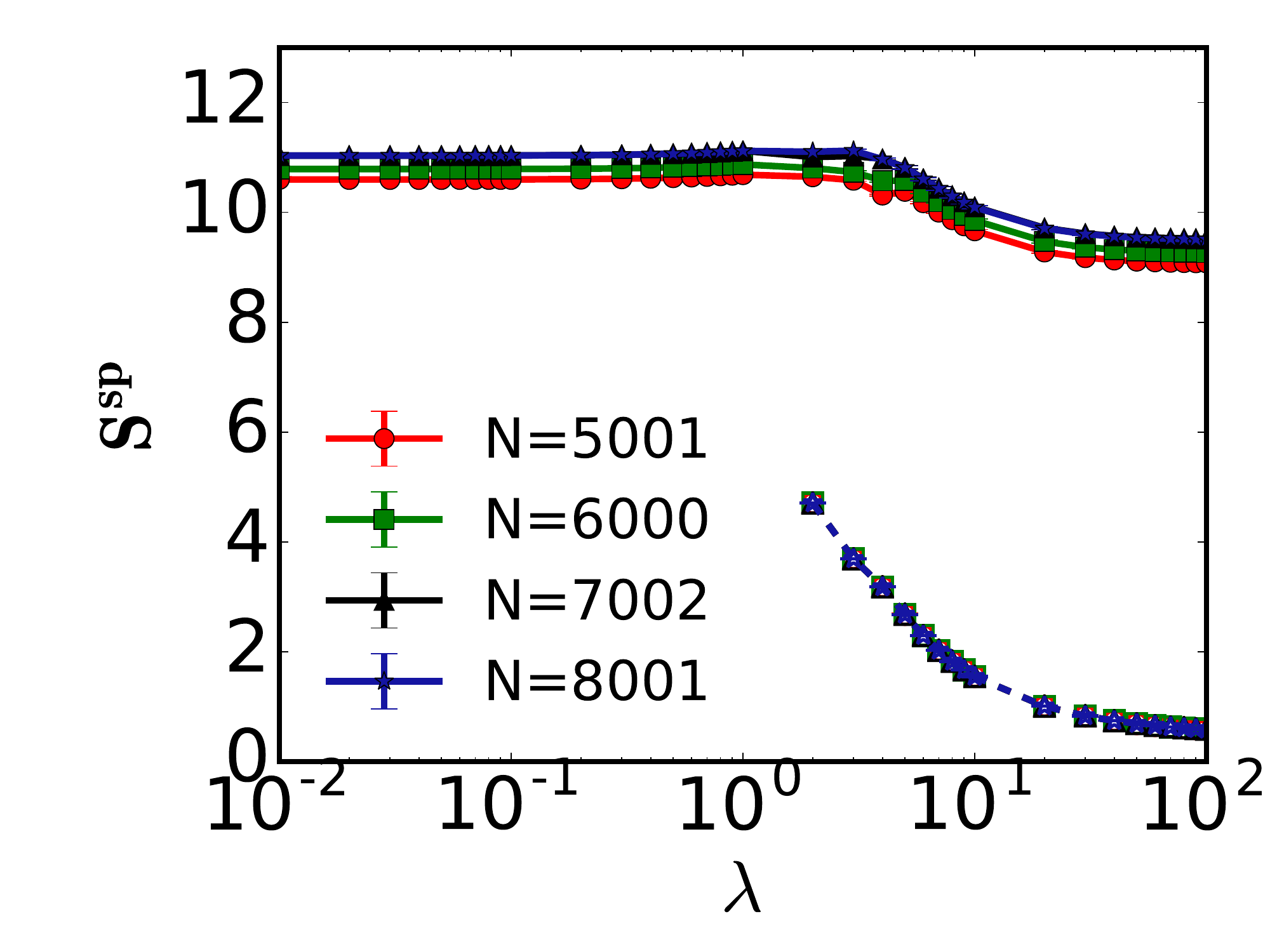}}
\stackunder{\hspace{-5cm}(c)}{\includegraphics[height=3.9cm, width=5.6cm]{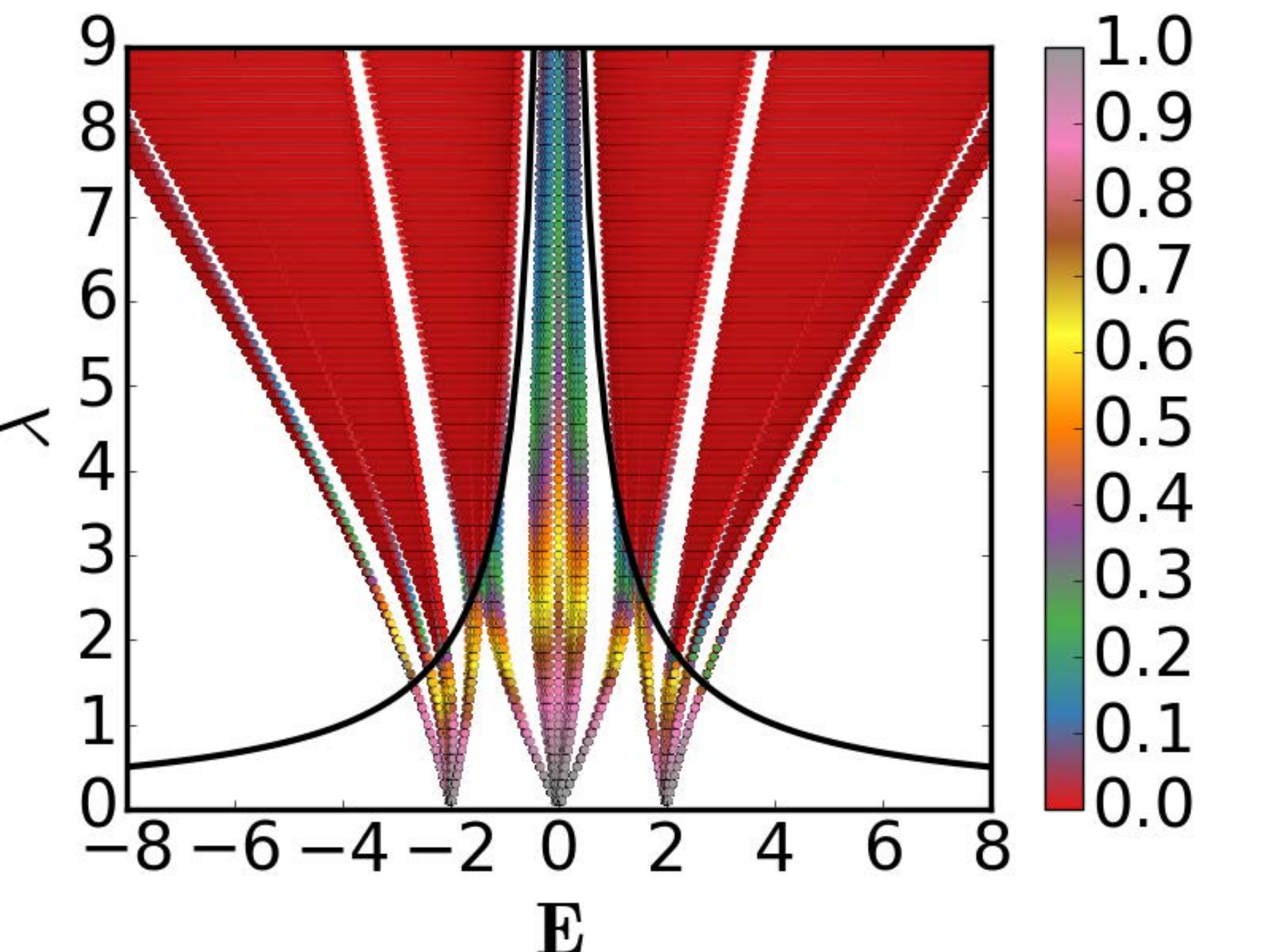}}
\caption{\label{fig9}\rev{In the antisymmetric case (a) IPR and PR
    averaged of all the single-particle eigenstates separated by the
    fractal mobility edge $\lambda = 4/|E|$ (solid lines with filled
    symbols for states in the inner region and dashed lines with open
    symbols for states in the outer region) with increasing strength
    of $AA$ potential $\lambda$ for $N=6000$ averaged over $50$ values
    of $\theta_p$. The shaded region signifies the transition of the
    eigenstates from extended to Anderson localization. (b) Single
    particle von Neumann entropy $S^\mathrm{sp}$ averaged over the
    eigenstates separated by the fractal mobility edge (solid lines
    with solid symbols for states in the inner region and dashed lines
    with open symbols for states in the outer region) with increasing
    strength of $AA$ potential $\lambda$ for various system
    sizes. Averaging has been done over $50$ values of $\theta_p$ for
    all cases. (c) The spectrum of the diamond lattice, where the
    color denotes the value of fidelity $F$ with respect to
    eigenstates at $\lambda=0.1$. Here the system size considered is
    $N=6000$. The black solid line in panel (c) shows the transition
    between multifractal and localized states, conjectured in a recent
    preprint~\cite{https://doi.org/10.48550/arxiv.2208.11930} using an
    analogy to the extended Harper problem (see Sec.~\ref{sec:level5}
    for more details).}}
\end{figure*}

\subsection{AA disorder}
The normalized participation ratio (PR)~\cite{PhysRevB.96.085119} is
closely related to the inverse participation ratio and is given by:
\begin{equation}
P_{k}=\left[N \sum_{n=1}^{\frac{N}{3}} \sum_{\alpha=u, c, d}\left|\psi_{k}( \rev{\alpha_n})\right|^{4}\right]^{-1}.
\end{equation}
It vanishes for a perfectly localized eigenstate and goes to unity for
a perfectly delocalized eigenstate. \rev{In Fig.~\ref{fig9}(a), we have
plotted the IPR, and the PR averaged over all the eigenstates in the inner part (shown with solid line) and the outer part (shown with dashed line) of the fractal mobility edge $\lambda = 4/|E|$~\cite{https://doi.org/10.48550/arxiv.2208.11930}, with
increasing strength of $\lambda$.} These two quantities together help
in the identification of the transition region. The transition from
the extended ($0<\text{PR}<1$) to the Anderson localized regime lies around
\rev{$\lambda\simeq 1.5$}. This is because, in the zero disorder limit, the
gaps between the flat bands are precisely of size $2$, and thus a
disorder strength of around \rev{$1.5-2$} allows an inter-band hybridization.

Another measure that provides an understanding of the extent of
localization in a system is the \rev{single-particle von Neumann entanglement}
entropy~\cite{PhysRevB.78.115114}. The von Neumann entropy associated
with site $\alpha$ of the $n^{th}$ unit cell in the $k^{th}$
eigenstate is given by~\cite{PhysRevB.77.014208}:
\rev{
\begin{equation}
\begin{split}
S_{k}^{\alpha_n}=&-\left|\psi_{k}(\alpha_n))\right|^{2} \log _{2}\left(\left|\psi_{k}(\alpha_n)\right|^{2}\right)\\
&-\left(1-\left|\psi_{k}(\alpha_n)\right|^{2}\right) \log _{2}\left(1-\left|\psi_{k}(\alpha_n)\right|^{2}\right).
\end{split}
\end{equation}}
For a delocalized eigenstate \rev{$\left|\psi_{k}( \alpha_n)\right|^{2}=1 /
N$} and hence \rev{$S_{k}^{\alpha_n} \approx \frac{1}{N} \log _{2}
N+\frac{1}{N}$} for large values of $N$ whereas for an eigenstate
localized on a single-site \rev{$S_{k}^{\alpha_n}=0$}. The contributions
from all sites for a particular eigenstate are given by
$S_{k}=\sum_{n,\alpha} S_{k}^{\alpha_n}$. Thus the average von
Neumann entropy over all the eigenstates is defined as:
\begin{equation}
S^{\mathrm{sp}}=\frac{\sum_{k=1}^{N} S_{k}}{N}.
\end{equation}
For large values of $N$, $S^{\mathrm{sp}} \approx\left(\log _{2}
N+1\right)$ in the delocalized phase whereas $S^{\mathrm{sp}} \approx
0$ in an extremely (single-site) localized phase.
\rev{Figure~\ref{fig9}(b) shows the single-particle entanglement
  entropy averaged over all the eigenstates in the inner part (shown
  with solid line) and the outer part (shown with dashed line) of the
  fractal mobility edge $\lambda =
  4/|E|$~\cite{https://doi.org/10.48550/arxiv.2208.11930}, with
  increasing strength of $\lambda$ for various system sizes. Here, we
  observe a system size dependence in $S^{\mathrm{sp}}$ for the states
  in the inner region with its magnitude being marginally less than
  its maximum value, which is a sign of the extended nature of the
  eigenstates. $S^{\mathrm{sp}}$ is largely system-size independent
  with its magnitude approaching $O(10^{-1})$, for the states outside
  the fractal mobility edge indicating Anderson localization.}

We have also plotted the fidelity or overlap between the eigenstates, which helps distinguish extended and localized regions in the spectrum. Fidelity between the $k^\text{th}$ eigenstates corresponding to two values of $\lambda$ is given by
\begin{equation}
F_{12}^{k}=\left|\left\langle\psi_{k}\left(\lambda_{1}\right) \mid \psi_{k}\left(\lambda_2\right)\right\rangle\right|^{2}.
\end{equation}
By choosing the first parameter $\lambda_1$ as the reference point,
the second parameter $\lambda_2$ is varied in Fig.~\ref{fig9}(c). We
have previously shown that all the states below $\lambda\approx 1.5$
have multifractal nature. We observe that for larger $\lambda$ the
magnitude of fidelity is close to zero for all the states (except the
central band) indicating localization. For the central band, for all
values of $\lambda$, the fidelity $0<F<1$. This indicates that though
the states are multifractal in the central band, the amplitude
distribution on the lattice sites does not remain fixed with
increasing strength of the potential $\lambda$.

\begin{figure*}
\centering
\stackunder{\hspace{-5cm}(a)}{\includegraphics[height=4.0cm, width=5.8cm]{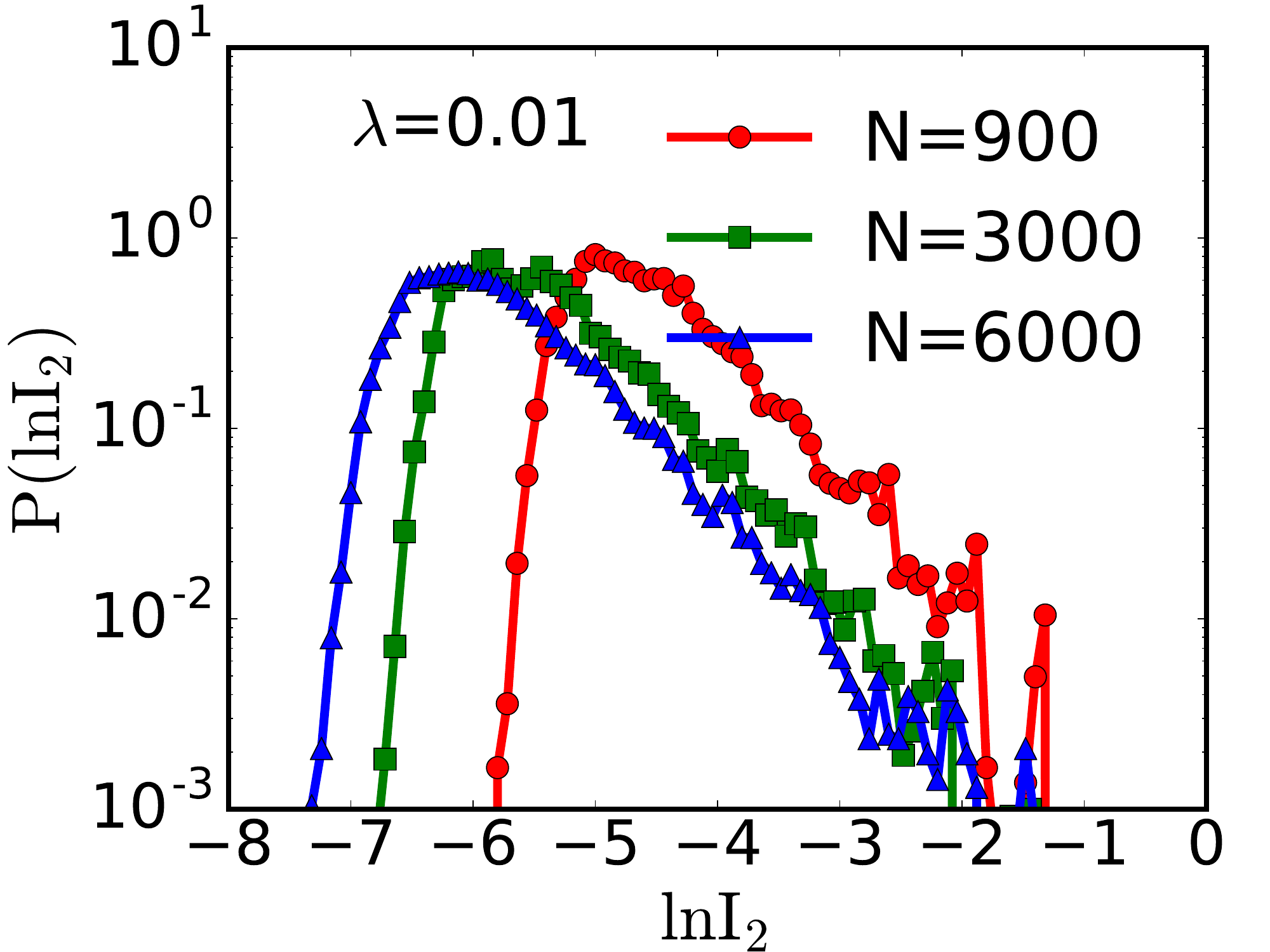}}\hspace{-0.1cm}
\stackunder{\hspace{-5cm}(b)}{\includegraphics[height=4.0cm, width=5.8cm]{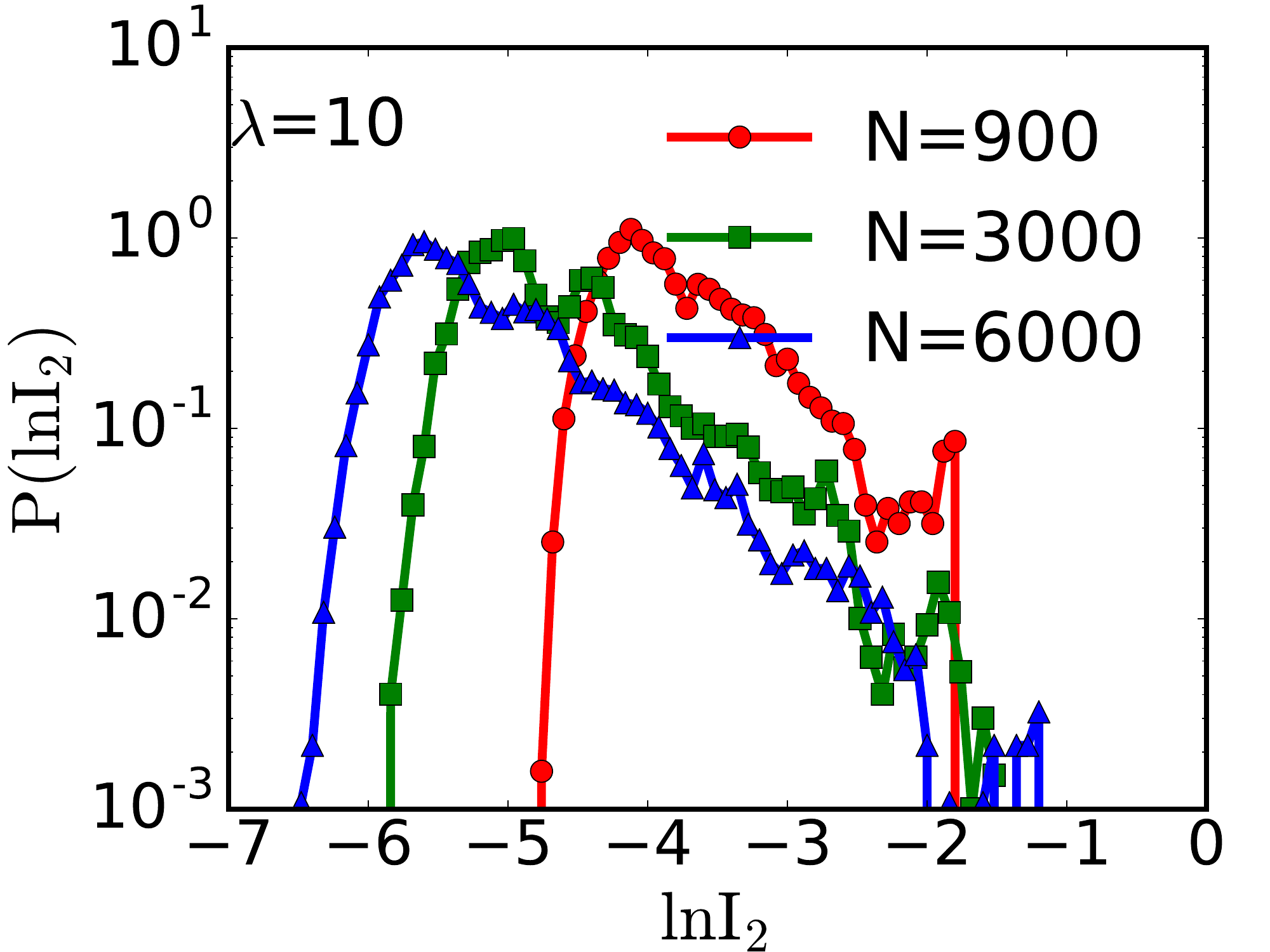}}\hspace{-0.1cm}
\stackunder{\hspace{-5cm}(c)}{\includegraphics[height=4.0cm, width=5.8cm]{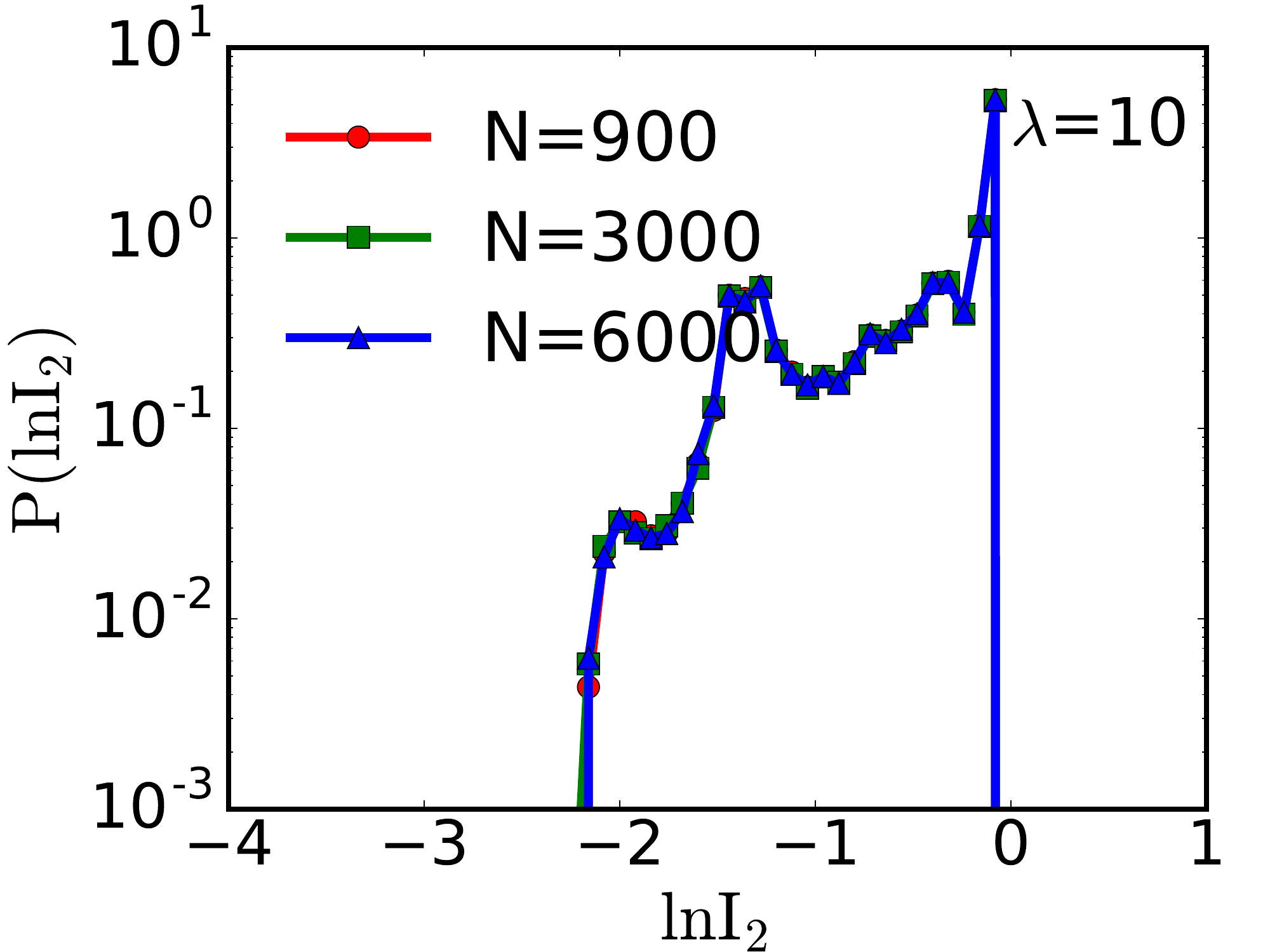}}

\vspace{-0.4cm}
\stackunder{\hspace{-5cm}(d)}{\includegraphics[height=4.0cm, width=5.8cm]{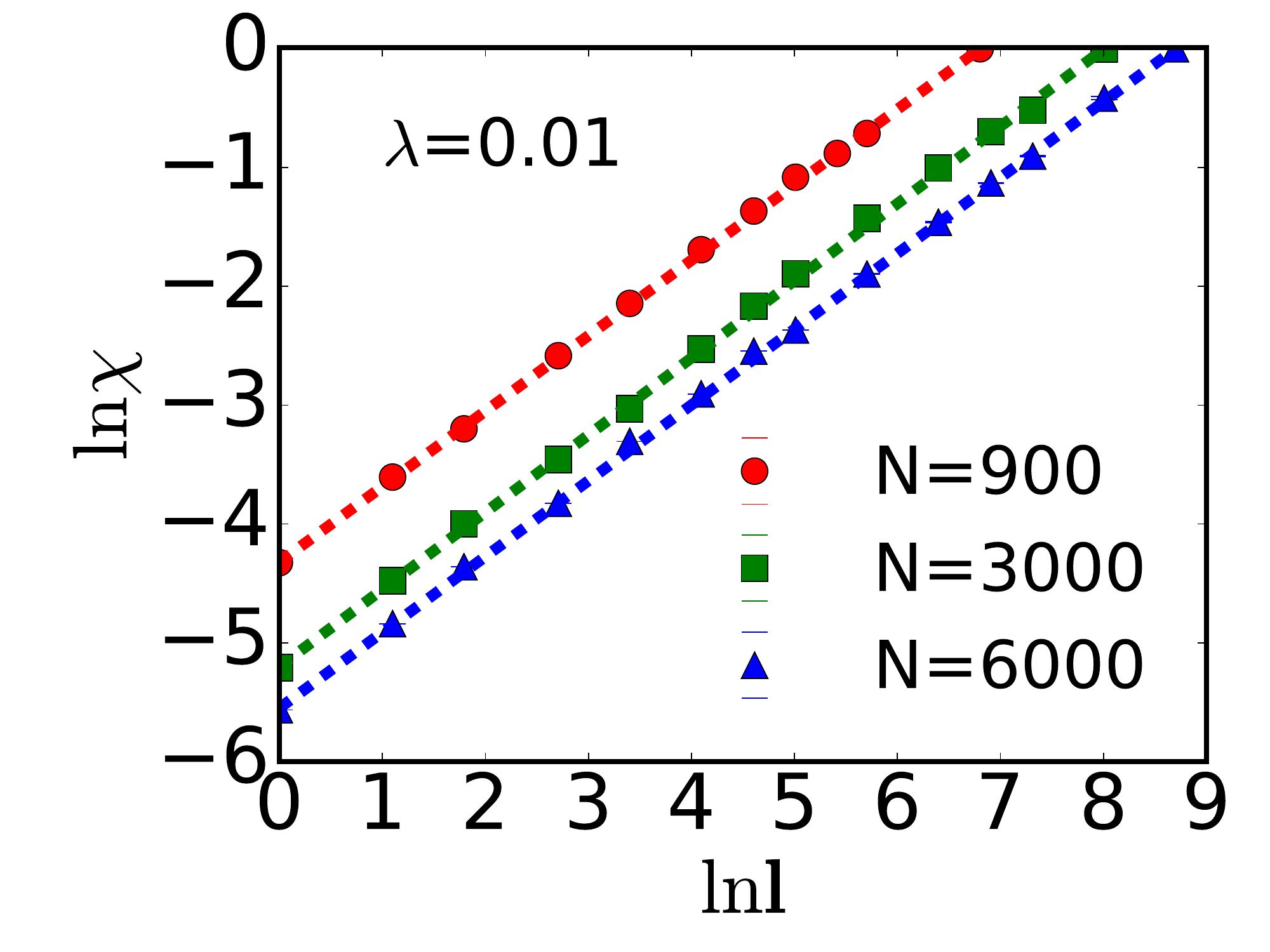}}\hspace{-0.1cm}
\stackunder{\hspace{-5cm}(e)}{\includegraphics[height=4.0cm, width=5.8cm]{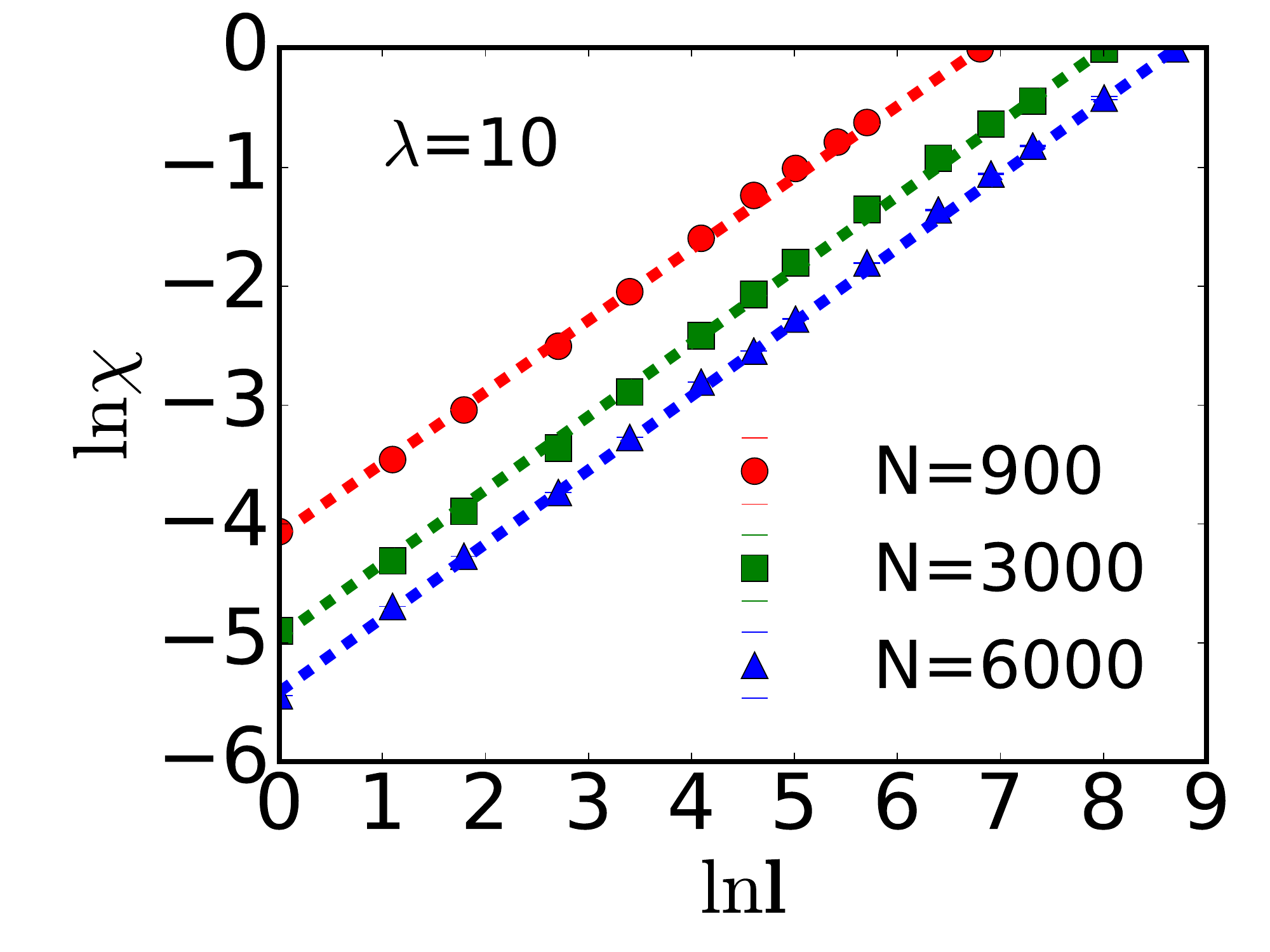}}\hspace{-0.1cm}
\stackunder{\hspace{-5cm}(f)}{\includegraphics[height=4.0cm, width=5.8cm]{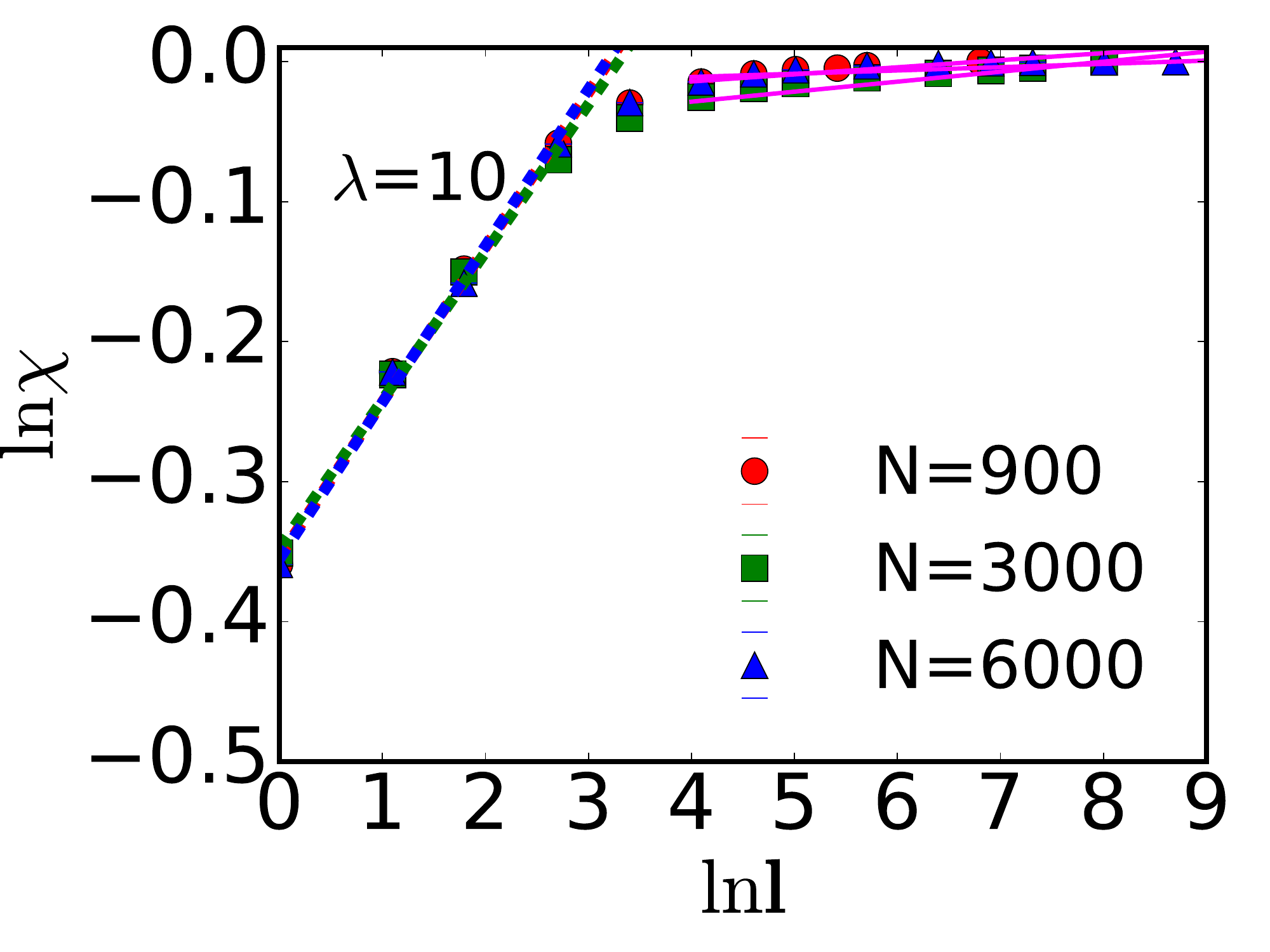}}
\caption{\label{fig10}\rev{In the anti-symmetric case, distribution of
    $P(\ln I_2)$ for different system sizes computed using (a)~all the
    eigenstates at $\lambda=0.01$ (b)~multifractal states comprising
    the inner region of the fractal mobility edge $\lambda = 4/|E|$,
    conjectured in~\cite{https://doi.org/10.48550/arxiv.2208.11930} at
    $\lambda=10$ and (c)~localized states that exist in the outer
    region of the fractal mobility edge at $\lambda=10$. The change of
    $\ln \chi$ as a function of $\ln l$ for different system sizes
    computed using (d)~all the eigenstates at $\lambda=0.01$
    (e)~multifractal eigenstates comprising the inner region of the
    fractal mobility edge $\lambda = 4/|E|$, at $\lambda=10$ and
    (f)~localized states that exists in the outer region of the
    fractal mobility edge at $\lambda=10$.}}
\end{figure*}

\rev{\subsubsection{Scaling Analysis} The scaling analysis of the
  distribution of the IPR logarithm $P(\text{ln} I_2)$ at the critical
  point shows invariance of shape or width with increasing system size
  $N$ ~\cite{PhysRevLett.84.3690}. After shifting the curves along the
  x-axis, they all lie on top of each other, forming a scale-invariant
  IPR distribution. We utilize this measure and analyze all the
  non-ergodic extended states at low $-\lambda=0.01$ (see
  Fig.~\ref{fig10}(a)) and those separated by the fractal mobility
  edge~\cite{https://doi.org/10.48550/arxiv.2208.11930}
  $\lambda_c=\frac{4}{|E|}$ at high $-\lambda=10$ (see
  Fig.~\ref{fig10}(b)), and observe that the distributions of IPR are
  indeed scale invariant. However for the states outside the
  fractality edge at a higher disorder strength (like $\lambda=10$),
  we observe that the distribution is independent of $N$, confirming
  that the states are localized (see Fig.~\ref{fig10}(c)).}

\rev{ We also perform a multifractal analysis of the wave functions
  using the box-counting method. For the AAH model, it was observed
  that the wave functions exhibit multifractal behaviour extending to
  all length scales at the critical point. For localized states,
  multifractal features are observed up to the localization length and
  for extended states they are observed up to the correlation
  length~\cite{Siebesma_1987}. We analyze the multifractal properties
  here by coarse-graining the system into boxes of length $l$. Given a
  normalized wave function $\left|\psi_{k}\right\rangle=$
  $\sum_{i=1}^N \psi_{k}(i)|i\rangle$ defined over a lattice of size
  $N$, we divide the lattice into $N / l$ segments of length
  $l$~\cite{Wang2016}:
\begin{equation}
\chi_{j}(q)=\sum_{p=1}^{N / l}\left[\sum_{i=(p-1) l+1}^{p l}|\psi_{k}(i)|^2\right]^{q}.
\end{equation}
The average is then considered over the total number of states in the
central band (CB) $j_\text{CB}$:
\begin{equation}
\chi(q)_\text{CB}=\frac{1}{j_\text{CB}} \sum_{j}^{j_\text{CB}} \chi_{j}(q),
\end{equation}
and the total number of states of the sidebands (SB) $j_\text{SB}$:
\begin{equation}
\chi(q)_\text{SB}=\frac{1}{j_\text{SB}} \sum_{j}^{j_\text{SB}} \chi_{j}(q).
\end{equation}
Multifractality is characterized by a power-law behavior of $\chi(q)
\sim(l / L)^{\tau(q)}$ with the exponent $\tau(q)$ determining the
multifractal dimension $D_{q}=\tau(q) /(q-1)$ where $q=2$ gives
$\tau(2)=D_{2}$. In Fig.~\ref{fig10}(d--f), we display the change of
$\ln \chi$ as a function of $\ln l$ for different system sizes. In the
critical regime, it is found that $\ln \chi$ is a linear function of
$\ln l$ described by a series of parallel lines for different $N$ with
the same slope $D_{2}$. We observe this behavior for the all states at
low$-\lambda=0.01$ (see Fig.~\ref{fig10}(d)) with $\rev{D_2}=0.63 \pm
0.005$ indicating that they are indeed non-ergodic extended. The same
is observed at high $-\lambda=10$, for the states comprising the inner
band of the fractal mobility
edge~\cite{https://doi.org/10.48550/arxiv.2208.11930}
$\lambda_c=\frac{4}{|E|}$ with $\rev{D_2}=0.61 \pm 0.005$ (see
Fig.~\ref{fig10}(e)). In the localized region, for lengths less than
the localization length $l<$ $l_{c}$, $\ln \chi$ is a linear function
of $\ln l$, which completely superposes together for different $N$
with an identical slope of $D_{2}$ - so one might naively conclude
that the system exhibits multifractal behaviour. However for $l>l_{c}$
the slopes decrease to $0$. The same can be observed for all the
states outside the fractal mobility edge at high $-\lambda$ (see
Fig.~\ref{fig10}(f)) with $D_{2} \approx 0.11$ (fitted with the dotted
curves) for $l<l_{c}$ while the slopes decrease to $0$ (fitted with
the magenta lines) for $l>l_{c}$, thus showing that in fact these
states are localized.}
\begin{figure*}
\centering
\stackunder{\hspace{-4.8cm}(a)}{\includegraphics[height=4.0cm, width=5.5cm]{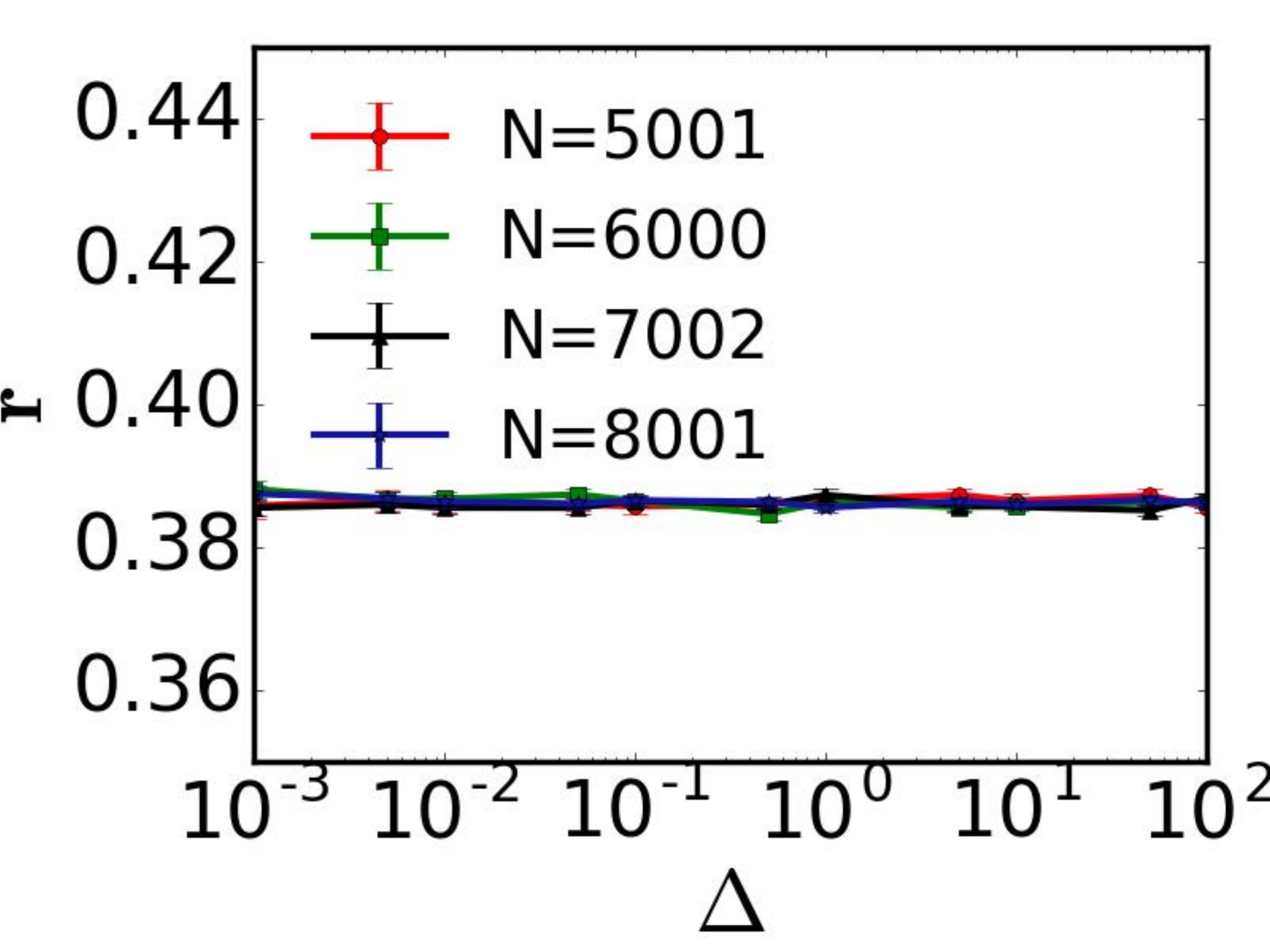}}\hspace{-0.1cm}
\stackunder{\hspace{-5.5cm}(b)}{\includegraphics[height=3.8cm, width=5.8cm]{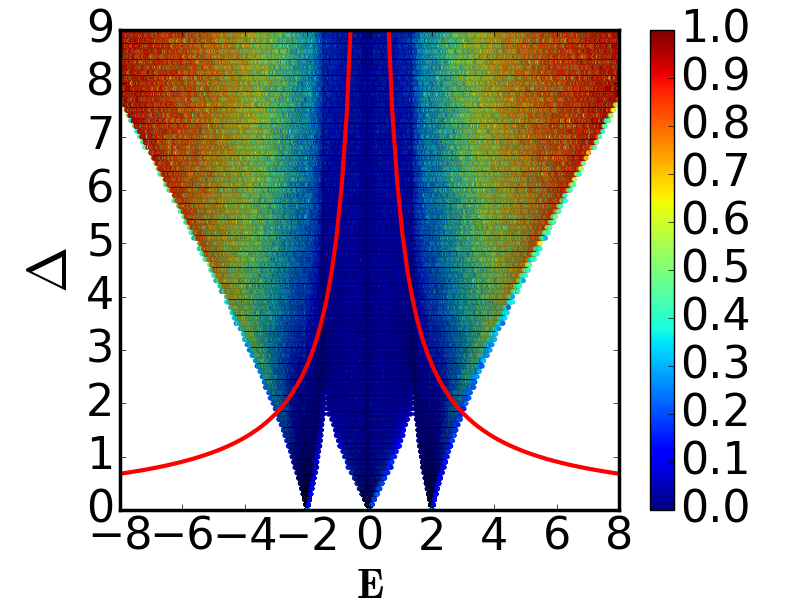}}\hspace{-0.1cm}
\stackunder{\hspace{-4.8cm}(c)}{\includegraphics[height=4.0cm, width=5.5cm]{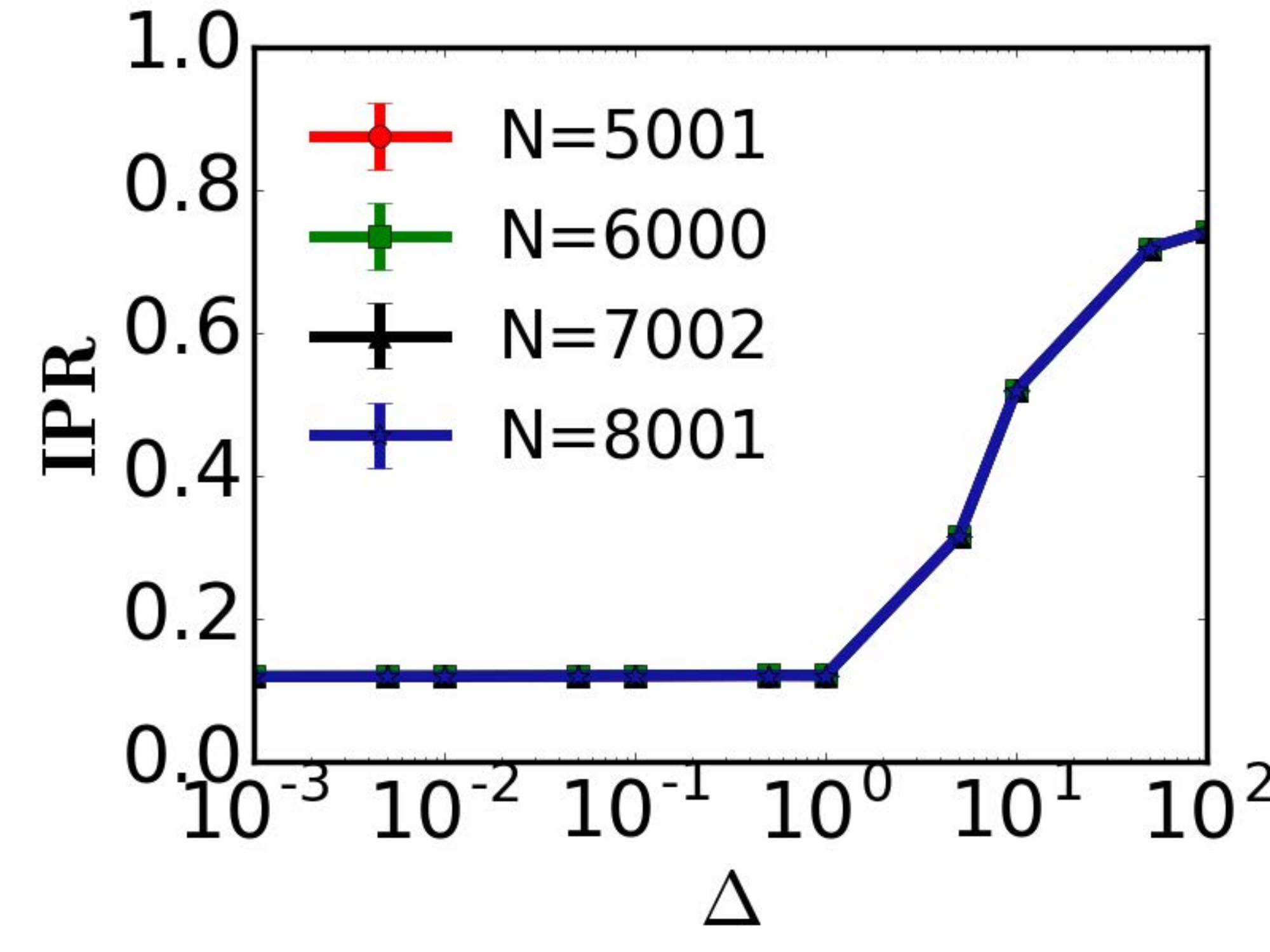}}
\vspace{-0.5cm}

\stackunder{\hspace{-4.8cm}(d)}{\includegraphics[height=4.0cm, width=5.5cm]{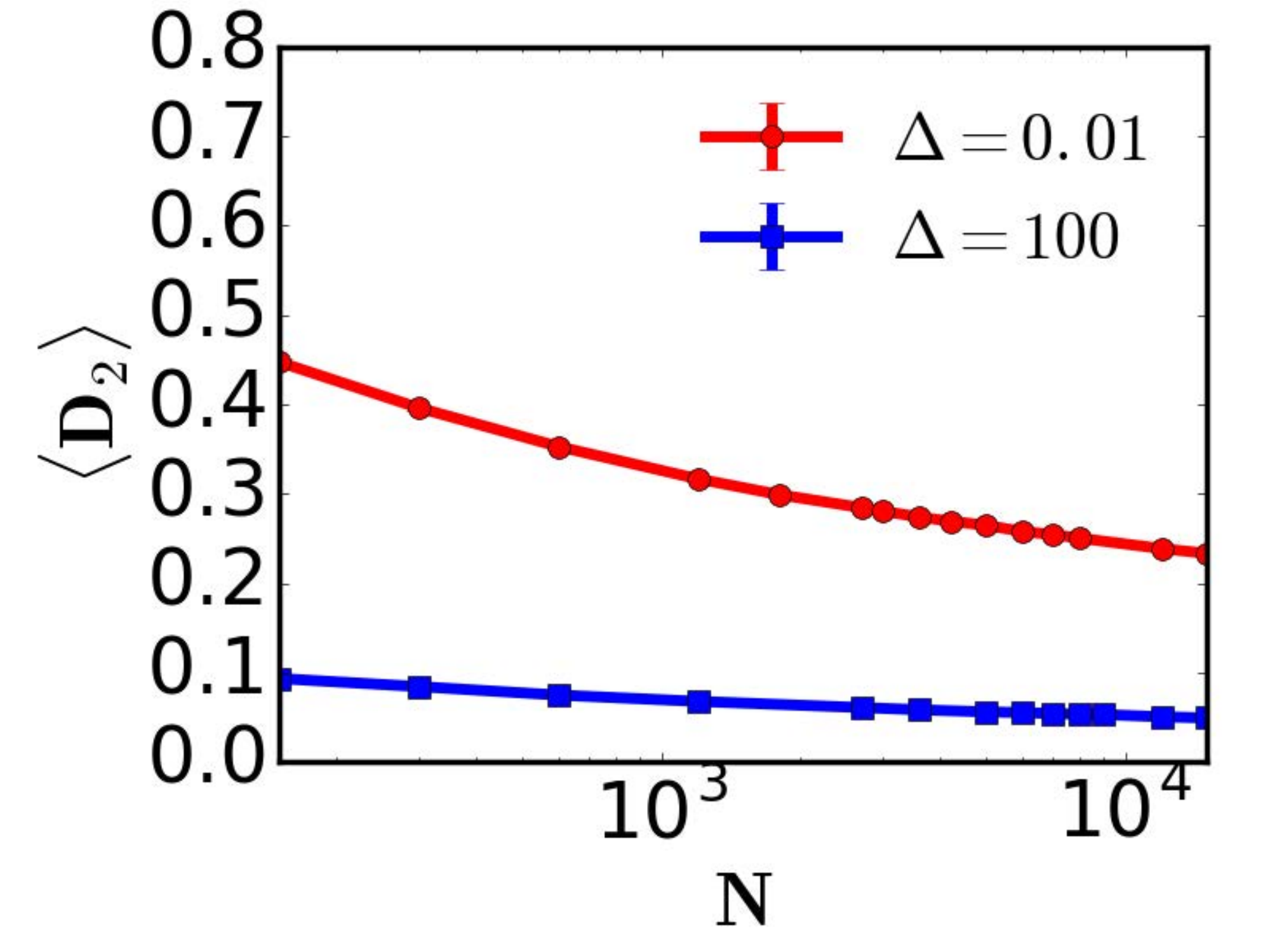}}\hspace{-0.1cm}
\stackunder{\hspace{-5.5cm}(e)}{\includegraphics[height=3.8cm, width=5.8cm]{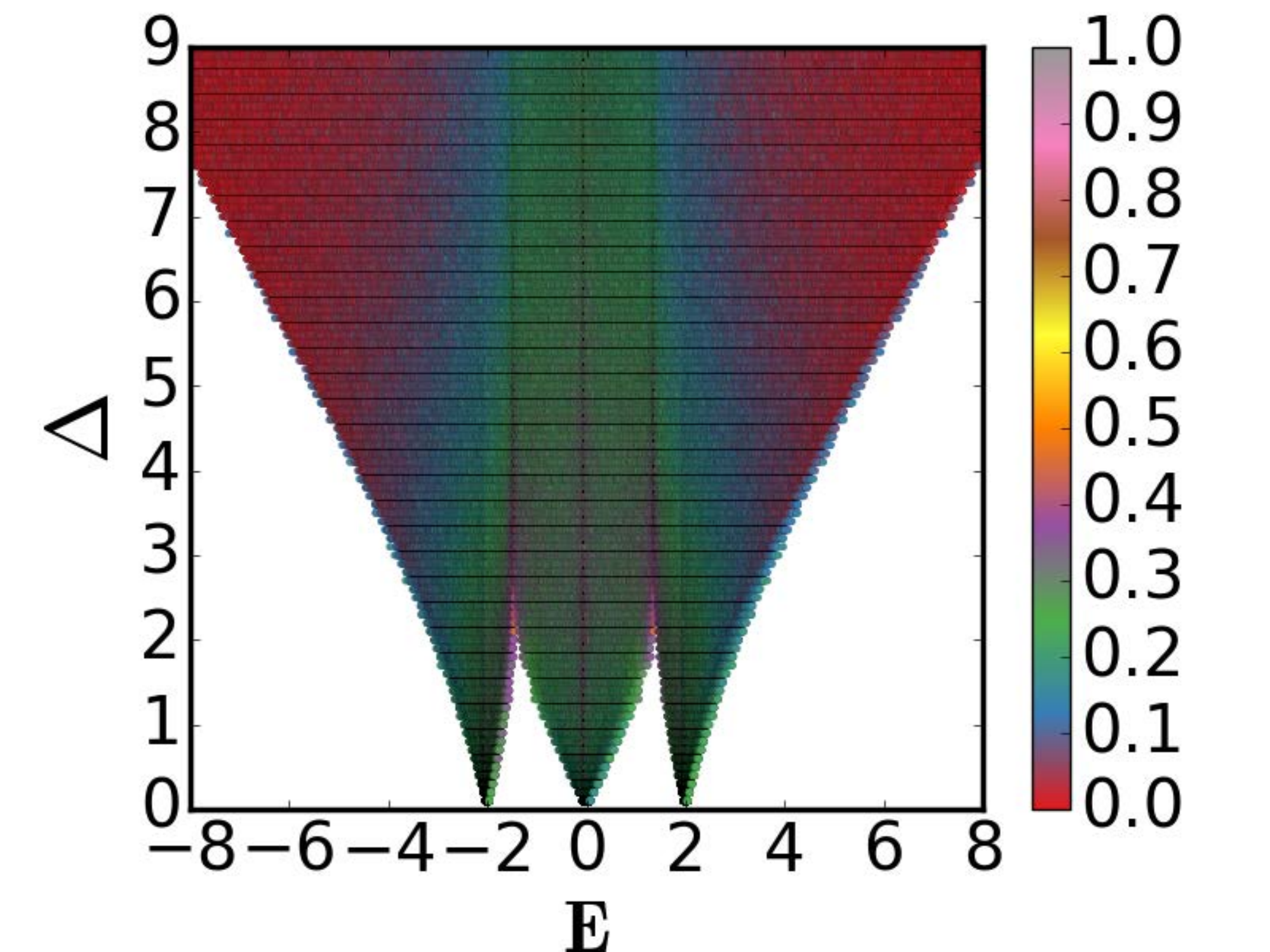}}\hspace{-0.1cm}
\stackunder{\hspace{-4.8cm}(f)}{\includegraphics[height=4.0cm, width=5.5cm]{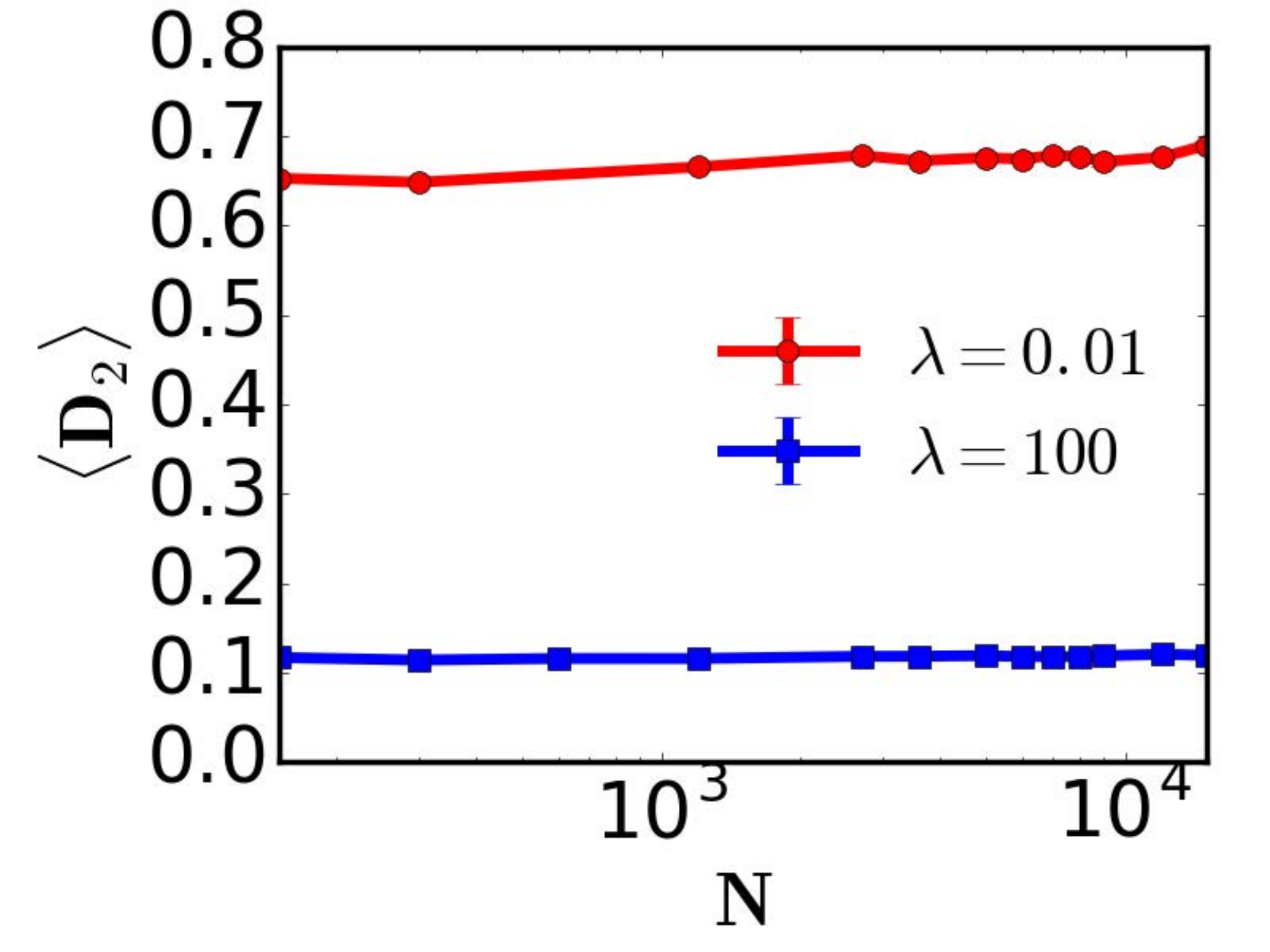}}
\caption{\label{fig11}In the anti-symmetric case (a) gap ratio
  $r$~\cite{PhysRevB.75.155111,PhysRevLett.110.084101} and (c) IPR
  averaged over all the eigenstates for uniform uncorrelated random
  disorder with increasing strength $\Delta$ for various system
  sizes. The averaging has been done over $50$ values of
  $\theta_p$. The spectrum of system size $N=6000$ with increasing
  strength of disorder $\Delta$, where the colour denotes the value of
  (b) IPR and (e) fractal dimension $D_2$ for all the single-particle
  eigenstates. $D_2$ averaged over all the eigenstates with increasing
  system size when applied disorder is (d) uniform uncorrelated random
  disorder and (f) quasiperiodic $AA$ disorder.}
\end{figure*}
\subsection{Uniform \rev{uncorrelated random} disorder}

We have also analyzed the application of the uniform \rev{uncorrelated
  random} disorder in an antisymmetric manner. Here $\Delta$ is the
disorder strength. We observe from Fig.~\ref{fig11}(a) that the
average gap ratio $r$~\cite{PhysRevB.75.155111,PhysRevLett.110.084101}
remains around the Poisson value ($\approx0.39$) at all strengths of
the disorder. Further from the energy-resolved IPR study (see
Fig.~\ref{fig11}(b)), it can be observed that all the eigenstates
exhibit low IPR below $\Delta\approx2$.  \rev{The presence of a
  mobility edge is also observed here.  Dividing
  Eq.~\eqref{eqn:1d_zeta^2} by $\Delta$ we get:
\begin{equation}
\frac{E}{\Delta}\left[E^2-4\right] c_n=w_n \frac{\Delta E}{2} c_n- t_n c_{n+1}- t_{n-1} c_{n-1},
\end{equation}
where $t_n = \zeta_n/\Delta$ are i.i.d. random numbers, homogeneously
distributed in a unit interval $|t_n|<1$, while $w_n =
4\zeta_n^2/\Delta^2$ are i.i.d. random numbers, whose distribution
$P(w_n) \sim \theta(1 - w_n)/(2\sqrt{w_n})$ is singular but integrable
due to the cut tail for $|w_n|>1$.  From this one can estimate the
finite-size mobility edge as the line where the localization length
$\xi(E)$, determining the eigenstate exponential decay
\begin{gather}
\label{eq:WF_exp-decay}
|\psi_E(r)| \sim e^{-|r-r_E|/\xi_E}
\end{gather}
with respect to the random energy-dependent maximum $r_E$, is of the
order of the system size $\xi(E) \simeq N$.}

\rev{
The expression for $\xi(E)$ can be estimated as (see, e.g.,~\cite{Shlyapnikov2007})
\begin{gather}
\label{eq:loc_length}
\xi_E \simeq F\left(\frac{E t_{typ}}{w_{typ}^2}\right) \frac{t_{typ}^2}{w_{typ}^2} \ ,
\end{gather}
where a smooth function $F(x)\simeq O(N^0)$~\rev{\cite{fn4}} can be
replaced by a constant as it changes by $10~\%$ from $x=0$ to
$x=\infty$ and the typical value $w_{typ}$ ($t_{typ}$) of the on-site
disorder $w_n$ (hopping $t_n$) is given by the typical value of the
distributions of $P(\ln w_n)$ ($P(\ln t_n)$).  In our case
\begin{gather}
w_{typ} =  e^{\langle \ln w_n\rangle} = e^{-2}, \quad t_{typ} = e^{\langle \ln t_n\rangle} =  e^{-1}\ ,
\end{gather}
i.e.,
\begin{gather}
\xi_E(N) \simeq \left(\frac{2 e}{\Delta |E|}\right)^2 F \ .
\end{gather}
Thus from $\xi_E\simeq N$ we get:
\begin{gather}
\Delta_c(E) = \frac{2 e}{E}\sqrt{\frac{F}{N}} \ .
\end{gather}
This demonstrates that the mobility edge, shown in Fig.~\ref{fig11}(b)
for $N=6000$ and $F=6000$ is a finite-size effect.  However, on the
other hand, it also shows why the state at exactly zero energy $E=0$
(which exists for odd $N$) will not localize. The latter is related to
the conserved chiral symmetry in the system, where the low energy
states may keep their delocalized nature even in the $1$d chiral
Anderson model.  In any case this regime deserves further detailed
investigations in future works.}

 From the IPR averaged over all the eigenstates (see
 Fig.~\ref{fig11}(c)), we observe that it is system size-independent,
 indicating localization. Thus while the averaged gap ratio and IPR
 suggest localization for the entire disorder range, the same is not
 observed from the energy-resolved IPR. The spectrum resolved fractal
 dimension $D_2$ is plotted in Fig.~\ref{fig11}(e). While the
 eigenstates below $\Delta\approx2$ and those belonging to the central
 band remain multifractal with $D_2\approx0.2$, they are less extended
 when the applied perturbation is $AA$ ($D_2\approx0.6$). Thus, we
 infer that the localization characteristics of the eigenstates in the
 low-disorder regime depends on the nature of the applied
 potential. On the other hand, when the strength of the disorder is
 sufficient, the different bands hybridize, conventional Anderson
 localization takes over, and the details of the form of the disorder
 are not important. Figure~\ref{fig11}(d) shows $D_2$ averaged over
 all the eigenstates when the uniform disorder is applied. The
 magnitude of $D_2$ decreases both in the low and high disorder regime
 with increasing system size, indicating a steady decline in the
 fraction of states exhibiting multifractality. In contrast, the
 fraction of states exhibiting the multifractal behaviour remains
 robust for the $AA$ potential Figure~\ref{fig11}(f). This suggests
 that the specific form of the $AA$ potential has an important role in
 ensuring multifractality.

\bibliography{dia}

\begin{thebibliography}{110}
\bibitem [{fn1()}]{fn1}%
  \BibitemOpen
  \href@noop {} {\ }\bibinfo {note} {\textcolor{black}{Here we note that the
  choice of periodic boundary conditions is not very important, as the
  quasiperiodic potential anyway has a jump at the boundary. {Unlike} some
  other models, we do not put the rational approximants ${M/N}$ of $b$ by
  taking certain system sizes ${N}$ from the above approximations}}\BibitemShut
  {NoStop}%
\bibitem [{fn4()}]{fn4}%
  \BibitemOpen
  \href@noop {} {\ }\bibinfo {note} {\textcolor{black}{In a standard {Anderson}
  model ${F} \simeq 100$, but we consider it as a constant fitting
  parameter.}}\BibitemShut {Stop}%
\end{thebibliography}

\end{document}